\documentclass[review,onefignum,onetabnum]{siamart251216}

\usepackage{amsmath}
\usepackage{mathtools}
\usepackage{physics}
\usepackage{subcaption}
\usepackage{cancel}
\usepackage{amssymb}

\newcommand{\ds}{\displaystyle}
\newcommand{\bs}{\boldsymbol}
\newcommand{\s}{\\}
\newcommand{\eps}{\varepsilon}
\newcommand{\mbf}{\mathbf}
\newcommand{\jac}{J_{\mbf u} \,}

\newcommand{\evs}[1]{\textcolor{red}{#1}}

\usepackage{lipsum}
\usepackage{amsfonts}
\usepackage{graphicx}
\usepackage{epstopdf}
\usepackage{algorithmic}
\ifpdf
\DeclareGraphicsExtensions{.pdf,.pdf,.png,.jpg}
\else
\DeclareGraphicsExtensions{.pdf}
\fi

\newsiamremark{remark}{Remark}
\newsiamremark{hypothesis}{Hypothesis}
\crefname{hypothesis}{Hypothesis}{Hypotheses}
\newsiamthm{claim}{Claim}

\newsiamthm{result}{Result}
\headers{Degenerate Turing and localised pattern}{E.~Villar-Sep\'{u}lveda and A.R.~Champneys} 

\title{Degenerate Turing bifurcation and the birth of localised patterns in activator-inhibitor systems\thanks{Submitted to SIAM 15th July 2022}
	\funding{This work was funded by ANID, Beca Chile Doctorado en el extranjero, \evs{number} 72210071.}}

\author{Edgardo Villar-Sep\'{u}lveda\thanks{Department of Engineering Mathematics, University of Bristol, Bristol, UK 
		(\email{rt20469@bristol.ac.uk}.}
	\and Alan Champneys\thanks{Department of Engineering Mathematics, University of Bristol, Bristol BS8 1TW, UK 
		(\email{a.r.champneys@bristol.ac.uk})}
}

\usepackage{amsopn}

\ifpdf
\hypersetup{
	pdftitle={Degenerate Turing bifurcation and the birth of localised patterns in activator-inhibitor systems},
	pdfauthor={E. Villar-Sep\'{u}lveda, A. Champneys}
}
\fi

\begin{document}
	
	\maketitle
	
	% REQUIRED
\begin{abstract}
    Precise conditions are provided for the existence and criticality of Turing bifurcations in a general class of activator-inhibitor reaction-diffusion equations on a one-dimensional infinite domain. The class includes generalised Schnakenberg and Brusselator models, as well as other models with cubic autocatalytic nonlinear terms. Previous numerical work suggests the existence of a bifurcation structure containing localised patterns due to the so-called homoclinic snaking mechanism. This paper provides explicit calculations to justify those results.
    
    Two distinct scalings of parameters that lead to tractable normal-form coefficients are considered in the limit that the diffusion ratio $\delta \to 0$. First, under a small-parameter scaling, the Turing bifurcation is shown to be always subcritical. Second, a large-parameter scaling reveals the Turing bifurcation to be supercritical, leading, by continuity, to the existence of a codimension-two degenerate bifurcation. The sign of a 5th-order normal form coefficient is also computed, which is shown to have the correct sign for the local birth of homoclinic snaking. 
    
    For the case of the Brusselator, two such codimension-two points can be found explicitly, as can the leading-order expression for the Maxwell point, in a parameter wedge about which localised patterns emerge. Numerical results are found to be consistent with the theory.
\end{abstract}
	
	% REQUIRED
	
	% REQUIRED
	
\section{Introduction}
    There has been a recent resurgence of interest in the application of Turing instability patterns within systems of partial differential equations modelling a wide variety of real-world phenomena, see e.g. \cite{Krause_review} and references therein. The basic form of the bifurcation, first described by Alan Turing 70 years ago \cite{Turing} leads to the onset of spatially periodic patterns as a result of a pitchfork bifurcation from a simple eigenvalue corresponding to a critical spatial wavenumber.
    
    As with any pitchfork bifurcation, a Turing instability can be either super- or sub-critical. Yet the onset of stable small amplitude spatially periodic patterns is a property of the supercritical case only. In contrast, the sub-critical case has been shown to be analogous to that of pattern formation bifurcation, as seen for example in the Swift-Hohenberg equation, see \cite{Kno} and references therein. Here there is a propensity to see localised patterns, which arise within the pinning region caused by the existence of a secondary re-stabilising fold of the bifurcating pattern state. Such patterns can be described by the theory of homoclinic snaking, see \cite{Woods_Intro} and references therein. That theory can be stated rigorously by appealing to so-called spatial dynamics where one poses the system on an infinite domain \cite{Beck}. Then the pattern-formation instability, also known as a Hamiltonian-Hopf bifurcation corresponds to the accumulation point of Turing bifurcations that would occur on a long finite domain \cite{Vec}.
    
    Recently, Al Saadi {\em et al.} \cite{FahadWoods} studied a class of activator-inhibitor models (summarised in Table \ref{tab:models} below) in one spatial dimension. In each, numerical evidence was found for a codimension-two Turing bifurcation point at which the criticality of the bifurcation switches as a second parameter is varied. These points were found to act as the organising centres for wedges in \evs{the} parameter space in which there are stable localised patterns, which can become further destabilised by Hopf bifurcations, and by crossing a so-called Belyakov-Devaney line, can transition into isolated spikes as described by semi-strong asymptotic analysis, \evs{see \cite{Gai}}. Similar behaviour has been found in other reaction-diffusion models, see \cite{FahadWoods,MeronIssue,DeWitt} and references therein.
    
    The normal form of a pattern-formation bifurcation on an infinite domain has been studied by several authors, see e.g.~\cite{Burke_NF,Nico,FahadWoods}, references therein and Sec.~\ref{sec:3} below. The computation of the third-order coefficient determines the criticality of the bifurcation. The main thrust of this paper is to study the codimension-two bifurcation where this coefficient vanishes upon \evs{the} variation of a second parameter. We shall call such points {\em degenerate} Turing bifurcations.
    
    If the fifth-order coefficient at the degenerate bifurcation \evs{is negative}, then one finds a curve emanating into the subcritical region in which a heteroclinic connection occurs within the normal form between the homogeneous steady state and the stable post-fold periodic state \cite{Woods}. A generic argument, made precise in \cite{Beck}, shows that such a connection will break up into a heteroclinic tangle in the presence of the terms that break the artificial symmetry of the normal form.  Within this tangle, one finds localised patterns organised into the characteristic homoclinic snake.
    
    Note that the case of the other sign of the fifth-order coefficient is of less physical relevance in pattern formation problems, but has been found in models of nonlinear waves, and leads to the existence of large-amplitude solitary waves with algebraically decaying tails \cite{IoossDias}.
    
    Terms that break the normal-form symmetry typically arise beyond all algebraic orders. Thus, the precise computation of the width of the snaking region is a delicate task. Particular calculations have been carried out for Swift-Hohenberg-like models, starting with the ground-breaking work of Kozyreff \& Chapman \cite{Kozyreff,Kozyreff2}; see also \cite{Dean,DeWitt}. Nevertheless, from the point of view of normal form theory, the break up of a non-structurally stable heteroclinic cycle into a heteroclinic tangle is the only ingredient required to prove the existence of a homoclinic tangle once the coefficients of the normal form have been shown to be of the correct signs. Such a break-up is generic under any perturbation that does not possess the normal form symmetry.
    
    The question remains though as to why so many reaction-diffusion systems, particularly many well-studied activator-inhibitor models, seem to possess degenerate Turing bifurcations, leading to the parameter wedge of localised patterns.
    
    The rest of the paper is outlined as follows. In Sec.~\ref{sec:main} we present the general model to be studied and give relevant results about its homogeneous equilibrium and conditions for it to go through a Turing bifurcation. Sec.~\ref{sec:3} develops a general theory for computing the coefficients of the normal form of an activator-inhibitor model undergoing a Turing bifurcation. Sections \ref{sec:4} and \ref{sec:5} consider two different asymptotic scaling of reaction parameters in the limit that the diffusion ratio $\delta$ between activator and inhibitor is small.  \evs{Specifically, Sec.~\ref{sec:4} shows that, in a limit when the constant feed rates in the reaction kinetics are both $\mathcal{O}(\delta)$, the Turing bifurcation is always} \evs{sub-critical}. In contrast, Sec.~\ref{sec:5} \evs{considers a limit where the feed and decay rates of the inhibitor are $\mathcal{O}(\delta^{-1})$ which is shown to lead to a transition point between sub- and super-critical Turing bifurcations for appropriate values of other parameters.} Section \ref{sec:6} illustrates the theory with some numerical results, and Sec.~\ref{sec:7} treats the special case of the Brusselator model, \evs{where} two degenerate Turing bifurcations can be found explicitly. Sec.~\ref{sec:8} draws conclusions and points to future work.

\section{Preliminaries} \label{sec:main}
    This paper concerns a class of 2-component reaction-diffusion equations posed on the real line that can be written in the form \evs{
    \begin{equation}
        \mbf u_t = \mbf f(\mbf u) + \mathbb D \, \mbf u_{xx}, \qquad \mbox{for } \mbf u = (u(t, x), v(t, x))^T \in {\mathbb R}^2 \label{vectorfield}
    \end{equation}
    for all $(t, x)\in {\mathbb R}^+_0\times \mathbb R$, $0 < \delta \ll 1$, where
    \begin{align*}
        \mbf f\begin{pmatrix}
            u
            \\
            v
        \end{pmatrix} = \begin{pmatrix}
        	a
        	\\
        	b
        \end{pmatrix} + \begin{pmatrix}
        	-c & d
        	\\
        	h & -d
        \end{pmatrix} \begin{pmatrix}
            u
            \\
            v
        \end{pmatrix} + \begin{pmatrix}
            1
            \\
            -1
        \end{pmatrix} u^2 \, v, \quad \mbox{and } \mathbb D = \begin{pmatrix}
            \delta^2 & 0
            \\
            0 & 1
        \end{pmatrix}
    \end{align*} }
    %  
    %  
    %  =  \mbf{c}+A \begin{pmatrix}
    %		u
    %		\\
    %		v
    %	\end{pmatrix}+\begin{pmatrix}
    %		1
    %		\\
    %		-1
    %	\end{pmatrix}u^2v+\begin{pmatrix}
    %		\delta^2 \nabla^2 u
    %		\\
    %		\nabla^2 v
    %	\end{pmatrix}, \quad {where } 
    %	\mbf c=\begin{pmatrix}
    %		a
    %		\\
    %		b
    %	\end{pmatrix}, \quad A=\begin{pmatrix}
    %		-c & d
    %		\\
    %		h & -d
    %	\end{pmatrix}
    %       \quad \mbox{and }
    %\mathbb D=\begin{pmatrix} \delta^2 & 0 \\ 0 & 1  \end{pmatrix}      
    %\end{align*}
    \evs{where $\delta>0$ and} all other parameters $a, b, c, d, h$ are assumed to be non-negative. As we will be interested in the onset of localised patterns, we shall ignore the effect of boundary conditions, assuming only that the {\em derivatives} $u_x$ and $v_x$ should tend to zero as $x \to \pm \infty$, so that the solution in the far field is assumed to approach a spatially homogeneous steady state. In practice, we shall approximate the infinite domain with a large finite domain subject to homogeneous Neumann boundary conditions on $u$ and $v$.
    
    Note that, under different choices of parameters, \eqref{vectorfield} contains several different known models that have been studied previously, see Table \ref{tab:models}.
    \begin{table}[ht]
        \centering
        \begin{tabular}{|c|c|c|}
            \hline
            \textbf{Author/Model name} & \textbf{Fixed parameters} & \textbf{Free parameters}
            \\
            \hline
            Schnakenberg \cite{Schnakenberg1979} & $c=1$, $d=h=0$, & $a,b$.
            \\
            \hline
            Brusselator \cite{Prigogine1968} & $h=c-1>0$, $b=d=0$ & $a,c$.
            \\
            \hline
            Glycolysis \cite{Tyson1975} & $c=1$, $a=h=0$ & $b,d$.
            \\
            \hline
            Sel'kov Schnakenberg \cite{Selkov1968} & $c=1$, $h=0$ & $a,b,d$.
            \\
            \hline
            Root-hair \cite{Payne2009} & $a=0$, $c>h>0$ & $b,c,d,h$.
            \\
            \hline
        \end{tabular}
        \caption{Well-known reaction-diffusion models that have the form \eqref{vectorfield}, with their corresponding parameter values (after \cite{FahadWoods})}
        \label{tab:models}
    \end{table}
    Note that the choice of the signs \evs{of the coefficients of the linear kinetic terms and that the coefficient of $v$ in the first equation is the negative of that of $u$ in the second equation}. These restrictions are sufficient to ensure that there is a unique homogeneous steady state (equilibrium point) of \eqref{vectorfield} given by 
    \begin{align}
        \mbf P = \left(u^*, v^*\right) = \left(\frac{a + b}{c - h}, \frac{(c - h)(b \, c + a \, h)}{(a + b)^2 + d \, (c - h)^2}\right). \label{eq:Pdef}
    \end{align}
    Furthermore, we shall henceforth assume that 
    \begin{align}
        c > h, \label{eq:1stquad}
    \end{align}
    which ensures that the equilibrium point is located in the first quadrant. A notable exception from the list in Table \ref{tab:models} is the Gray-Scott system, which can be written in the form \eqref{vectorfield}, but \evs{without the restriction on the final column of the matrix of coefficients of linear kinetic terms.}
    $a_{2,2} \neq -a_{2,1}$. This results in there being three homogeneous equilibria for most parameter values, which leads to completely different kinds of bifurcation diagrams (see \cite{FahadGrayScott} and references therein).
    
    Models of the \evs{form} \eqref{vectorfield} subject to \eqref{eq:1stquad} were studied extensively in \cite{FahadWoods} using numerical and asymptotic methods, and a certain ubiquitous two-parameter bifurcation structure of localised patterns was observed. The birth point of the localised patterns is a point at which the fundamental pattern-formation or Turing bifurcation changes from being \evs{super-} to sub-critical. Furthermore, a distinction can be drawn between parameter regions where there are stable localised patterns (localised modulations of a pattern with an internal wavelength) and those in which there are stable isolated spikes. The division between these regions can be described by a so-called Belyakov-Devaney (BD) transition in which there is a transition between complex spatial eigenvalues and real ones (see \cite{Nico2} for an explanation and an asymptotic unfolding).
    
    But the question that remains is why there should be \evs{this} kind of degenerate Turing bifurcation. The analytic answer to this question forms the subject of this paper.
    
    \subsection{Necessary \evs{conditions} for Turing instability}
    
    From the standard theory of Turing bifurcations, see \cite{Krause_review,Murray2001,Satnoianu2000,Hoang2013}, we know that there are two conditions that must be fulfilled in order to ensure the occurrence of a Turing bifurcation.
    
    \begin{description}
        \item[T1:] $\mbf P$ must be a stable equilibrium point of \eqref{vectorfield} in the absence of diffusion. 
        \item[T2:] There exists a positive wavenumber $k^*>0$ such that $\jac \mbf f(\mbf P) - (k^*)^2\mathbb D$ has a zero eigenvalue $\lambda(k^*)=0$, such that
        \begin{equation}
            \left.\dfrac{\dd \lambda}{\dd k}\right|_{(k, \lambda) = \left(k^*, 0\right)}=0, \qquad \evs{\left.\dfrac{\dd^2 \lambda}{\dd k^2}\right|_{(k, \lambda) = \left(k^*, 0\right)} < 0.} \label{secondcondition}
        \end{equation}
    \end{description}
    We start with general conditions under which T1 is satisfied.
    
    \begin{result}[Linear stability of $\mbf P$]\label{thm:bigthm}
        Under the condition \eqref{eq:1stquad}, $\mbf P$, given by \eqref{eq:Pdef} is a \evs{linearly} stable equilibrium point of the system \eqref{vectorfield} in the absence of diffusion if and only if
        \begin{align}
            \frac{2 \, (a + b)(a \, h + b \, c)}{(a + b)^2 + d \, (c - h)^2}< \frac{(a + b)^2 + d \, (c - h)^2}{(c - h)^2} + c. \label{turingcond}
        \end{align}
    \end{result}
    \begin{proof}
        Note that
        \begin{align}
            \jac \mbf f(\mbf P) &= \begin{pmatrix}
                \frac{2 \, (a + b) (a \, h + b \, c)}{(a + b)^2 + d \, (c - h)^2} - c & \frac{(a + b)^2}{(c - h)^2} + d
                \\
                h - \frac{2 \, (a + b) (a \, h + b \, c)}{(a + b)^2 + d \, (c - h)^2} & -\frac{(a + b)^2}{(c - h)^2} - d
            \end{pmatrix}, \label{eq:DFP}
        \end{align}
        then
        \begin{align*}
            \tr(\jac \mbf f(\mbf P))&=\frac{2 \, (a + b) (a \, h + b \,  c)}{(a + b)^2 + d \, (c - h)^2} - \frac{(a + b)^2 + d \, (c - h)^2}{(c - h)^2} - c,
            \\ 
            \det(\jac \mbf f(\mbf P)) &= \frac{(a + b)^2}{c - h} + d \, (c - h)>0.
        \end{align*}
        Therefore, as the eigenvalues of a $2\times 2$ matrix are given by
        \begin{align*}
            \lambda_\pm = \frac{\tr(\jac \mbf f(\mbf P))\pm \sqrt{\left(\tr(\jac \mbf f(\mbf P))\right)^2 - 4 \, \det\left(\jac \mbf f(\mbf P)\right)}}{2},
        \end{align*}
        then a necessary and sufficient condition for these eigenvalues to have a negative real part is that $\tr(\jac \mbf f(\mbf P)) < 0$ and $\det(\jac \mbf f(\mbf P)) > 0$. Therefore, as $\det(\jac \mbf f(\mbf P)) > 0$ for all the parameter values we are taking into account, then $\mbf P$ is a stable equilibrium point in the absence of diffusion if and only if $\tr(\jac \mbf f(\mbf P)) < 0$, which \evs{is equivalent to}
        \begin{align*}
            \frac{2 \, (a + b) (a \, h + b \, c)}{(a + b)^2 + d \, (c - h)^2} < \frac{(a + b)^2 + d \, (c - h)^2}{(c - h)^2} + c,
        \end{align*}
        \evs{concluding} the proof.
    \end{proof}
    
    We next give conditions under which T2 is satisfied
    \begin{result}[Codimension-one Turing bifurcation]\label{thm:bigthm2}
        Assume the hypotheses of \ref{thm:bigthm}, and the further condition 
        \begin{align}
            \frac{2 \, (a + b)(a \, h + b \, c)}{(a + b)^2 + d \, (c - h)^2} > c. \label{positivesubdet}
        \end{align}
        Then there exists $\delta > 0$ such that there is a Turing bifurcation at $\mbf P$ corresponding to wavenumber $k = k^*$ that satisfies
        \begin{equation}
            k^4 = \frac{(a + b)^2 + d \, (c - h)^2}{\delta^2 (c - h)}. \label{kfourth}
        \end{equation}
        Furthermore, for each $0 < \delta < 1$, a Turing bifurcation occurs on the codimension-one surface in the $(a,b,c,d,h)$-parameter space, given by
        \begin{equation}
            \frac{2 \, (a + b) (a \, h + b \, c)}{(a + b)^2 + d \, (c - h)^2} - \delta^2 \left(\frac{(a + b)^2}{(c - h)^2} + d\right) - c = 2 \, \delta \, \sqrt{\frac{(a + b)^2 + d \, (c - h)^2}{c - h}}. \label{eq:TuringCond}
        \end{equation}
     \end{result}
    
\begin{figure}
        \centering
        (a) \hspace{6cm} (b)
        \\
        \includegraphics[width = 0.45\textwidth]{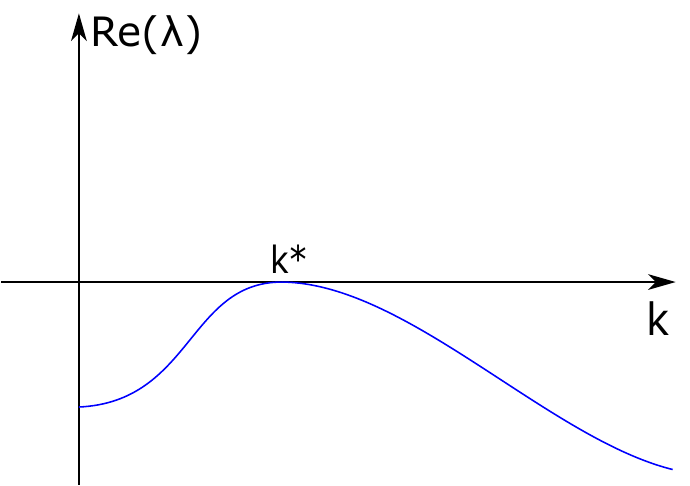}
        \includegraphics[width = 0.45\textwidth]{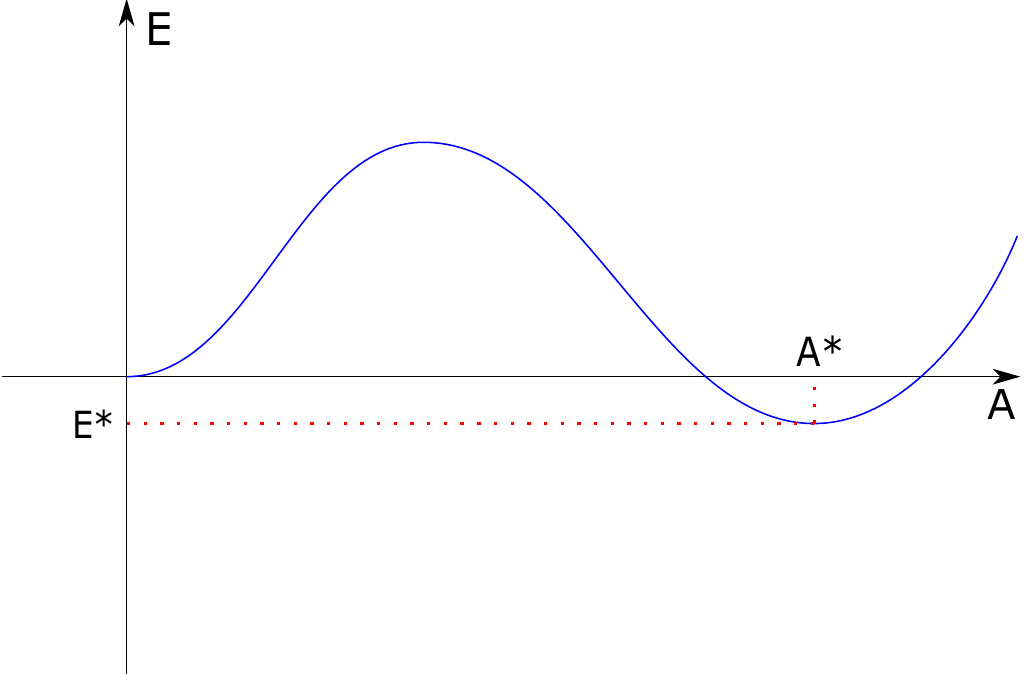}
          
        \caption{(a) Dispersion relation at a Turing bifurcation. The graph shows the largest eigenvalue $\lambda(k)$ of $\jac \mbf f(\mbf P) - k^2 \, \mathbb D$ that touches the $k$-axis for some value $k^*$. (b) Graph of \evs{$E(A)$} given by \eqref{eq:Ez}, \evs{for $\varepsilon = 0$, $C_3 > 0$, and $C_5 < 0$,} showing minima at \evs{$A = 0$} and another positive value \evs{$A = A^*$}. The Maxwell point occurs when \evs{$E(A^*) = 0$}, where the two minima have the same energy.}
        \label{fig:dispersionrelatbif}
    \end{figure}

    \begin{proof}
        From the studies on Turing bifurcations in general systems, \cite{Satnoianu2000,Hoang2013,Murray2001,General_Conditions}, a necessary \evs{condition} in order for there to be a Turing bifurcation upon adding diffusion to a system with a stable equilibrium point $\mbf P$, is that the Jacobian matrix $\jac \mbf f(\mbf{P})$ should have an unstable principal submatrix. Looking at the form of $\jac \mbf f(\mbf{P})$ in \eqref{eq:DFP} we see that the bottom right component is negative, so the only condition that can hold in this case is that the top left component of that matrix is positive. Thus, we arrive at the condition \eqref{positivesubdet} in order to ensure the existence of $0<\delta<1$ such that the conditions for having a Turing bifurcation at $\mbf P$ hold.
    
        Next, we note, again from \cite{Satnoianu2000,Hoang2013,Murray2001,General_Conditions}, that condition T2 for a problem posed on an infinite domain, can be translated into the existence of a wavenumber $k=k^*>0$ which simultaneously satisfies $\det\left(\jac \mbf f(\mbf P) - k^2 \, \mathbb D\right) = 0$, and  $k^*$ is a maximum point of the dispersion relation (locus of real eigenvalues of $\det\left(\jac \mbf f(\mbf P) - k^2 \, \mathbb D\right)$ as a function of $k$); see Fig.~\ref{fig:dispersionrelatbif}(a). Applying this to $\jac \mbf f(\mbf P)$, leads to the two conditions 
        \begin{align}
            k^2 \left(\delta^2 \left(\frac{(a + b)^2}{(c - h)^2} + d\right)-\frac{2 (a + b) (a h + b c)}{(a + b)^2 + d (c - h)^2} + c\right) + \delta^2 k^4 + \frac{(a + b)^2}{c - h} + d (c - h) &= 0, \label{zerodet}
            \\
            \frac{\frac{2 (a + b) (a h + b c)}{(a + b)^2 + d (c - h)^2} - \delta^2 \left(\frac{(a + b)^2}{(c - h)^2} + d\right) - c}{2 \delta^2} &= k^2. \label{zeroderivative}
        \end{align}
        From here, notice that we can get the expression for both the wavenumber $k$ and the bifurcation curve in \evs{the} parameter space. Specifically, assuming that \eqref{zeroderivative} holds, \evs{note} that \eqref{zerodet} is equivalent to \eqref{kfourth} which when back substituted into \eqref{zeroderivative} gives the required condition \eqref{eq:TuringCond}. \evs{Note that the nondegeneracy condition that $\lambda$ should be a maximum with respect to $k$ is easily shown because Re$\left(\lambda\left(k^2\right)\right)$ is a parabola with Re$(\lambda(0)) < 0$.}
    \end{proof}
    
    \paragraph{Remark.} Note (e.g.~from \cite{Nico2}) that replacing $k^2$ by $-k^2$ in \eqref{zerodet} and \eqref{zeroderivative} gives the conditions for a BD transition. Specifically, we find
    \begin{equation}
        \frac{2 \, (a + b) (a \, h + b \, c)}{(a + b)^2 + d \, (c - h)^2} - \delta^2 \left(\frac{(a + b)^2}{(c - h)^2} + d\right) - c = - 2 \, \delta \, \sqrt{\frac{(a + b)^2 + d \, (c - h)^2}{c - h}}. \label{eq:BDCond}
    \end{equation}
    
    \subsection{Super-critical, sub-critical and degenerate Turing bifurcation} \label{sec:maxwell}
    
    A key property of Turing bifurcations is that they give rise to non-homogeneous steady states that vary periodically in space with wavenumber close to \evs{$k^*$}. A key question is whether these bifurcations are super or sub-critical. It is common to study the nature of the bifurcation using amplitude equations \cite{Haragus}, derived using a form of Lyapunov-Schmidt reduction. \evs{It} can be shown that the study of such equations in the context of Turing or pattern-formation bifurcation is equivalent to the study of periodic orbits of the appropriate normal form of a Hamiltonian-Hopf (or reversible 1:1 resonance) bifurcation \cite{IoPe}. See for example \cite{Burke_NF}, who show just such an equivalence in the context of the Swift-Hohenberg model.
    
    Within a \evs{weakly nonlinear formalism, we suppose that a Turing bifurcation happens at a parameter value $\varepsilon = 0$ and seek an amplitude $A = A(t)$ of a periodic pattern with wavenumber $k^*$, where $A = \mathcal O\left(\varepsilon^{1/2}\right)$. It is supposed that such an amplitude is governed}, to leading order, by an ordinary differential equation of the form
    \begin{align}
        \frac{\dd A}{\dd t} &= C_{1,1} \, \varepsilon \, A + C_{3,0} \, A^3 + \mathcal O\left(A^5, \varepsilon \, A^3\right). \label{eq:pitchfork}
    \end{align}
    Here $C_{1,1}\neq 0$ ensures that a bifurcation happens as $\varepsilon$ varies, and $C_{3,0}$ is the cubic coefficient \evs{that determines the criticality of the bifurcation} (The notation for the coefficients $C_{i,j}$ will become apparent in \ref{sec:3} below). Without loss of generality, by redefinition of \evs{$\varepsilon$}, we shall set $C_{1,1}=1$ in this discussion and drop the second subscript on $C_3$.
    
    \evs{Note that \eqref{eq:pitchfork} is the general normal form equation of a pitchfork bifurcation, which is known to be sub-critical (resp. super-critical) if $C_3>0$ (\evs{resp.} $C_3<0$). For the sub-critical (\evs{resp.} super-critical) case, two unstable (resp. stable) orbits with amplitude $\approx \sqrt{|\varepsilon/C_3|}$ bifurcate from the trivial equilibrium for $\varepsilon<0$ (resp. $\varepsilon > 0$).} 
    
    \evs{Points of the Turing bifurcation manifold in which $C_3 = 0$ are the degenerate Turing bifurcations we are interested in}. In these cases, the appropriate amplitude equation can be written as
    \begin{align}
        \frac{\dd A}{\dd t}& = - f\left(A, \varepsilon, C_3\right) + \mathcal O\left(\varepsilon^2, A^7\right), \nonumber
        \\
        & \mbox{where} \quad
        f\left(A, \varepsilon, C_3\right) = - \varepsilon \, A - C_3 \, A^3 - C_5 \, A^5. \label{eq:5thorderNF}
    \end{align}
    
    For each value of $\varepsilon$ and $C_3$, it is useful to define an energy 
    \begin{align}
        \evs{E(A) = \int_0^A f\left(s, \varepsilon, C_3\right)\dd s = - \varepsilon \, \frac{A^2}{2} - C_3 \, \frac{A^4}{4} - C_5 \, \frac{A^6}{6}}. \label{eq:Ez}
    \end{align}
    Notice that $E$ is an even function, and since $A$ represents the amplitude of a periodic orbit, it is sufficient in what follows to consider only non-negative values of $A$. Note that $E(A)$ has critical points at
    \begin{align*}
        A = 0, \quad A = A^*_\pm = \sqrt{\frac{- C_3 \pm \sqrt{C_3^2 - 4 \, \varepsilon \, C_5}}{2 \, C_5}}.
    \end{align*}
    From the theory of normal forms, a so-called Maxwell point occurs when the two minima occur for the same value of $E$, \evs{and would correspond to the existence of a heteroclinic connection between flat and periodic patterned states.} Note that
    \begin{align*}
        E''(A) &= - \varepsilon - 3 \, C_3 \, A^2 - 5 \, C_5 \, A^4.
    \end{align*}
    Therefore, $E''(0) = - \varepsilon$, and $E''\left(A^*_\pm\right) = 4 \, \varepsilon + (C_3/C_5) \left(- C_3\pm \sqrt{C_3^2 - 4 \, \varepsilon \, C_5}\right)$.
    
    Because the energy is a smooth function, we need to have a maximum between two minima and the lowest critical point is $A=0$, then we must have $\varepsilon < 0$. \evs{Also, as we need the highest critical point to be a minimum, then $C_5<0$.} With this, we note that for both critical points $A^*_\pm$ to exist, we must have $C_3>0$, which\evs{,} in turn\evs{,} implies $0< A^*_+ < A^*_-$, with $A^*_-$ being a minimum of $E$ and $A^*_+$ a maximum. Thus, we seek the condition in which $E\left(A^*_-\right) = E(0) = 0$. We find that
    \begin{multline*}
        E^* := E\left(A_-^*\right) = \frac{\left(\sqrt{C_3^2 - 4 \varepsilon \, C_5} + C_3\right) \left(8 \eps \, C_5 - C_3 \left(\sqrt{C_3^2 - 4 \eps  \, C_5} + C_3\right)\right)}{48 \, C_5^2},
    \end{multline*}
    which has a zero when
    \begin{align}
        \eps = \frac{3 \, C_3^2}{16 \, C_5}, \quad \mbox{for } \eps<0, \quad C_3 > 0, \evs{\quad C_5<0,} \label{eq:Maxwell}
    \end{align}
    which gives us the expression for the Maxwell point, see also Fig.~\ref{fig:dispersionrelatbif}(b). 
    
    Note that the above calculation can also be performed in the case that $C_5 > 0$) but, in that case, there will be no Maxwell point for \evs{$\varepsilon<0$}. Instead, the non-trivial solutions $A^*_\pm$ exist for $C_3<0$, \evs{$\varepsilon>0$}, which gives rise to a very different kind of codimension-two bifurcation of the underlying PDE system, see \cite{IoossDias}. In what follows then, we shall seek specific conditions under which there is a Turing bifurcation whose normal form has $C_3=0$ with $C_5<0$.

\section{Computation of Turing bifurcation normal form} \label{sec:3}
    We present a general method for \evs{the} computation of a Turing bifurcation normal form for two-component reaction-diffusion systems, of which \eqref{vectorfield} is just a specific example, thus generalising the results in \cite{Burke_NF,Nico,FahadWoods}. Future work will show how to extend this analysis to $n$-component systems, and provide a general computational framework \cite{Edgardo_NFcomp}.
    
    Consider the general system of two reaction-diffusion equations given by
    \begin{align}
        \begin{pmatrix}
            u
            \\
            v
        \end{pmatrix}_t &= \begin{pmatrix}
            f(u, v, \varepsilon) + D_1 \, u_{xx}
            \\
            g(u, v, \varepsilon) + D_2 \, v_{xx}
        \end{pmatrix}. \label{normalformeq}
    \end{align}
    where $f, g$ \evs{are real differentiable functions of their arguments, up to at least order 5,}
    %that is of their arguments, that is $f,g \in \mathcal C^d$\footnote{\evs{$f\in \mathcal C^d$ if and %only if $f$ can be differentiated continuously at least $d$ times}}, with $d\geq 5$,
    %
    $\varepsilon \in \mathbb R$ is a parameter of the system, and $D_{1,2} \geq 0$ are the diffusion rates associated with $u$ and $v$, respectively. Furthermore, assume that \eqref{normalformeq} has a homogeneous steady state $\mbf P(\varepsilon) = (p(\varepsilon), q(\varepsilon))^\intercal$, which is asymptotically stable in the absence of diffusion. Without loss of generality, suppose that a smooth, parameter-dependent change of coordinates has been performed so that $(p, q) = (0, 0)$. Furthermore, suppose that when $\eps = 0$\evs{,} the origin goes through a Turing bifurcation with \evs{a} unique spatial wavenumber $k > 0$. For ease of notation\evs{,} we shall let $\mbf P := \mbf P(\varepsilon) = (0, 0)$.
    
    For each $\varepsilon$ sufficiently small, Taylor expansion up to order $d$ gives
    \begin{align*}
        f(u, v, \varepsilon) &= \sum_{i = 1}^d \frac{1}{i!} \left(u \, \frac{\partial}{\partial u} + v \, \frac{\partial}{\partial v}\right)^i f(0, 0, \varepsilon) + \mathcal O(u, v, \varepsilon),
        \\
        g(u, v, \varepsilon) &= \sum_{i = 1}^d \frac{1}{i!} \left(u \, \frac{\partial}{\partial u} + v \, \frac{\partial}{\partial v}\right)^i g(0, 0, \varepsilon) + \mathcal O(u, v, \varepsilon).
    \end{align*}
    For simplicity of notation, we shall henceforth denote partial derivatives via 
    \begin{align*}
        f_{i_1 i_2} := \frac{\partial^{i_1 + i_2} f}{\partial u^{i_1} \, \partial v^{i_2}}(0, 0, 0), \qquad g_{i_1 i_2} := \frac{\partial^{i_1 + i_2} g}{\partial u^{i_1} \, \partial v^{i_2}}(0,0,0).
    \end{align*}
    
    Following the usual procedure, see e.g.~\cite{Haragus}, we want to find a leading-order equation for the amplitude $A(t)$ of a small-amplitude bifurcating spatially periodic solution in the form
    \begin{align}
        \partial_t A &= h^{[1]}(A)+ h^{[2]}(A)+\ldots, \label{eqA}
    \end{align}
    where each function $h^{[i]}(A)$, for $i=1,2,\ldots$ is a real linear function of $A^i$. In particular in the light of \eqref{eq:5thorderNF} 
    $$
        \partial_t A = \eps \, C_{1, 1} \,  A +  C_3 \, A^3 + C_5 \, A^5 + \mbox{h.o.t.}
    $$
    To find such an amplitude equation, it is customary to make a complex ansatz of the form $(u, v)^\intercal \sim B \, \exp(ikx) \, \bs \phi + \mbox{c.c.}$ where $\bs \phi$ is a constant 2D vector, $B$ is a complex amplitude with $|B| = A$ and $k$ is the critical wavenumber at the Turing bifurcation; that is, $k$ satisfies $\det\left(\jac \mbf f(\mbf P) - k^2 \, \mathbb D - \lambda I\right)$ and \eqref{secondcondition}.
    $$
        \mathbb D = \begin{pmatrix}
            D_1 & 0
            \\
            0 & D_2
        \end{pmatrix} \quad \mbox{ and }  \jac \mbf f(\mbf P) = \left.\begin{pmatrix}
            f_{10} & f_{01}
            \\
            g_{10} & g_{01}
        \end{pmatrix}\right|_{\mbf P}.
    $$
    Without loss of generality, by translation in $x$ if necessary, we shall adopt instead the equivalent real ansatz 
    \begin{align*}
    	\begin{pmatrix}
    		u
    		\\
    		v
    	\end{pmatrix} &= A(t) \, \begin{pmatrix}
    		\bar u
    		\\
    		\bar v
    	\end{pmatrix} \, \cos(kx).
    \end{align*}
    
    The key to the method is to find a suitable change of coordinates
    \begin{align}
    	\begin{pmatrix}
    		u
    		\\
    		v
    	\end{pmatrix} = \mbf W^{[1]}(A) + \mbf W^{[2]}(A) + \ldots, \label{varchange}
    \end{align}
    where $(u, v)^\intercal$ solves \eqref{normalformeq} and  $A$ solves an equation of the form \eqref{eqA}. In particular, we are interested in computing the $\varepsilon$-independent terms $C_3$ and $C_5$, which can be carried out by considering \eqref{varchange} term by term when $\varepsilon = 0$.
    
    \subsection{Order 1}
    
        At first order, \eqref{normalformeq} becomes
        \begin{align*}
        	\partial_A \mbf W^{[1]} \, h^{[1]} &= \left(\jac \mbf f(0, 0) + \mathbb D \, \partial_{xx}\right) \mbf W^{[1]}.
        \end{align*}
        \evs{Since there is no $\varepsilon$-independent linear term in the normal form, we choose $h^{[1]}=0$. Hence,} we need to solve 
        \begin{align*}
        	\begin{pmatrix}
        		f_{10} + D_1 \, \partial_{xx} & f_{01}
        		\\
        		g_{10} & g_{01} + D_2 \, \partial_{xx}
        	\end{pmatrix} \mbf W^{[1]} &= 0.
        \end{align*}
        The general solution is given by
        \begin{align*}
        	\mbf W^{[1]} &= A \, \begin{pmatrix}
        		\phi_1
        		\\
        		\phi_2
        	\end{pmatrix} \, \cos(kx), \quad \mbox{ where } \phi_1 = \hat \alpha \, f_{01}, \quad \phi_2 = \hat\alpha \left(k^2 \, D_1 - f_{10}\right),
        \end{align*}
        and $\hat \alpha$ is an arbitrary non-zero scalar, which can be chosen so that $\bs \phi =(\phi_1, \phi_2)$ is a unit vector.
    
    \subsection{Order 2}
        At \evs{this} order, equation \eqref{normalformeq} is given by
        \begin{multline}
        	\partial_A \mbf W^{[1]} \, h^{[2]} + \cancelto{0}{\partial_A \mbf W^{[2]} \, h^{[1]}}
        	\\ 
        	= \left(\jac \mbf f(0, 0) + \mathbb D \, \partial_{xx}\right) \mbf W^{[2]} + \frac{u_1^2}{2!}\begin{pmatrix}
        		f_{20}
        		\\
        		g_{20}
        	\end{pmatrix} + \frac{2}{2!} u_1v_1\begin{pmatrix}
        		f_{11}
        		\\
        		g_{11}
        	\end{pmatrix} + \frac{v_1^2}{2!} \begin{pmatrix}
        		f_{02}
        		\\
        		g_{02}
        	\end{pmatrix}, \label{secondorder}
        \end{multline}
        where the subscripts in the variables $u$ and $v$ stand for the order of each term in the expansion \eqref{varchange}. Note that the right-hand side of \eqref{secondorder} has no secular terms. Therefore we can choose $h^{[2]} = 0$. Therefore, we need to solve
        \begin{multline}
            \begin{pmatrix}
            	f_{10} + D_1 \, \partial_{xx} & f_{01}
            	\\
            	g_{10} & g_{01} + D_2 \, \partial_{xx}
            \end{pmatrix} \mbf W^{[2]} = - \frac{A^2}{2}\left(\frac{\phi_1^2}{2}\begin{pmatrix}
            	f_{20}
            	\\
            	g_{20}
            \end{pmatrix} + \phi_1 \, \phi_2\begin{pmatrix}
            	f_{11}
            	\\
            	g_{11}
            \end{pmatrix}\right.
            \\ 
            \left.\quad + \frac{\phi_2^2}{2} \begin{pmatrix}
            	f_{02}
            	\\
            	g_{02}
            \end{pmatrix} \right)\left(1+\cos(2kx)\right), \label{secondorderequation}
        \end{multline}
        \evs{where we have used $2\cos^2(kx) = (1+ \cos(2kx)).$} The solution will, in general, be composed of a homogeneous part and a particular integral solution. \evs{Because the homogeneous part has already been solved at first order}, we are interested only in the solution that is orthogonal to $\ker\left(\jac \mbf f(0, 0) + \mathbb D \, \partial_{xx}\right)$, which has the following form
        \begin{align*}
        	\mbf W^{[2]} = A^2 \left(\begin{pmatrix}
        		\alpha_1
        		\\
        		\alpha_2
        	\end{pmatrix} + \begin{pmatrix}
        		\beta_1
        		\\
        		\beta_2
        	\end{pmatrix} \, \cos(2kx)\right).
        \end{align*}
        Then $\left(\alpha_1, \alpha_2\right)^\intercal$ solves
        \begin{align*}
        	\begin{pmatrix}
        		f_{10} & f_{01}
        		\\ 
        		g_{10} & g_{01}
        	\end{pmatrix}\begin{pmatrix}
        		\alpha_1
        		\\ 
        		\alpha_2
        	\end{pmatrix}= -\frac{1}{4}\left(\phi_1^2\begin{pmatrix}
        		f_{20}
        		\\ 
        		g_{20}
        	\end{pmatrix} + 2 \, \phi_1 \, \phi_2 \begin{pmatrix}
        		f_{11}
        		\\ 
        		g_{11}
        	\end{pmatrix} + \phi_2^2 \begin{pmatrix}
        		f_{02}
        		\\ 
        		g_{02}
        	\end{pmatrix}\right)
        \end{align*}
        and $\left(\beta_1, \beta_2\right)^\intercal$ solves 
        \begin{align*}
        	\begin{pmatrix}
        		f_{10} - 4 \, k^2 \, D_1 & f_{01}
        		\\ 
        		g_{10} & g_{01} - 4 \, k^2 \, D_2
        	\end{pmatrix}\begin{pmatrix}
        		\beta_1
        		\\ 
        		\beta_2
        	\end{pmatrix} = -\frac{1}{4}\left(\phi_1^2\begin{pmatrix}
        		f_{20}
        		\\ 
        		g_{20}
        	\end{pmatrix} + 2 \, \phi_1 \, \phi_2\begin{pmatrix}
        		f_{11}
        		\\ 
        		g_{11}
        	\end{pmatrix} + \phi_2^2\begin{pmatrix}
        		f_{02}
        		\\ 
        		g_{02}
        	\end{pmatrix}\right).
        \end{align*}
        The assumption that $k > 0$ is the only wavenumber such that $\det\left(\jac \mbf f(0, 0) - k^2 \, \mathbb D\right) = 0$, implies that the \evs{left}-hand-side of both these systems is non-singular and so these systems have a unique solution. 
    
    \subsection{Order 3}
        \evs{At this} order\evs{,} equation \eqref{normalformeq} gives
        \begin{multline}
        	\partial_A \mbf W^{[1]} \, h^{[3]} + \cancelto{0}{\partial_A \mbf W^{[2]} \, h^{[2]}} + \cancelto{0}{\partial_A \mbf W^{[3]} \, h^{[1]}}
        	\\
        	= \left(\jac \mbf f(0, 0) + \mathbb D \, \partial_{xx}\right) \mbf W^{[3]}
        	\\ 
        	+ \frac{2}{2!} \, u_1 \, u_2\begin{pmatrix}
        		f_{20}
        		\\ 
        		g_{20}
        	\end{pmatrix} + \frac{1}{2!}\left(2 \, u_1 \, v_2 + 2 \, u_2 \, v_1\right)\begin{pmatrix}
        		f_{11}
        		\\ 
        		g_{11}
        	\end{pmatrix} + \frac{2}{2!} \, v_1 \, v_2 \begin{pmatrix}
        		f_{02}
        		\\ 
        		g_{02}
        	\end{pmatrix}
        	\\
        	+\frac{1}{3!} \, u_1^3 \begin{pmatrix}
        		f_{30}
        		\\ 
        		g_{30}
        	\end{pmatrix} + \frac{3}{3!} \, u_1^2 \, v_1\begin{pmatrix}
        		f_{21}
        		\\ 
        		g_{21}
        	\end{pmatrix} + \frac{3}{3!} \, u_1 \, v_1^2\begin{pmatrix}
        		f_{12}
        		\\ 
        		g_{12}
        	\end{pmatrix} + \frac{1}{3!} \, v_1^3 \begin{pmatrix}
        		f_{03}
        		\\ 
        		g_{03}
        	\end{pmatrix}. \label{thirdorder}
        \end{multline}
        \evs{In this equation} there will, in general, be secular terms because the right-hand side of \eqref{thirdorder}, will have terms that are constant vectors multiplied by $\cos(kx)$. The Fredholm alternative then implies that for a solution, we require 
        \begin{align*}
        	\left(\Im(\jac \mbf f(0, 0) + \mathbb D \, \partial_{xx})\right)^\perp = \ker\left(\left(\jac \mbf f(0, 0) + \mathbb D \, \partial_{xx}\right)^*\right).
        \end{align*}
        The image of this operator is obtained from $\bs \psi$, the vector that spans the kernel of its adjoint:
        \begin{align*}
            \begin{pmatrix}
        		f_{10} + D_1 \, \partial_{xx} & g_{10}
        		\\ 
        		f_{01} & g_{01} + D_2 \, \partial_{xx}
        	\end{pmatrix}\bs \psi &= \mbf 0,
        \end{align*}
        from which we can choose the following function:
        \begin{align*}
        	\bs \psi = \begin{pmatrix}
        		\psi_1
        		\\ 
        		\psi_2
        	\end{pmatrix} \, \cos(kx), \qquad \mbox{with} \quad \psi_1 = \hat \beta \, g_{10} \mbox{ and } \psi_2 = \hat \beta \, \left(k^2 \, D_1 - f_{10}\right).
        \end{align*}
        Here, $\hat\beta$ is an arbitrary non-zero scalar that can be chosen so that $\bs \psi \cdot \bs \phi = 1$. For simplicity of presentation, we shall suppose $\hat \alpha = \hat \beta=1$, but it can be important in numerical evaluation to take other values (see the remark at the end of this section).
        
        Therefore, a sufficient condition for the solvability of the third-order equation is that the inner product between the right-hand side of \eqref{thirdorder} and $\bs \psi$ is equal to zero. We choose the natural inner product for $\bs a, \bs b\in \mathbb R^2$ and $f, g:\left[0, \frac{2\pi}{k}\right]\to \mathbb R$ continuous functions, namely
        \begin{align*}
        	\left \langle f(x) \, \mbf a, g(x) \, \mbf b \right \rangle &= \frac{k}{\pi} \int_0^{\frac{2\pi}{k}} f(x) \, g(x) \, \dd x \, (\mbf a \cdot \mbf b).
        \end{align*}
        Thus, we get
        \begin{multline}
            \left \langle \bs \psi, \left(\jac \mbf f(0, 0) + \mathbb D \, \partial_{xx}\right)\mbf W^{[3]}\right \rangle = \left \langle \bs \psi, \partial_A \mbf W^{[1]} \, h^{[3]} \right \rangle - \left \langle \bs \psi, u_1 \, u_2 \begin{pmatrix}
            	f_{20}
            	\\ 
            	g_{20}
            \end{pmatrix}\right \rangle
            \\
            - \left \langle \bs \psi, \left(u_1 \, v_2 + u_2 \, v_1\right) \begin{pmatrix}
            	f_{11}
            	\\
            	g_{11}
            \end{pmatrix} \right \rangle 
            - \left \langle \bs \psi, v_1 \, v_2\begin{pmatrix}
            	f_{02}
            	\\ 
            	g_{02}
            \end{pmatrix}\right \rangle - A^3 \, \left \langle \bs \psi, \mbf B \left(3 \, \cos(kx) + \cos(3kx)\right) \right \rangle, \label{thirdordereq}
        \end{multline}
        where
        \begin{align*}
        	\mbf B = \frac{1}{4}\left(\frac{\phi_1^3}{3!} \begin{pmatrix}
        		f_{30}
        		\\ 
        		g_{30}
        	\end{pmatrix} + \frac{\phi_1^2\phi_2}{2} \begin{pmatrix}
        		f_{21}
        		\\ 
        		g_{21}
        	\end{pmatrix} + \frac{\phi_1\phi_2^2}{2} \begin{pmatrix}
        		f_{12}
        		\\ 
        		g_{12}
        	\end{pmatrix} + \frac{\phi_2^3}{3!} \begin{pmatrix}
        		f_{03}
        		\\ 
        		g_{03}
        	\end{pmatrix}\right).
        \end{align*}
        Setting this expression to $0$, we find that
        		%	\begin{multline*}
        			%		h^{[3]}(A)=\frac{A^3}{\psi_1\phi_1+\psi_2\phi_2}\begin{pmatrix}
        				%			\psi_1
        				%			\s 
        				%			\psi_2
        				%		\end{pmatrix}\left( \phi_1\left(\alpha_1+\frac{\beta_1}{2}\right)\begin{pmatrix}
        				%			f_{20}
        				%			\s 
        				%			g_{20}
        				%		\end{pmatrix} + \phi_1 \left(\alpha_2+\frac{\beta_2}{2} \right)\begin{pmatrix}
        				%			f_{11}
        				%			\s 
        				%			g_{11}
        				%		\end{pmatrix}\right.
        			%		\s 
        			%		\left. + \phi_2 \left(\alpha_1+\frac{\beta_1}{2}\right)\begin{pmatrix}
        				%			f_{11}
        				%			\s 
        				%			g_{11}
        				%		\end{pmatrix} + \phi_2 \left(\alpha_2+\frac{\beta_2}{2}\right)\begin{pmatrix}
        				%			f_{02}
        				%			\s 
        				%			g_{02}
        				%		\end{pmatrix}\right)
        			%		\s 
        			%		+ \frac{3 A^3}{\psi_1\phi_1+\psi_2\phi_2} \begin{pmatrix}
        				%			\psi_1
        				%			\s 
        				%			\psi_2
        				%		\end{pmatrix}\cdot \mbf B,
        			%	\end{multline*}
        		%	which can be written as
        $h^{[3]}(A) = C_3 \, A^3$, where
        \begin{multline*}
        	C_3 = \frac{1}{\psi_1 \, \phi_1 + \psi_2 \, \phi_2}\begin{pmatrix}
        		\psi_1
        		\s 
        		\psi_2
        	\end{pmatrix}\cdot \left(\phi_1\left(\alpha_1 + \frac{\beta_1}{2}\right)\begin{pmatrix}
        		f_{20}
        		\s 
        		g_{20}
        	\end{pmatrix} + \phi_2 \left(\alpha_2 + \frac{\beta_2}{2}\right)\begin{pmatrix}
        		f_{02}
        		\s 
        		g_{02}
        	\end{pmatrix}\right.
        	\s 
        	\left. + \left(\phi_1 \left(\alpha_2 + \frac{\beta_2}{2} \right) + \phi_2\left(\alpha_1 + \frac{\beta_1}{2}\right)\right)\begin{pmatrix}
        		f_{11}
        		\s 
        		g_{11}
        	\end{pmatrix} \right)
        	+ \frac{3}{\psi_1 \, \phi_1 + \psi_2 \, \phi_2} \begin{pmatrix}
        		\psi_1
        		\s 
        		\psi_2
        	\end{pmatrix}\cdot \mbf B,
        \end{multline*}
        which can be written explicitly as
        \begin{multline}
        	C_3 = \frac{f_{01}}{f_{01} \, g_{10} + \left(k^2 \, D_1 - f_{10}\right)^2}\left(\alpha_1 + \frac{\beta_1}{2}\right)\left(g_{10} \, f_{20} + \left(k^2 \, D_1 - f_{10}\right) g_{20}\right)
        	\s 
        	+ \frac{k^2 \, D_1 - f_{10}}{f_{01} \, g_{10} + \left(k^2 \, D_1 - f_{10}\right)^2} \left(\alpha_2 + \frac{\beta_2}{2}\right)\left(g_{10} \, f_{02} + \left(k^2 \, D_1 - f_{10}\right) g_{02}\right)
        	\s 
        	+ \frac{g_{10} \, f_{11} + \left(k^2 \, D_1 - f_{10}\right) g_{11}}{f_{01} \, g_{10} + \left(k^2 \, D_1 - f_{10}\right)^2} \left(f_{01} \left(\alpha_2 + \frac{\beta_2}{2} \right) + \left(k^2 \, D_1 - f_{10}\right)\left(\alpha_1 + \frac{\beta_1}{2}\right)\right)
        	\s 
        	+ \frac{3}{f_{01} \, g_{10} + \left(k^2 \, D_1 - f_{10}\right)^2} \begin{pmatrix}
        		g_{10}
        		\s 
        		k^2 \, D_1 - f_{10}
        	\end{pmatrix}\cdot \mbf B, \label{C2D}
        \end{multline}
                    
    \subsection{Higher-order terms}
    
        The above procedure can be continued up to arbitrary order. For our present purposes\evs{,} it is necessary to compute \evs{it} up to order \evs{five}. The details are relegated to Appendix A.
        
        %With this, we are ready to start analyzing the Turing %bifurcations we find in the models %\eqref{vectorfield}. In the next two sections, we are %going to perform different scale changes to our parameters in order to study the Turing bifurcations %we can find in those models.

        % HERE I AM
    	
    \subsection{Unfolding}
    
        Everything we have done so far is a linear process; we went order by order finding the coefficients of a polynomial in $A$. However, these are only the terms that are relevant to the bifurcation point at $\varepsilon = 0$. To complete the calculation, we have to include the lowest-order parameter-dependent term; which comes in at second-order. In particular, we want $A$ to solve:
        \begin{align*}
            \partial_t A = h^{[1, 1]}(A, \varepsilon) + h^{[1, 0]}(A, \varepsilon) + h^{[2, 0]}(A, \varepsilon) + \ldots,
        \end{align*}
        where $h^{[i, j]}(A, \eps)$ is a linear function in $A^i \eps^j$.
        
        To make this possible, we need to add an extra term to our change of variables:
        \begin{align*}
            \begin{pmatrix}
                u
                \\ 
                v
            \end{pmatrix} = \mbf W^{[1, 1]}(A, \varepsilon) + \mbf W^{[1, 0]}(A, \varepsilon) + \mbf W^{[2, 0]}(A, \varepsilon) + \ldots,
        \end{align*}
        Note that $\mbf W^{[i, 0]}(A, \varepsilon) := \mbf W^{[i]}(A)$ and $h^{[i, 0]}(A, \varepsilon) := h^{[i]}(A)$, for $i = 1, 2, 3$ were already found with our previous procedure. To compute $\mbf W^{[1, 1]}(A, \varepsilon)$ we \evs{just need to} consider the terms in \eqref{normalformeq} that have a product between $A$ and $\varepsilon$ when we perform a Taylor expansion in both $f$ and $g$, with respect to $A$ and $\varepsilon$. For this purpose, it is useful to introduce the notation
        \begin{align*}
            f_{i_1i_2i_3} = \frac{\partial^{i_1 + i_2 + i_3} f}{\partial u^{i_1}\partial v^{i_2}\partial \varepsilon^{i_3}}(0, 0, 0),
            \qquad
            g_{i_1i_2i_3} = \frac{\partial^{i_1 + i_2 + i_3} g}{\partial u^{i_1}\partial v^{i_2}\partial \varepsilon^{i_3}}(0, 0, 0).
        \end{align*}
        Therefore, at order 1 in $A$ and $\varepsilon$, \eqref{normalformeq} becomes
        \begin{multline*}
            \cancelto{0}{\partial_A \mbf W^{[1, 1]} \, h^{[1, 0]}} + \partial_A \mbf W^{[1, 0]} \, h^{[1, 1]}
            = \left(\jac \mbf f(0, 0, 0) + \mathbb D \, \partial_{xx}\right)\mbf W^{[1, 1]}
            \s
            + u_1 \, \varepsilon \begin{pmatrix}
                f_{101}
                \s 
                g_{101}
            \end{pmatrix} + v_1 \, \varepsilon\begin{pmatrix}
                f_{011}
                \s 
                g_{011}
            \end{pmatrix}
        \end{multline*}
        which is equivalent to:
        \begin{multline}
            \left(\jac \mbf f(0, 0, 0) + \mathbb D \, \partial_{xx}\right)\mbf W^{[1, 1]} =
                \partial_A \mbf W^{[1,0]} \, h^{[1,1]} - A \, \varepsilon \, \phi_1 \, \cos(kx) \begin{pmatrix}
                    f_{101}
                    \s 
                    g_{101}
                \end{pmatrix}
                \s 
                - A \, \varepsilon \, \phi_2 \, \cos(kx)\begin{pmatrix}
                    f_{011}
                    \s 
                    g_{011}
                \end{pmatrix}.
             \label{unfolding}
        \end{multline}
        Now, note that the right-hand side of \eqref{unfolding} may have some secular terms, so we need to use the Fredholm alternative to find $h^{[1, 1]}$. In particular:
        \begin{multline*}
            \left \langle \bs \psi, \left(\jac \mbf f(0, 0, 0) + \mathbb D \, \partial_{xx}\right)\mbf W^{[1, 1]}\right \rangle = \left \langle \bs \psi, \partial_A \mbf W^{[1, 0]} h^{[1, 1]}\right \rangle
            \s 
            - A \, \varepsilon \, \phi_1 \left \langle \bs \psi, \cos(kx) \begin{pmatrix}
                f_{101}
                \s 
                g_{101}
            \end{pmatrix} \right \rangle - A \, \varepsilon \, \phi_2 \left \langle \bs \psi, \cos(kx)\begin{pmatrix}
                f_{011}
                \s 
                g_{011}
            \end{pmatrix}\right \rangle,
        \end{multline*}
        which is equivalent to
        \begin{multline*}
            \left \langle \bs \psi, \left(\jac \mbf f(0, 0, 0) + \mathbb D \, \partial_{xx}\right)\mbf W^{[1,1]}\right \rangle = h^{[1,1]}\begin{pmatrix}
                \psi_1
                \s 
                \psi_2
            \end{pmatrix} \cdot \begin{pmatrix}
                \phi_1
                \s 
                \phi_2
            \end{pmatrix}
            \s 
            - A \, \varepsilon \, \phi_1 \begin{pmatrix}
                \psi_1
                \s 
                \psi_2
            \end{pmatrix}\cdot\begin{pmatrix}
                f_{101}
                \s 
                g_{101}
            \end{pmatrix} - A \, \varepsilon \, \phi_2 \begin{pmatrix}
                \psi_1
                \s 
                \psi_2
            \end{pmatrix}\cdot \begin{pmatrix}
                f_{011}
                \s 
                g_{011}
            \end{pmatrix}.
        \end{multline*}
        Therefore, when we equal this expression to zero, we find 
        \begin{align*}
            h^{[1, 1]}(A, \varepsilon) = C_{1, 1} \, A \, \eps,
        \end{align*}
        where
        \begin{align*}
            C_{1, 1} = \frac{1}{\psi_1 \, \phi_1 + \psi_2 \, \phi_2}\begin{pmatrix}
                \psi_1
                \s 
                \psi_2
            \end{pmatrix}\cdot \left(\phi_1 \begin{pmatrix}
                f_{101}
                \s 
                g_{101}
            \end{pmatrix} + \phi_2\begin{pmatrix}
                f_{011}
                \s 
                g_{011}
            \end{pmatrix}\right).
        \end{align*}
        
        \paragraph{Remark.} It is straightforward to see how these expressions change when taking $\hat \alpha\neq 1$ or $\hat \beta\neq 1$. In particular, the choice of $\hat \beta$ does not change anything in the calculation, whilst a different choice of $\hat \alpha$ would make $C_3$ become $\hat \alpha^2 \, C_3$, and leave $C_{1, 1}$ as it is. \evs{This} change in $C_3$ is not important since $\hat \alpha^2 \geq 0$, which implies that this will not generate any extra changes of sign in $C_3$.

\section{Small parameter limit} \label{sec:4}
    Returning to \eqref{vectorfield}, \evs{we start by assuming the inhibitor kinetic parameters $a$, $b$ and $d$ are small.} Then, a distinguished limit can be found under the following scaling of parameters as $\delta \to 0^+$:
    \begin{align}
        (a, b, c, d, h) = \left(\delta \, \alpha, \delta \, \beta, \gamma, \delta^2 \, \eta, \xi\right), \quad \mbox{where $0 \leq \alpha, \beta, \gamma, \eta, \xi = \mathcal O(1)$, and $\gamma > \xi$.} \label{eq:small}
    \end{align}
    Specifically, we can prove the following result.
    \begin{result}
        Under the parameter scaling \eqref{eq:small},
        the Turing bifurcation condition \eqref{eq:TuringCond} for system \eqref{vectorfield} can be expressed as
        \begin{equation}
            2 \, (\alpha + \beta)(\alpha \, \xi + \beta \, \gamma) - \gamma \, (\alpha + \beta)^2 -\gamma \, \eta \, (\gamma - \xi)^2 = \mathcal{O} \left(\delta^2\right). \label{eq:smallDelta}
        \end{equation}
        Moreover, the Turing bifurcation is sub-critical for sufficiently small $\delta$ within this scaling. Furthermore, the parameter values for which a BD-transition occurs obey
        precisely the same asymptotic estimate \eqref{eq:smallDelta}.
    \end{result}
    \begin{proof}
        Considering the parameter scaling stated in the result, note that the condition \eqref{turingcond} becomes
        \begin{align}
        	\gamma < \frac{2 \, (\alpha + \beta) (\alpha \, \xi + \beta \, \gamma)}{(\alpha + \beta )^2 + \eta \, (\gamma - \xi)^2} < \frac{\delta^2 \left((\alpha + \beta)^2 + \eta \, (\gamma - \xi)\right)}{(\gamma - \xi)^2} + \gamma. \label{firstturingcond}
        \end{align}
        Next, from \eqref{zeroderivative}, we can see that
        \begin{align}
        	k^2 &= \delta^{-2} \, \frac{1}{2} \left[\frac{2 \, (\alpha + \beta) (\alpha \, \xi + \beta \, \gamma)}{(\alpha + \beta)^2 + \eta \, (\gamma - \xi)^2} - \gamma\right] + {\mathcal O}(1). \label{firstksquared}
        \end{align}
        First, consider the case where $k^2$ is positive in order to ensure that we are dealing with a Turing bifurcation, not a BD point. \evs{Note though that the rescaled version of condition \eqref{eq:TuringCond} implies}
        \begin{align*}
        	\frac{2 \, (\alpha + \beta) (\alpha \, \xi + \beta \, \gamma)}{(\alpha + \beta)^2 + \eta \, (\gamma - \xi)^2} - \delta^4 \, \frac{(\alpha + \beta)^2 + \eta \, (\gamma - \xi)^2}{(\gamma - \xi)^2} - \gamma &> 0
        \end{align*}
        But note, from \eqref{firstksquared}, that this inequality must be fulfilled by default for $0 < \delta \ll 1$ sufficiently small. Next, taking the rescaled version of \eqref{eq:TuringCond}, \evs{we can seek a solution for small $\delta$}\evs{,} in which case we find a function $\nu(\delta)$ such that
        \begin{align}
        	\frac{2 \, (\alpha + \beta)(\alpha \, \xi + \beta \, \gamma)}{(\alpha + \beta)^2 + \eta \, (\gamma - \xi)^2} - \gamma = \nu(\delta) \, \delta^2, \label{equationforeta}
        \end{align}
        and
        \begin{align*}
        	\nu(0) = 2 \, \sqrt{\frac{(\alpha + \beta)^2 + \eta \, (\gamma - \xi)^2}{\left(\gamma -\xi\right)}}>0, 
        \end{align*}
        which gives us the condition \eqref{eq:smallDelta} stated in result.
            
        Now, let us consider the condition for a BD-transition to occur. In this case, the only thing that changes is the sign of $k^2$. In particular, note that expression \eqref{eq:TuringCond} comes from getting the positive square root of $k^4$. Therefore, the BD-line is defined by the same equation with a negative sign. In particular, we have that
        \begin{align*}
            \frac{2 \, (\alpha + \beta)(\alpha \, \xi + \beta \, \gamma)}{(\alpha + \beta)^2 - \eta(\gamma - \xi)^2} - \gamma &= - \nu(\delta) \, \delta^2,
        \end{align*}
        This means that, as $\delta\to 0^+$, the leading-order expression for the codimension-one set in the $(\alpha, \beta, \gamma, \eta, \xi)$-parameter space for the BD-transition is identical to that of the Turing bifurcation.
        
        Finally, consider the statement about the criticality of the Turing bifurcation. Carrying out the calculation outlined in the previous section, and making the appropriate substitutions in \eqref{C2D}, one finds that the rescaled third-order coefficient can be written in the form
        \begin{align}
        	C_3 = \frac{b_0 + \mathcal O(\delta)}{c_0 + \mathcal O(\delta)}, \label{eq:C3}
        \end{align}
        where
        \begin{align*}
        	b_0 &= - 4 \, \gamma^5 \, \nu \, (\alpha + \beta) (\gamma - \xi)^9 \, (\alpha \, \xi + \beta \, \gamma)^3 < 0,
        	\\ 
        	c_0 &= - 72 \, \gamma^4 \, (\alpha + \beta)^2 (\gamma - \xi)^{10} (\alpha \, \xi + \beta \, \gamma)^2 < 0.
        \end{align*}
        Therefore, within the scaling given in the result statement, the Turing bifurcation is always sub-critical for sufficiently small $\delta > 0$.
    \end{proof}
    
    Note that slightly simpler expressions are available for the special cases of the Schnakenberg and the Brusselator models because, in those cases, $\eta = 0$. 

    \paragraph{Schnakenberg model} In this case, the general result is still valid, but the proof gets simplified since the scaling \eqref{eq:small} becomes $(a, b, c, d, h) = \left(\delta \, \alpha, \delta \, \beta, 1, 0, 0\right)$. With this, from equation \eqref{zeroderivative}, we can see that
    \begin{align*}
    	k^2 = \frac{- \delta^4 \, (\alpha + \beta)^2 + \frac{2 \, \beta}{\alpha + \beta} - 1}{2 \, \delta^2} = \frac{\beta - \alpha}{2 \, \delta^2 \, (\alpha + \beta)} + \mathcal{O}\left(\delta^2\right),
    \end{align*}
    which is positive if and only if $\beta > \alpha$. On the other hand, from \evs{the scaled version of} equation \eqref{kfourth}, we have that $k^2 = \alpha + \beta$. Therefore, for \evs{$0 < \delta \ll 1$ sufficiently small}, there must exist a positive function $\hat \nu(\delta)$\evs{, such} that
    $$
    	\beta - \alpha = \hat{\nu}(\delta) \, \delta^2, \qquad \evs{\mbox{where } \hat \nu(0) = 2 \, (\alpha + \beta)^2}.
    $$
    Finally, the third-order coefficient associated with this bifurcation is given by \eqref{eq:C3}, with
    \begin{align*}
        b_0 = - 512 \, \alpha^7 \, \nu<0, \qquad 
        c_0 = - 512 \, \alpha^4 \, \left(8 \, \alpha^4 + \nu^2\right).
    \end{align*}
    And so the Turing bifurcation is clearly sub-critical for $\delta>0$ sufficiently small. 
    	
    \paragraph{Brusselator model}  Again, in this case, the general result is still valid, but the proof gets simplified in a different way since the scaling \eqref{eq:smallDelta} becomes $(a, b, c, d, h) = \left(\delta \, \alpha, 0, \gamma, 0, \gamma - 1\right)$.
    With this, we can see that
    \begin{align*}
        k^2 &= \frac{- \alpha^2 \, \delta^4 + \gamma - 2}{2 \, \delta^2} \approx \frac{\gamma - 2}{2 \, \delta^2} + \mathcal O(1).
    \end{align*}
    On the other hand, the expression for $k^2$ from \eqref{kfourth} reduces to  $k^2 = \alpha$. Therefore, there must exist a positive function $\nu(\delta)$, defined for all $0 < \delta \ll 1$, such that
    \begin{align*}
        \gamma = 2 + \nu(\delta) \, \delta^2,
    \end{align*}
    and $\ds{\lim_{\delta\to 0^+} \nu(\delta) = 2 \, \alpha}$. Taking this into account, the third-order coefficient associated with the Turing bifurcation is given by \eqref{eq:C3}, with 
    \begin{align*}
        b_0 = - 16 \, \alpha^4 \, \varepsilon < 0, \qquad 
        c_0 = - 16 \, \left(2 \, \alpha^2 \, \varepsilon^2 + \alpha^4\right) < 0.
    \end{align*}
    showing the Turing bifurcation to be sub-critical \evs{for a sufficiently small $\delta$} in this case, also.

\section{Large parameter limit} \label{sec:5}
    \evs{In contrast, we next suppose that the kinetic parameters $b$ and $d$ for the inhibitor get large whilst $a$ still gets small as $\delta \to 0^+$. We find a distinguished limit under the scaling of parameters for small $\delta$:}
    \begin{align}
        (a, b, c, d, h) = \left(\delta \, \alpha , \delta^{-1} \, \beta, \gamma, \delta^{-1} \, \eta, \xi\right). \label{eq:largescale}
    \end{align}
        
    \begin{result} \label{thm:largeparamlim}
        Consider the parameter scaling \eqref{eq:largescale}, for $\beta > 0$, and $\gamma > \xi$. Then, in the limit $\delta \to 0^+$, there exists a codimension-two manifold within the codimension-one parameter set of Turing bifurcations, given by 
    	\begin{align}
    	  6 \, \beta^2 \, \nu + 32 \, \beta^4 - 27 \, \nu^2 = 0, \quad \mbox{with } \nu = 2 \, \beta   \, (\gamma - \xi)^{3/2} + \mathcal{O}(\delta) \label{eq:codim2cond}
    	\end{align}
        at which the criticality of the Turing bifurcation changes from  super- to sub-critical. 
    \end{result}
    
    \paragraph{Remark.} Note that this result applies to all models in Table \ref{tab:models} except for the Brusselator, which has $b=0$\evs{.} In that case a different kind of scaling is available, \evs{which enables explicit the 3rd and 5th-order normal form coefficients to be expressed in particularly simple form}. The special case of the Brusselator is treated in Sec.\ref{sec:7} below. In the case of models such as the Schnackenberg, Glycolysis and Sel'kov-Schnakenberg that have $h=0$, Result \ref{thm:largeparamlim} can be improved by using a slightly different large-parameter scaling as $\delta \to 0^+$. \evs{See Section \ref{sec:7} below for the details}.
    
    \begin{proof}
        Consider the parameter scaling \eqref{eq:largescale}, for which the Turing bifurcation condition \eqref{turingcond} can be expressed as 
    	\begin{align*}
    		\gamma < \frac{2 \left(\alpha \, \delta^2 + \beta\right) \left(\alpha \, \delta^2 \, \xi + \beta \, \gamma\right)}{\delta \left(\alpha^2 \, \delta^3 + \eta \, (\gamma - \xi)^2\right) + 2 \, \alpha \, \beta \, \delta^2 + \beta^2} < \gamma + \frac{1}{\delta^2} \left(\frac{\left(\alpha \, \delta^2 + \beta\right)^2}{(\gamma - \xi)^2} + \delta \, \eta \right).
    	\end{align*}
        As $\delta \to 0^+$, this expression is given approximately by
    	\begin{align*}
    		0 < \gamma < \frac{\beta^2}{\delta^2 (\gamma - \xi)^2} + \mathcal{O}(1)
    	\end{align*}
    	which is clearly true for $0<\delta \ll 1$ sufficiently small.
    
        Now, from \eqref{eq:TuringCond}, joining all the terms that depend on $\delta$, we can see that there must exist a positive function $\nu = \nu(\delta)$, such that
    	\begin{align*}
    		\gamma \, (\gamma - \xi)^2 - \beta^2 = \nu(\delta), \quad \mbox{such that} 
    		\lim_{\delta \to 0^+} \nu(\delta) = 2 \, \beta \, (\gamma - \xi)^{3/2}.
    	\end{align*}
        Hence\evs{,} we can write
        $$
            \xi = \gamma - \frac{\sqrt{\gamma \left(\beta^2 + \nu\right)}}{\gamma},
        $$
        Then, from \eqref{zerodet}, we have that the leading-order term is given by:
    	\begin{align*}
    		- \gamma \, \nu^2 \sqrt{\gamma \left(\beta^2 + \nu\right)} + 4 \, \beta^2 \, \nu^2 + 8 \, \beta^4 \, \nu + 4 \, \beta^6 = 0.
    	\end{align*}
        Hence, up to leading order in $\delta$, we obtain
    	\begin{align}
            \gamma = \frac{2 \, \sqrt[3]{2} \, \beta^{4/3} \left(\beta^2 + \nu\right)}{\nu^{4/3}}. \label{gammaeq}
    	\end{align}
        Thus, by substitution of the above expressions for $\xi$ and $\gamma$, into the expression \eqref{C2D}, keeping only leading-order terms in $\delta$, we find that the third-order coefficient of the Turing bifurcation is given by
    	\begin{align*}
    		C_3 = \frac{b_0 + \mathcal O(\delta)}{c_2 \, \delta^{2} + \mathcal O \left(\delta^3\right)},
    	\end{align*}
    	where
    	\begin{align*}
    		b_0 &= - 4096 \, \beta^{26} \, \left(\beta^2 + \nu\right)^2 \left(2 \, \beta^2 + \nu\right) \left(6 \, \beta^2 \, \nu + 32 \, \beta^4 - 27 \, \nu^2\right),
    		\\ 
    		c_2 &= - 36864 \, \sqrt[3]{2} \, \beta^{64/3} \, \nu^{8/3} \, \left(\beta^2 + \nu\right)^2 \left(2 \, \beta^2 + \nu\right)^2.
    	\end{align*} 
        From the sign of the above expression for $C_3$, we find that the Turing bifurcation is sub-critical if
    	\begin{align*}
    		6 \, \beta^2 \, \nu + 32 \, \beta^4 - 27 \, \nu^2 > 0,
    	\end{align*}
    	and super-critical if
    	\begin{align*}
    		6 \, \beta^2 \, \nu + 32 \, \beta^4 - 27 \, \nu^2 < 0.
    	\end{align*}
        Therefore, in all the models where the parameters $\beta, \xi$ are positive, we can have a super-critical or sub-critical bifurcation, depending on the values of the parameters.
        
        In fact, with this scaling, it is evident that we will be able to find a codimension-two bifurcation approximately in the points of the Turing bifurcation curve that fulfill the condition \eqref{eq:codim2cond} as stated. 
    
    	Furthermore, note that when $0 < \delta \ll 1$ is sufficiently small, we have that
    	\begin{align*}
    		\gamma \, (\gamma - \xi)^2 - \beta^2 \approx 2 \, \beta \, (\gamma - \xi)^{3/2}
    		\\
    		\iff \beta \approx  \left(\sqrt{\frac{\gamma}{\gamma - \xi} + 1} - 1\right) (\gamma - \xi)^{3/2}.
    	\end{align*}
    	Therefore,
    	\begin{align*}
    		\varepsilon &\approx 2 \left(\sqrt{\frac{\gamma}{\gamma - \xi} + 1} - 1\right) (\gamma -\xi)^3
    	\end{align*}
    	In conclusion, taking into account all our conditions, then we have that the Turing bifurcation gets super-critical when $\delta \to 0^+$ if and only if
    	\begin{align*}
    		\xi \left(13 \, \sqrt{\frac{\gamma}{\gamma - \xi} + 1} + 14\right) < \gamma \left(13 \, \sqrt{\frac{\gamma}{\gamma - \xi} + 1} + 6\right).
    	\end{align*}
    \end{proof}

\section{Application to specific models} \label{sec:6}
    Here\evs{,} we illustrate the results of the previous two sections in each of the \evs{models} in Table \ref{tab:models}\evs{.} In each case\evs{,} we use the large parameter scaling of Result \ref{thm:largeparamlim}. 

    \subsection{Root Hair Model}
\begin{figure}
    		\centering
    		\begin{subfigure}[b]{0.45\textwidth}
    			\centering
    			\includegraphics[width=\textwidth]{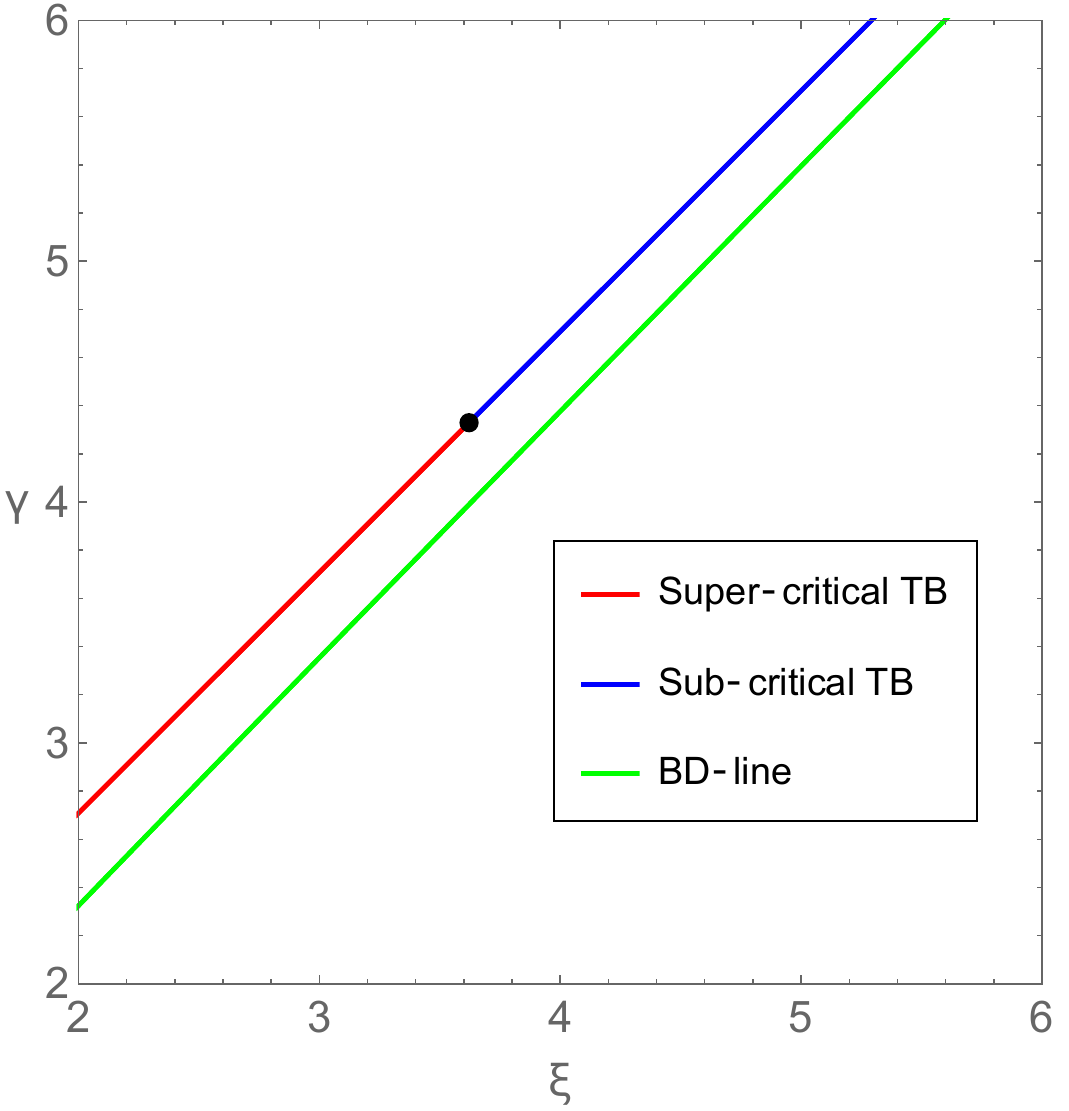}
    			\caption{$(\xi,\gamma)$ plane.}
    			\label{fig:gammaxi}
    		\end{subfigure}
    		\hfill
    		\begin{subfigure}[b]{0.45\textwidth}
    			\centering
    			\includegraphics[width=\textwidth]{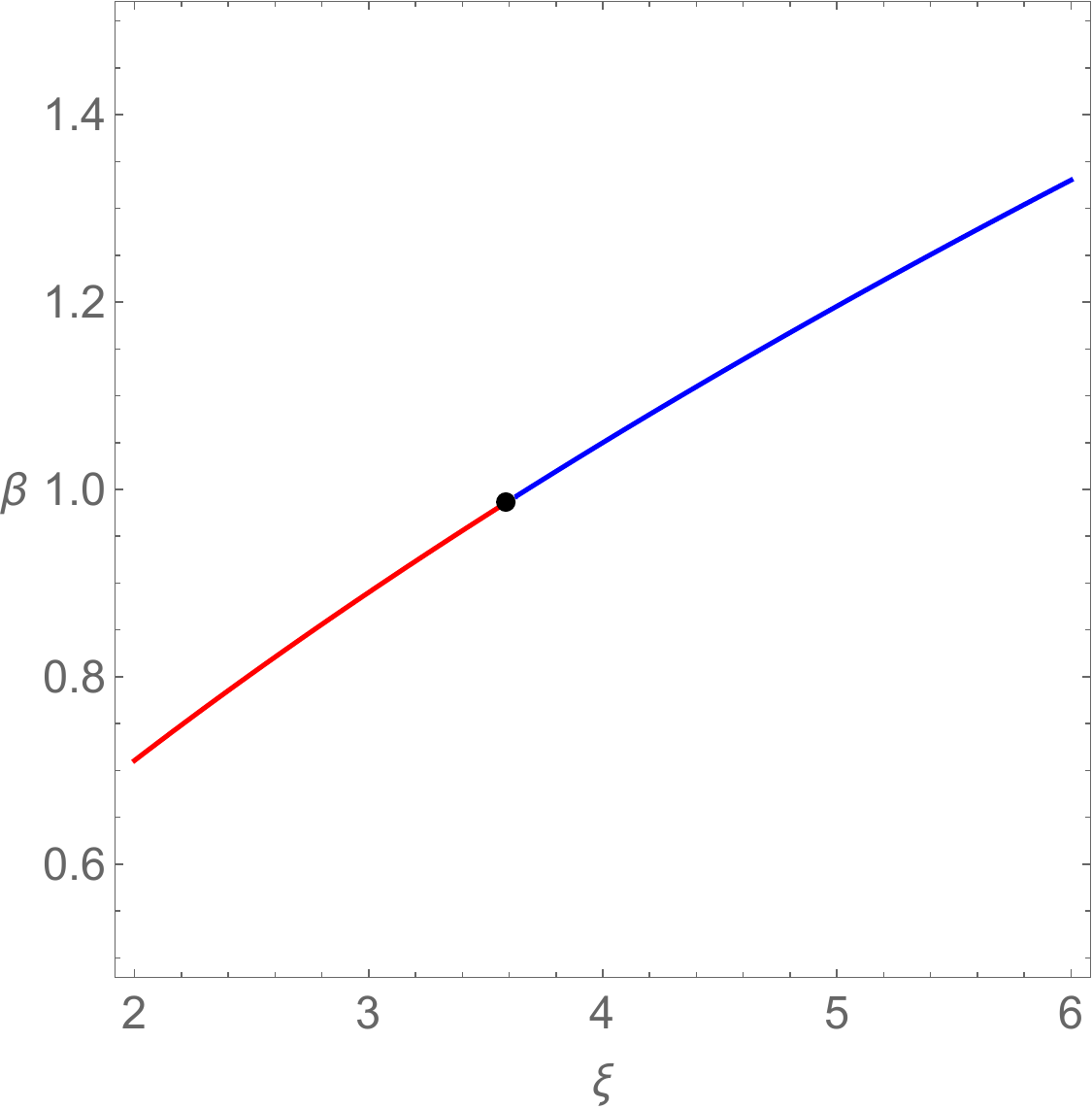}
    			\caption{$(\xi,\beta)$ plane.}
    			\label{fig:betaxi}
    		\end{subfigure}
            \caption{Bifurcation diagrams associated with the Turing bifurcation in the general system \eqref{vectorfield} after the scaling of parameters, setting $(\alpha, \eta, \delta) = (0, 0.9, 0.001)$ and taking in (a), $\beta = \dfrac{\sqrt{\sqrt{2} \, \xi + 2} - 1}{2^{3/4}}$, leaving $\xi$ and $\gamma$ free, while in (b), $\gamma = \xi + \dfrac{1}{\sqrt{2}}$, leaving $\xi$ and $\beta$ free. A dot marks the point where the bifurcation changes its criticality in the limit $\delta \to 0^+$.} 
    		\label{fig:transition}
    	\end{figure}

        To be concrete, consider the particular case where $\gamma = \xi + \frac{1}{\sqrt{2}}$, so that $2 \, (\gamma - \xi)^2 = 1$. Then, we have 
    	\begin{align*}
    		\beta &= \frac{\sqrt{\sqrt{2} \, \xi + 2} - 1}{2^{3/4}}\evs{.}
    	\end{align*}
    	\evs{Thus,} the Turing bifurcation is super-critical if
    	\begin{align*}
    		0\leq \xi < \frac{39 \, \sqrt{\frac{97}{2}}}{128} + \frac{265}{128 \, \sqrt{2}} \approx 3.58583,
    	\end{align*}
    	and sub-critical if
    	\begin{align*}
    		\xi > \frac{39 \, \sqrt{\frac{97}{2}}}{128} + \frac{265}{128 \, \sqrt{2}}\approx 3.58583.
    	\end{align*}
    	In particular, with this scaling, if we consider
    	\begin{align*}
            \beta = \frac{\sqrt{\sqrt{2} \, \xi + 2} - 1}{2^{3/4}}, \eta = 0.9, \mbox{ and } \delta = 0.001,
    	\end{align*}
        then we get the two-parameter bifurcation plot shown in \evs{Figure} \ref{fig:transition}\evs{,} in which we find a degenerate Turing bifurcation within this scaling. \evs{We plot the various bifurcation curves in both the $(\xi,\gamma)$ and $(\xi,\beta)$ parameter planes. In both, the region of existence of Turing patterns is the upper right-half of the diagram. The small amplitude bifurcation of the pattern is in the direction below the curve from the blue portion of the Turing curve. See \cite{FahadWoods} for an extensive two-parameter bifurcation diagram including the identification of regions where localised patterns exist close to the degenerate Turing bifurcation.}
	
	\subsection{Schnakenberg model}	
        In this case, the scaling leading to Result \ref{thm:largeparamlim} does not enable us to find a degenerate Turing bifurcation within the large-parameter limit because we only have one non-zero parameter in the limit $\delta \to 0$. Specifically, the relevant scaling is given by
    	\begin{align*}
    		(a, b, c, d, h) = \left(\delta \, \alpha, \delta^{-1} \, \beta, 1, 0, 0\right).
    	\end{align*}
    	Therefore, when grouping all the terms that depend on $\delta$, there must exist a function $\nu = \nu(\delta)$ such that the Turing bifurcation expression, \eqref{eq:TuringCond}, becomes
    	\begin{align*}
    	    \beta^2 + 2 \, \beta - 1 + \nu \, \delta^2 &= 0,
    	\end{align*}
    	where
    	\begin{align*}
    	    \lim_{\delta\to 0^+} \nu(\delta) = 3 \, \alpha \, \beta^2 + 4 \, \alpha \, \beta + \alpha.
    	\end{align*}
    	Furthermore, when $0 < \delta \ll 1$, this expression gets reduced to
        $$
            \beta = \sqrt{2} - 1 > \evs{0.}
        $$
    	Finally, the third-order coefficient of the Turing bifurcation is given by
    	\begin{align*}
    		C_3 = \frac{b_0 + \mathcal O(\delta)}{c_2 \, \delta^2 + \mathcal O\left(\delta^4\right)},
    	\end{align*}
    	where
    	\begin{align*}
    		b_0 &= - 4 \, (\beta - 1) \, \beta^4 \, (867 \, \beta - 359),
    		\\ 
    		c_2 &= - 288 \, (5 - 12 \, \beta )^2.
    	\end{align*}
        \evs{When} we evaluate $\beta=\sqrt{2}-1$, we get
    	\begin{align*}
    		b_0 \approx 0.00849496 > 0 \quad \mbox{and} \quad  c_0 \approx -0.249567 < 0
    	\end{align*}
        which means that the Turing bifurcation is always supercritical within the large-parameter scaling, as $\delta\to 0^+$. \evs{In contrast, we know the bifurcation is sub-critical as $\delta\to 0^+$ in the small-parameter scaling of Sec.~\ref{sec:4}. Therefore, by continuity, we can compute a path of Turing bifurcations as $\beta$ varies for small $\delta$, upon which we must find a degenerate Turing bifurcation.}

	\subsection{Glycolysis and Sel'kov-Schnakenberg models} 
        In this case, it is helpful to take a slightly different scaling to the general case, for the large-parameter limit:
    	\begin{align*}
    		(a, b, c, d, h) = \left(\delta \, \alpha, \delta^{-1} \, \beta, 1, \delta^{-2} \, \eta, 0\right).
    	\end{align*}
        This is done because we have only three parameters $a, b, d$, so it is useful to scale $d$ in such a way that it now plays a role.
    	
    	Therefore, when grouping all the terms that depend on $\delta$, there must exist a function $\nu = \nu(\delta)$ such that the Turing bifurcation expression, \eqref{eq:TuringCond}, becomes
    	\begin{align*}
    	    2 \, \beta^2 - \tau^2 \, (\tau + 1)^2 + \nu \, \delta^2 = 0,
    	\end{align*}
    	where $\tau = \sqrt{\left(\beta + \alpha \, \delta\right)^2 + \eta}$, and
    	\begin{align*}
    	    \lim_{\delta\to 0^+} \nu(\delta) = - 2 \,  \alpha \, \beta \, \left(3 \, \sqrt{\beta^2 + \eta} + 2 \, \beta^2 + 2 \, \eta\right).
    	\end{align*}
        The leading order part of this expression can be reduced to 
    	\begin{align*}
    	    \beta = \frac{\tau \, (\tau + 1)}{\sqrt{2}}>0.
    	\end{align*}
    	We highlight here that the Turing bifurcation curve depends on $\delta$ only in the Sel'kov-Schnakenberg model. For the glycolysis model, the bifurcation curve is $\delta$-invariant.
    	
        \evs{Furthermore,} the expression for the third-order coefficient can be written \evs{as}
    	\begin{align*}
    		C_3 = \frac{b_0 + \mathcal O\left(\delta^2\right)}{c_2 \, \delta^2 + \mathcal O\left(\delta^4\right)},
    	\end{align*}
    	where
    	\begin{align*}
    		b_0 &= - 4 \, \tau^{13} \, (\tau + 1)^2 \, (\tau \, (2 \, \tau \, (\tau \, (\tau + 2) (8 \, \tau + 23) - 14) - 27) + 4),
    		\\ 
    		c_2 &= - 144 \, \tau^{12} \, (\tau + 1)^2,
    	\end{align*}
        This implies that the Turing bifurcation is sub-critical in the limit as $\delta\to 0^+$, provided
    	\begin{align*}
    		2 \, (\tau + 2) (8 \, \tau + 23) \, \tau^3 + 4 < \tau \, (28 \, \tau + 27)\evs{,}
    	\end{align*}
        which can be rewritten as 
        \begin{align*}
        	0 < \sqrt{\beta^2 + \eta} < 0.138407 \ldots,
        \end{align*}
    	and super-critical when 
    	\begin{align*}
    		0.138407 \ldots < \sqrt{\beta^2 + \eta} < \frac{2 - \sqrt{2}}{\sqrt{2}}.
    	\end{align*}
        \evs{Moreover,} there is no Turing bifurcation in this limit when $\sqrt{\beta^2 + \eta} > \frac{2 - \sqrt{2}}{\sqrt{2}}$.
    	
\begin{figure}
    		\centering
    		\includegraphics[width=0.45\textwidth]{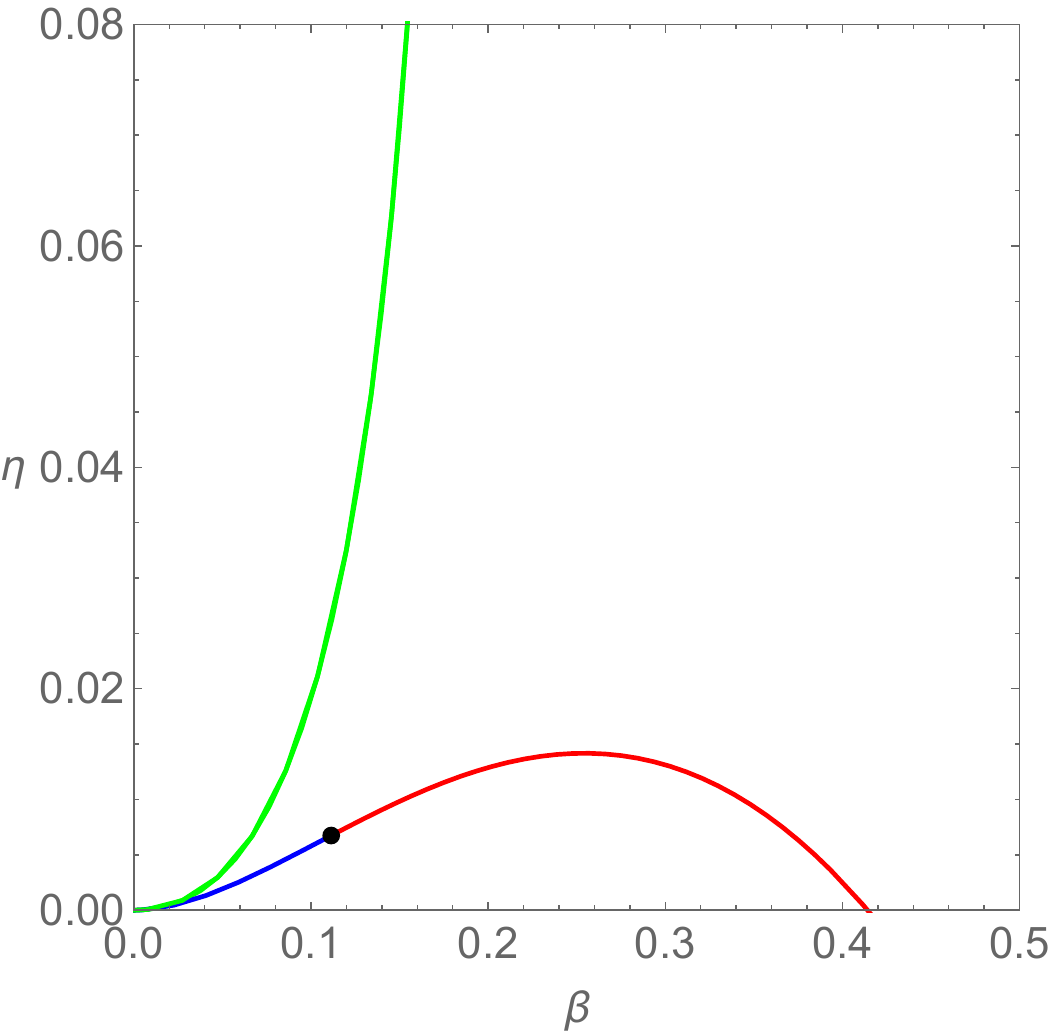}
    		\caption{Similar to Fig.~\ref{fig:transition} but for the Glycolysis and Selkov-Schnakenberg models when $(\alpha,\delta)=(1,0.001)$, in the $(\beta,\eta)$-plane.}
    		\label{fig:turingglyco}
    	\end{figure}

    	This implies that there is only one change of criticality of the Turing bifurcation which\evs{,} in the limit $\delta\to 0^+$, occurs when 
    	\begin{align*}
    		\sqrt{\beta^2 + \eta}\approx 0.138407.
    	\end{align*}
    	For instance, if we take the particular case when $\alpha = 1$ and $\delta = 0.001$, we get the two-parameter bifurcation diagram shown in Fig.~\ref{fig:turingglyco}. 
	
\section{Explicit formulae for the Brusselator} \label{sec:7} 
    In the case of the Brusselator, we can compute many expressions exactly under a different scaling.
    \begin{result}
        Consider the following parameter scaling for the Brusselator model:
        \begin{align}
            (a, c) = \left(\delta^{-1} \, \alpha, \gamma\right) \label{eq:brus_scale}
        \end{align}
        The codimension-one Turing bifurcation curve in the $(\alpha, \gamma)$-plane is given by $\gamma = (\alpha + 1)^2 + 1$, and is independent of  $0 < \delta < 1$. Moreover, there are two points on this curve, $(\alpha_{1, 2}, \gamma_{1, 2})$ with $\alpha_1<\alpha_2$ \evs{and $\gamma_1<\gamma_2$}, whose locations are also $\delta$-independent, in which the criticality changes such that the bifurcation is sub-critical for $\alpha < \alpha_1$ and $\alpha > \alpha_2$, whilst it is super-critical for $\alpha_1 < \alpha < \alpha_2$. \evs{Furthermore,} the normal form coefficient $C_5$ is negative at each of these codimension-two bifurcation points, for all $0 < \delta < 1$.
    \end{result}
        
    \begin{proof}
        Under the scaling \eqref{eq:brus_scale}, the condition \eqref{eq:TuringCond} for
        a Turing bifurcation is reduced to 
        \begin{align*}
            \gamma = \left(\alpha + 1\right)^2 + 1, 
        \end{align*}
        which is  independent of $\delta$, as required. 
        
        Next, explicit computation of the third-order normal-form coefficient gives 
        \begin{align*}
            C_3 = \frac{b_0}{36 \, \delta^2 \left(\delta^2 - 1\right)},
        \end{align*}
        where
        \begin{align*}
            b_0 &= - \alpha \, (\alpha + 2) (\alpha \, (8 \, \alpha - 21) + 4).
        \end{align*}
        Thus\evs{,} the Turing bifurcation is sub-critical if and only if $\alpha \, (8 \, \alpha - 21) + 4 > 0$ \evs{for $0<\delta<1$}. This implies there are two critical $\alpha$-values at which there are degenerate Turing bifurcations
        \begin{align}
            \alpha_1 = \frac{1}{16} \left(21 - \sqrt{313}\right)\approx 0.206762, \qquad \alpha_2 = \frac{1}{16} \left(21 + \sqrt{313}\right)\approx 2.41824. \label{eq:alpha12}
        \end{align}
        with the bifurcation \evs{being} super-critical for $\alpha \in (\alpha_1, \alpha_2)$ and sub-critical otherwise.
    \end{proof}
    
\begin{figure}
        \centering
        \includegraphics[width = 0.45\textwidth]{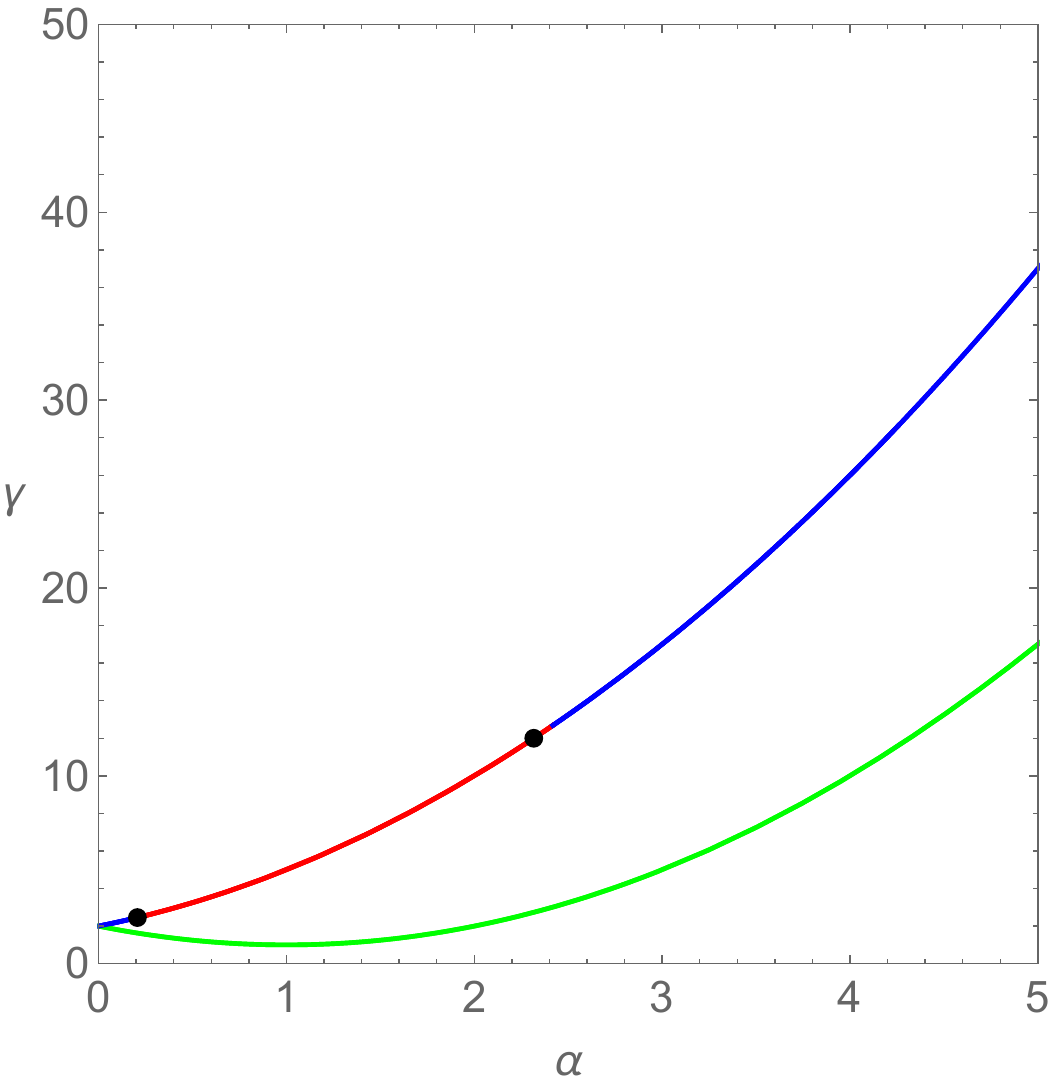}
        \caption{Similar to Fig~\ref{fig:transition} but for the Brusselator model.}
        \label{fig:2transitions}
    \end{figure}

    The bifurcation diagram in the case that $\delta = 0.1$ is shown in \evs{Figure} \ref{fig:2transitions}, where we can see all our theoretical findings visually.
    
\begin{figure}
        \begin{subfigure}[b]{\textwidth}
            \centering
            \includegraphics[width = 0.6\textwidth]{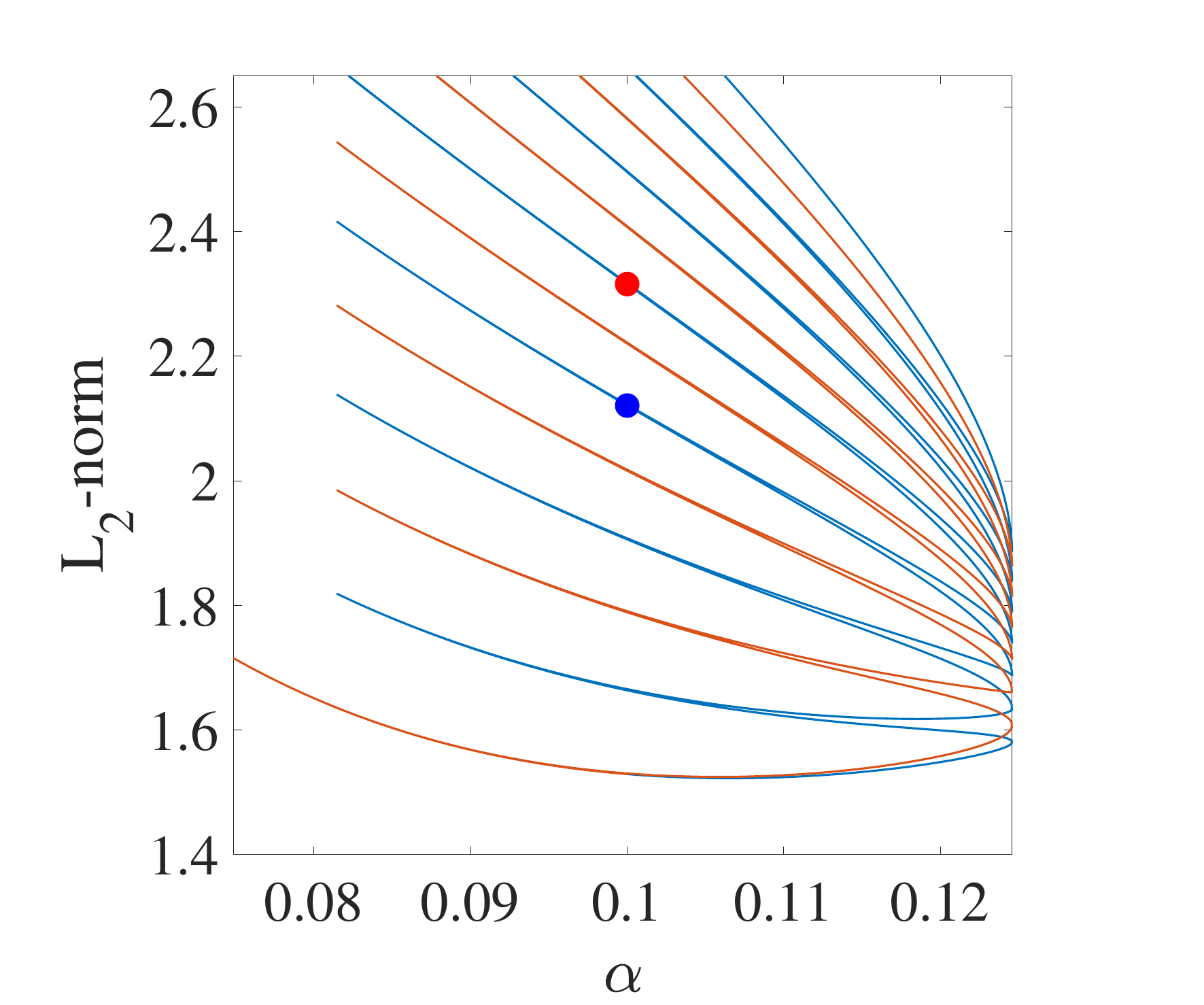}
            \caption{}
            \label{fig:leftsnaking}
        \end{subfigure}
        \\
        \begin{subfigure}[b]{0.45\textwidth}
            \includegraphics[width = \textwidth]{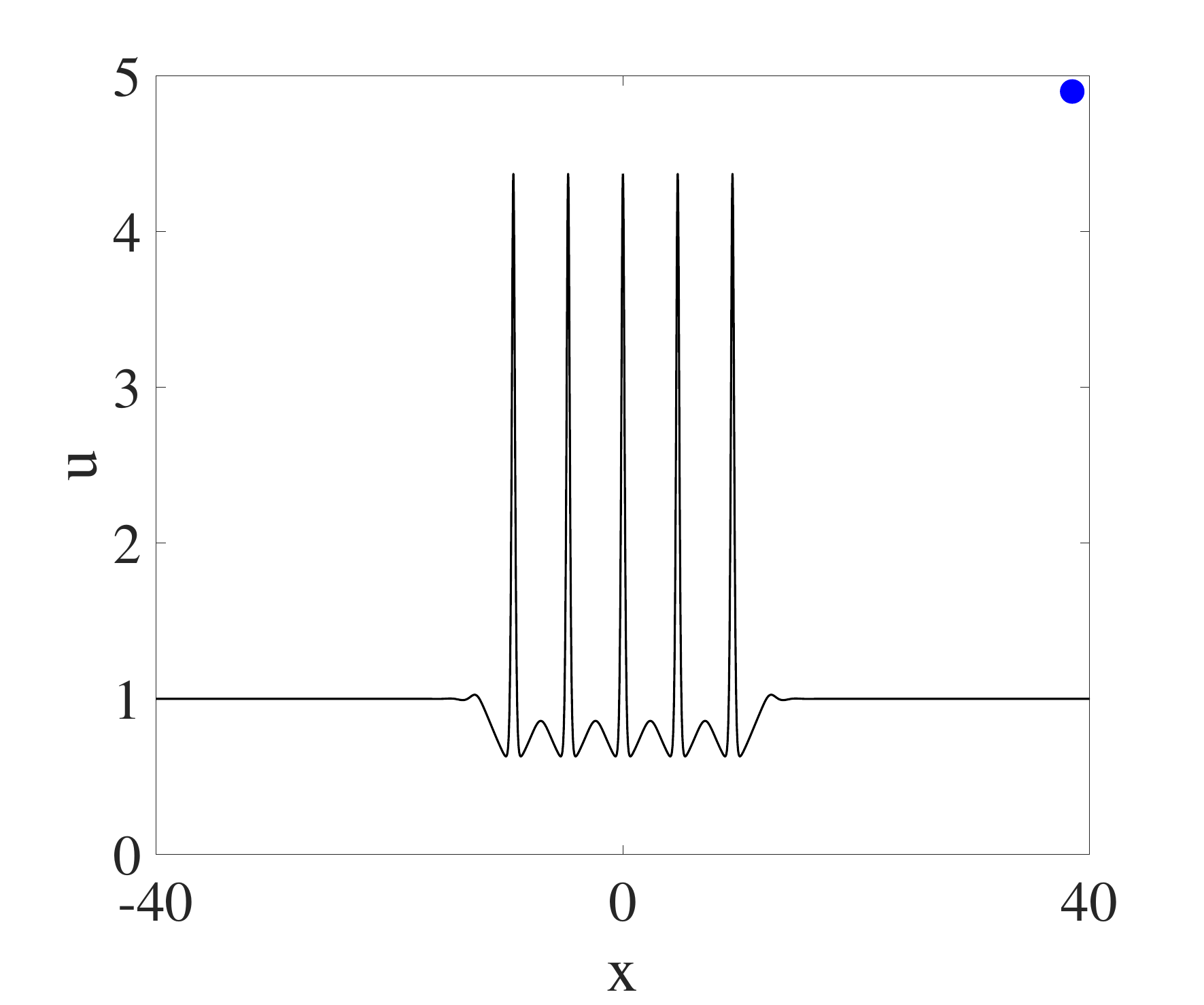}
            \caption{}
        \end{subfigure}
        \begin{subfigure}[b]{0.45\textwidth}
            \includegraphics[width = \textwidth]{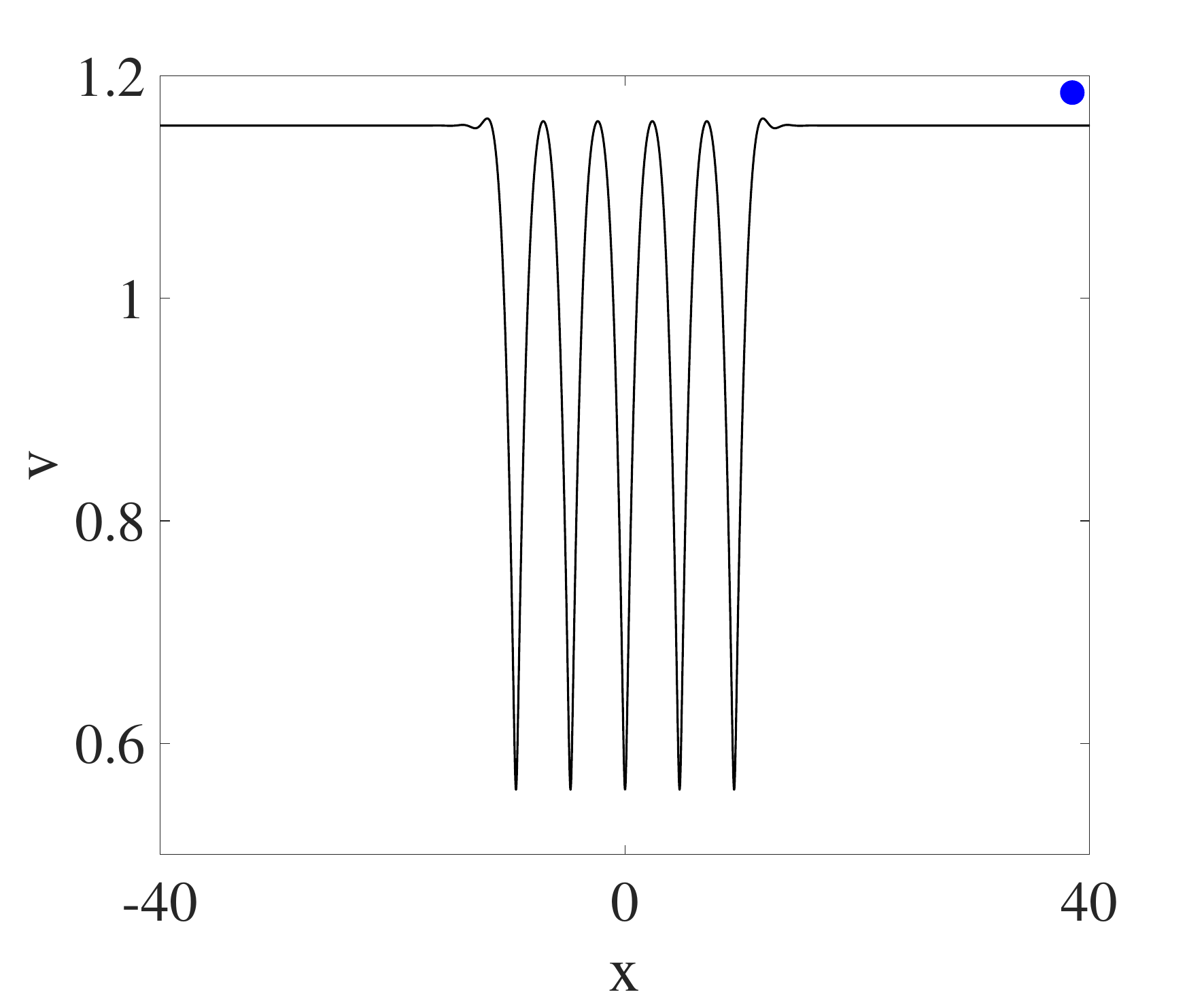}
            \caption{}
        \end{subfigure}
        \\
        \begin{subfigure}[b]{0.45\textwidth}
            \includegraphics[width = \textwidth]{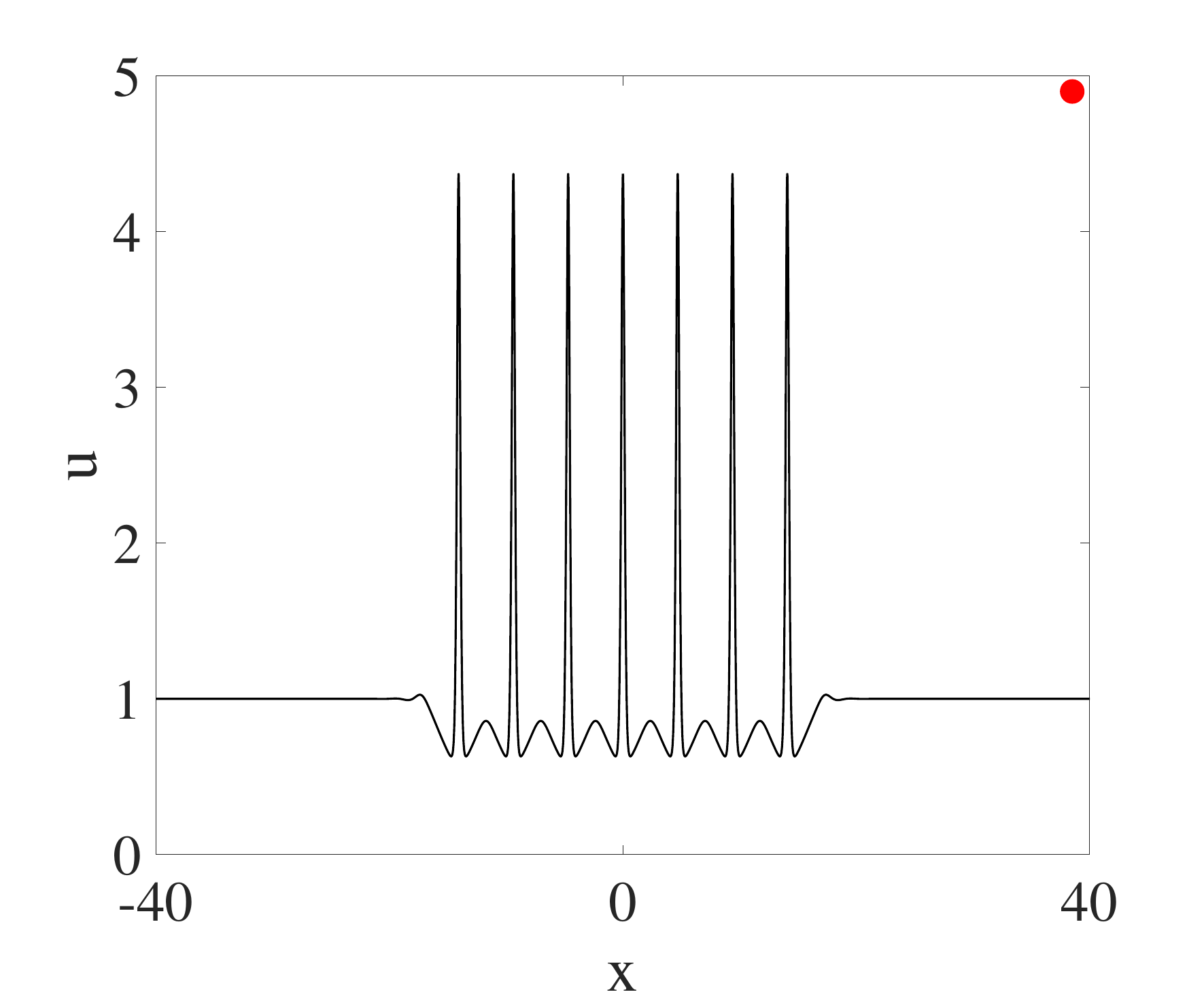}
            \caption{}
        \end{subfigure}
        \begin{subfigure}[b]{0.45\textwidth}
            \includegraphics[width = \textwidth]{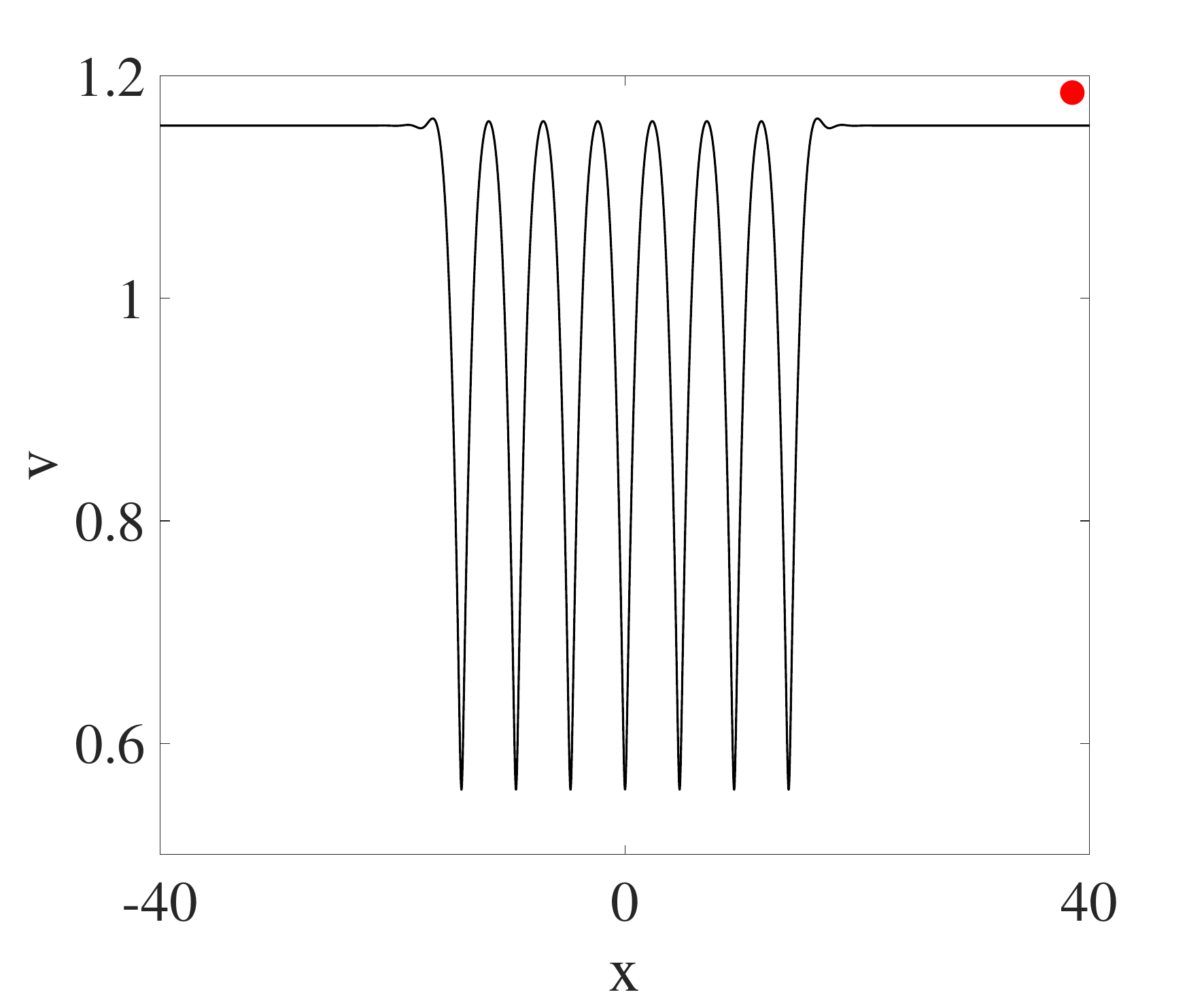}
            \caption{}
        \end{subfigure}
        \caption{(a) Homoclinic snaking for the Brusselator model on the subcritical side of the left-hand codimension-two point in Figure \ref{fig:2transitions} for $\delta = 0.1$ (b)-(e) Shapes of the variables $u$ and $v$ in space for points marked in (a).}
        \label{fig:snakings}
    \end{figure}

\begin{figure}[tbp]
        \begin{subfigure}[b]{\textwidth}
            \centering
            \includegraphics[width = 0.7\textwidth]{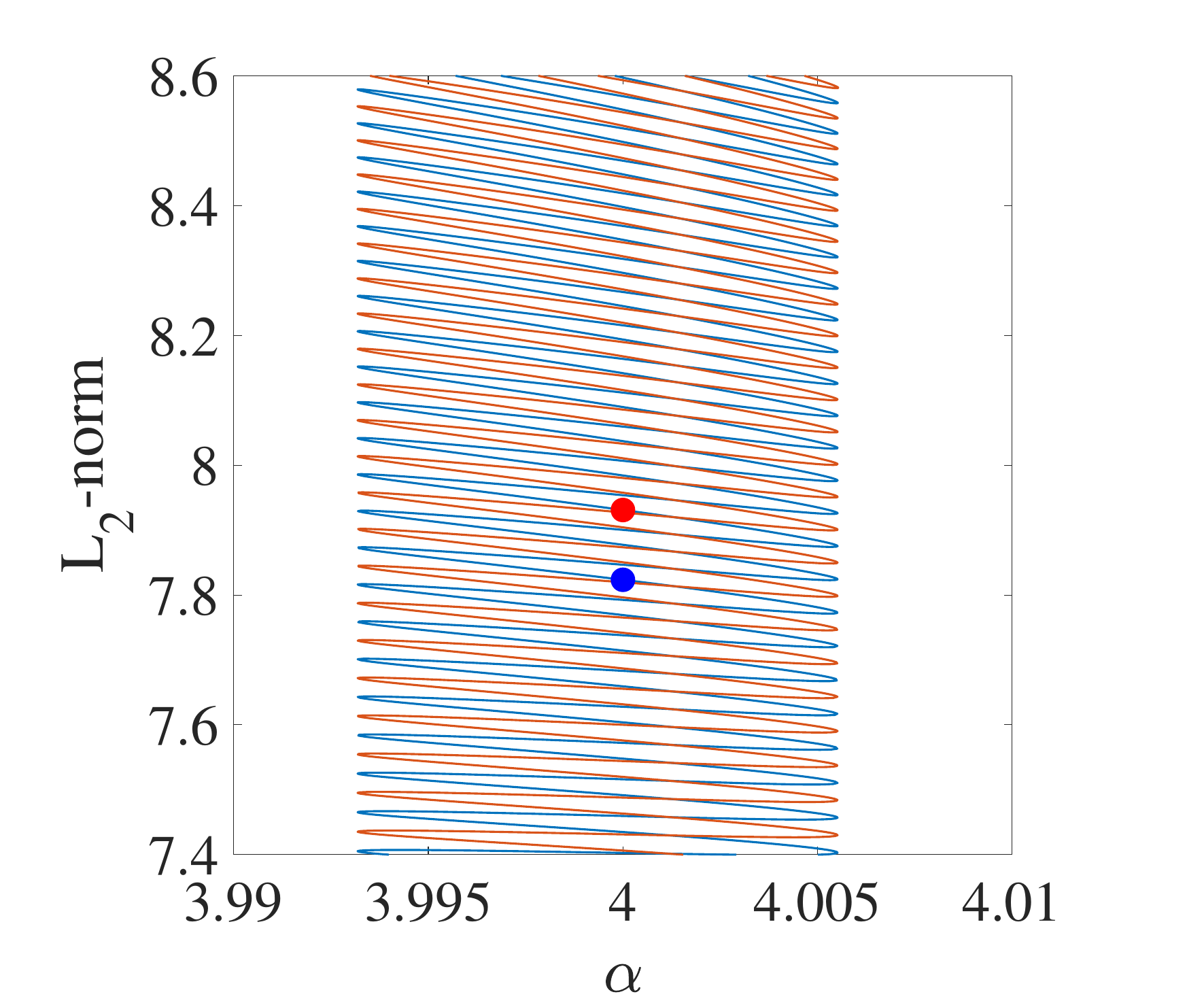}
            \caption{}
            \label{fig:rightsnaking}
        \end{subfigure}
        \\
        \begin{subfigure}[b]{0.45\textwidth}
            \includegraphics[width = \textwidth]{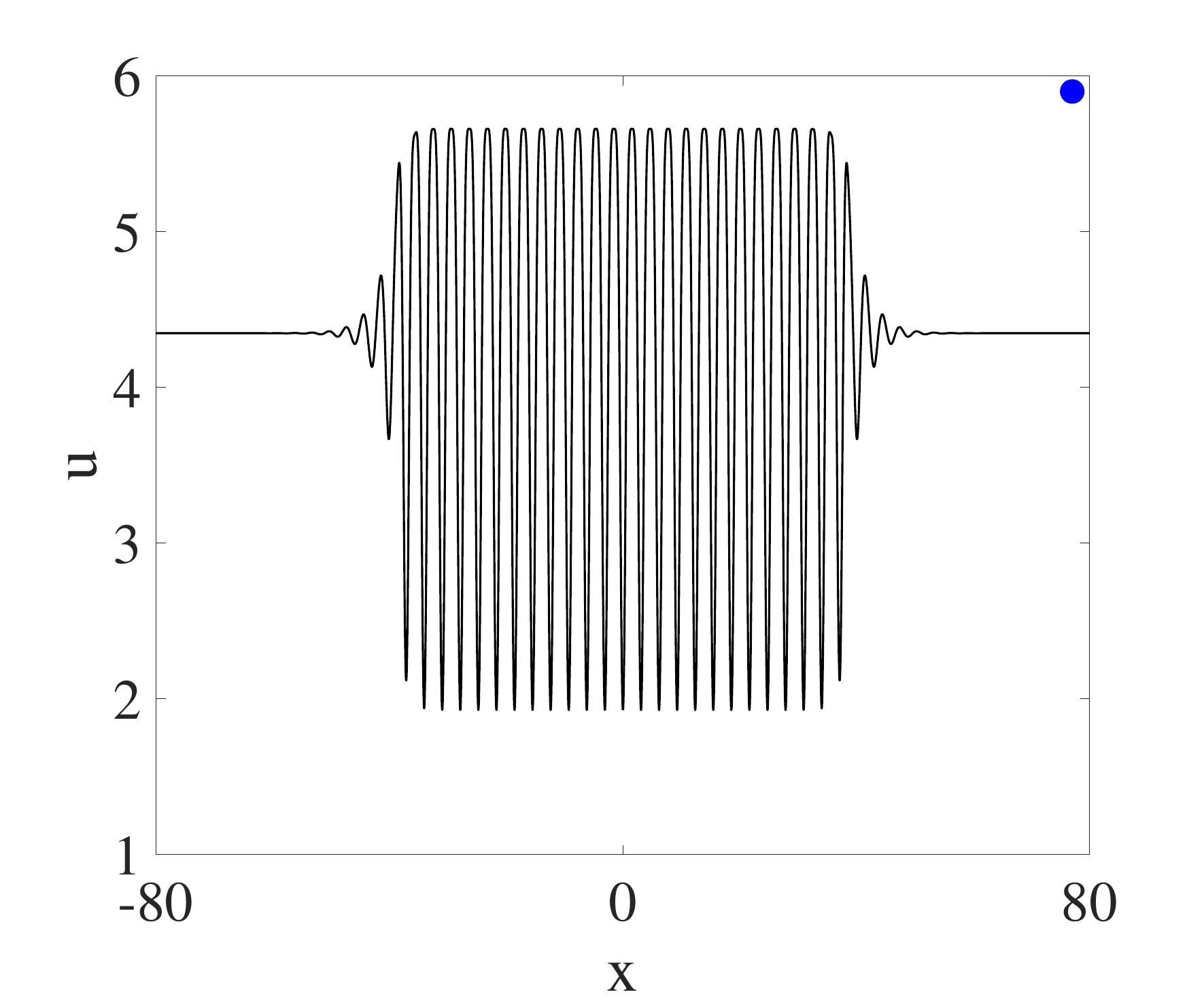}
            \caption{}
        \end{subfigure}
        \begin{subfigure}[b]{0.45\textwidth}
            \includegraphics[width = \textwidth]{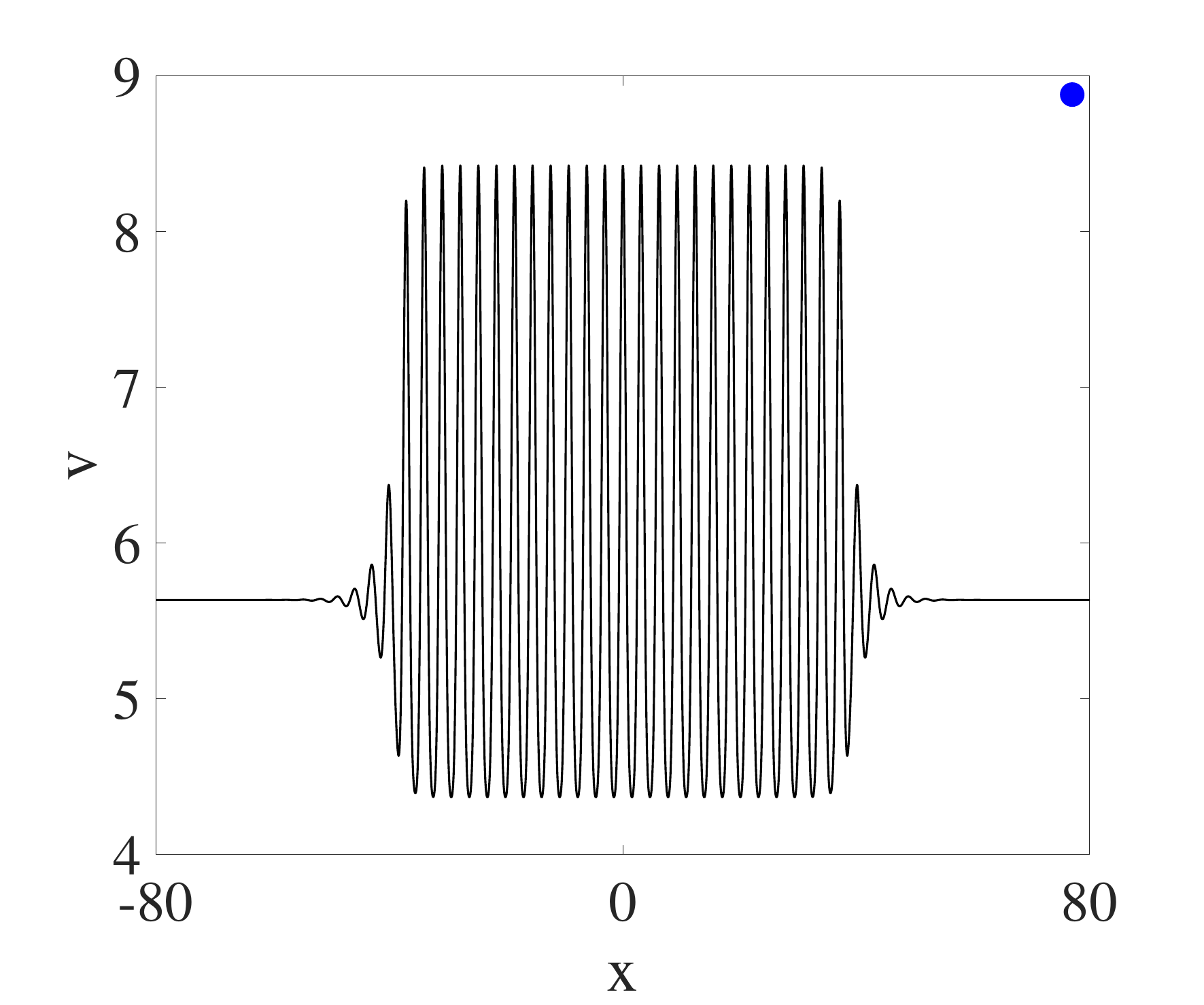}
            \caption{}
        \end{subfigure}
        \\
        \begin{subfigure}[b]{0.45\textwidth}
            \includegraphics[width = \textwidth]{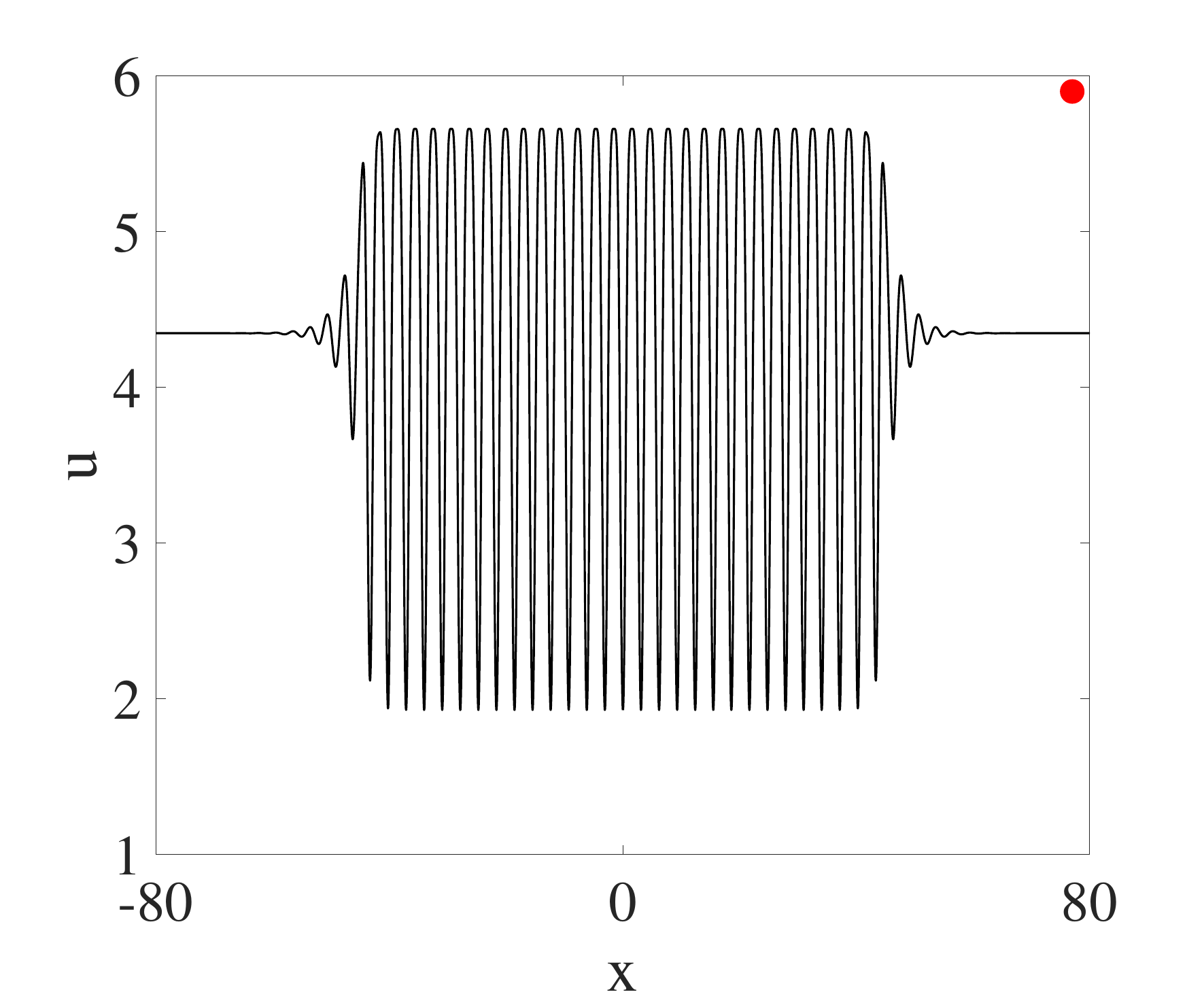}
            \caption{}
        \end{subfigure}
        \begin{subfigure}[b]{0.45\textwidth}
            \includegraphics[width = \textwidth]{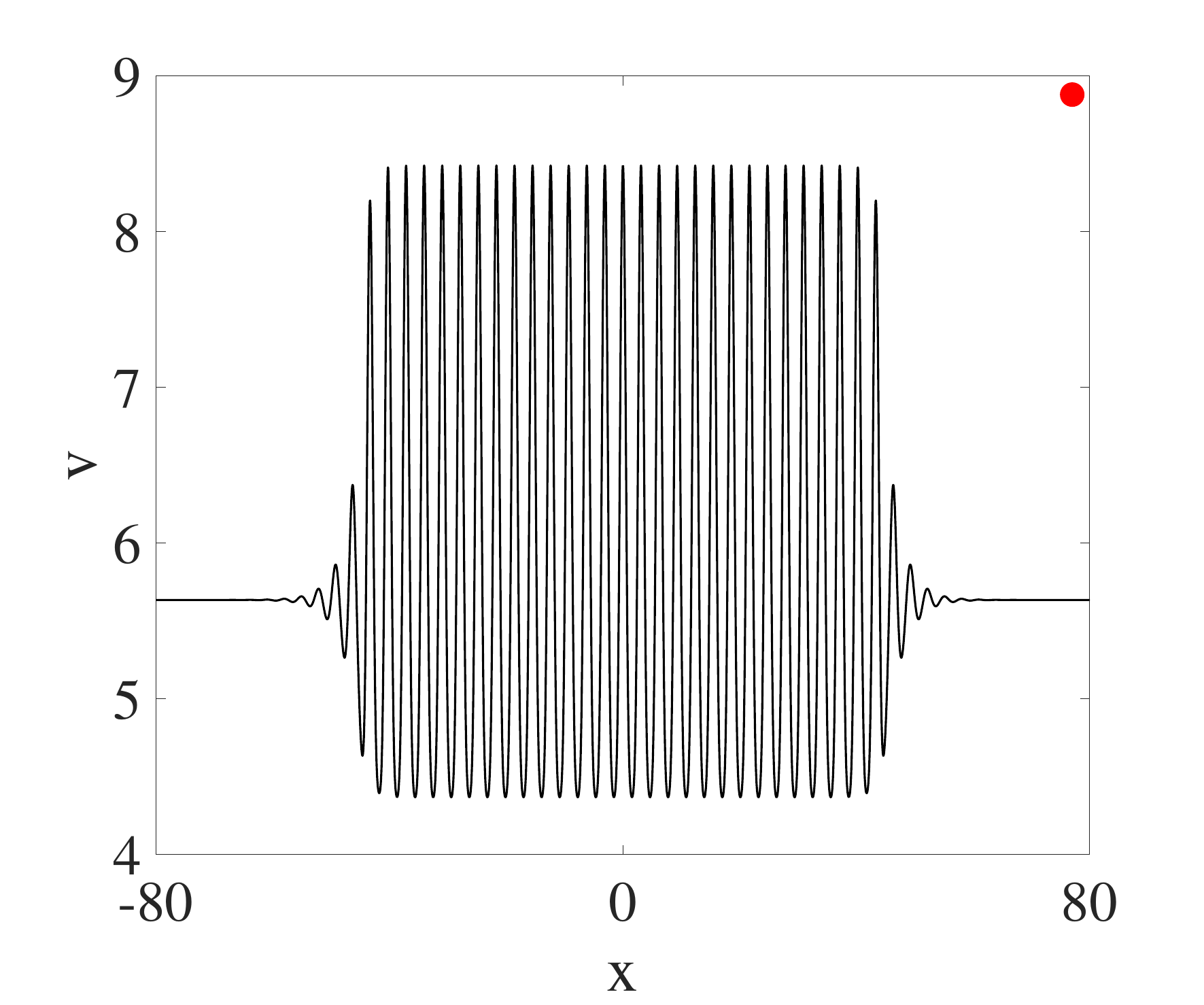}
            \caption{}
        \end{subfigure}
        \caption{Similar to Fig.~\ref{fig:snakings} but for $\alpha$-values beyond the right-hand degenerate Turing bifurcation in Fig.~\ref{fig:2transitions}, for $\delta = 0.92$.}
        \label{fig:rightsnakings}
    \end{figure}

    Figures \ref{fig:leftsnaking} and \ref{fig:rightsnaking} illustrate that the two codimension-two points give rise to two independent regions of localised patterns organised in snaking bifurcation curves. The values of $\delta$ differ in the two cases illustrated only for numerical convenience; Fig.~\ref{fig:leftsnaking} illustrates the localised patterns that emerge from the degenerate Turing bifurcation for $\alpha < \alpha_1$, and Fig.~\ref{fig:rightsnaking} illustrates the case for $\alpha > \alpha_2$ (where $\alpha_{1, 2}$ are defined in \eqref{eq:alpha12}). Solutions on each of the two primary intertwined curves of each snake, represent a homoclinic orbit in the spatial dynamics that has an increasing number of spikes, as the curve grows in norm; see Figs.~\ref{fig:leftsnaking}(b)-(e) and \ref{fig:rightsnaking}(b)-(e).

\begin{figure}
        \centering
        \begin{subfigure}[b]{0.45\textwidth}
            \centering
            \includegraphics[width=\textwidth]{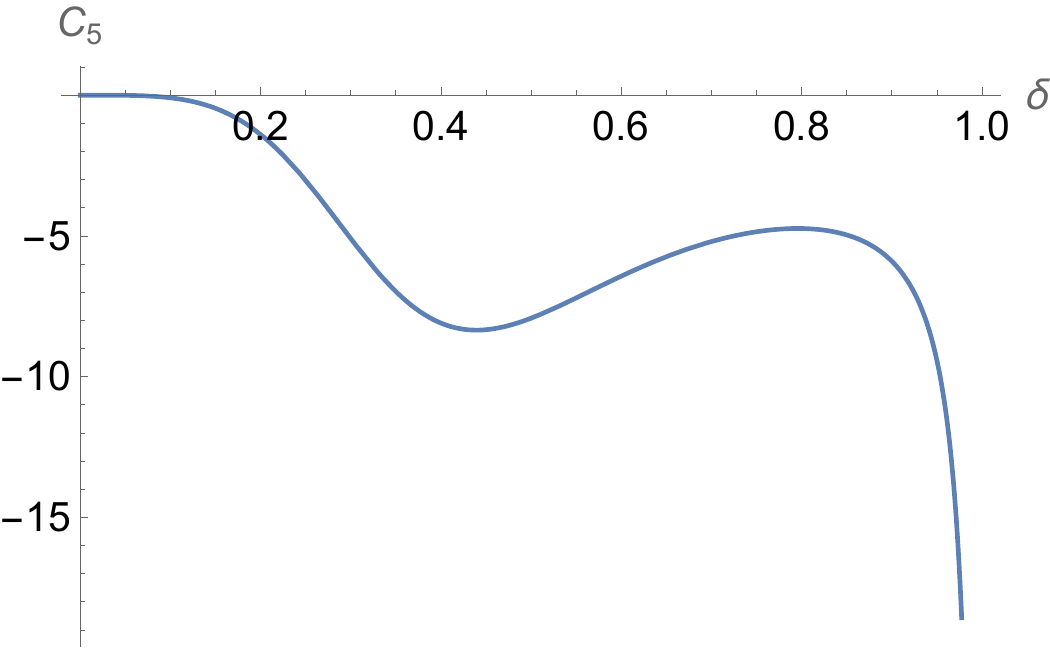}
            \caption{}
            \label{fig:5smallerpoint}
        \end{subfigure}
        \hfill
        \begin{subfigure}[b]{0.45\textwidth}
            \centering
            \includegraphics[width=\textwidth]{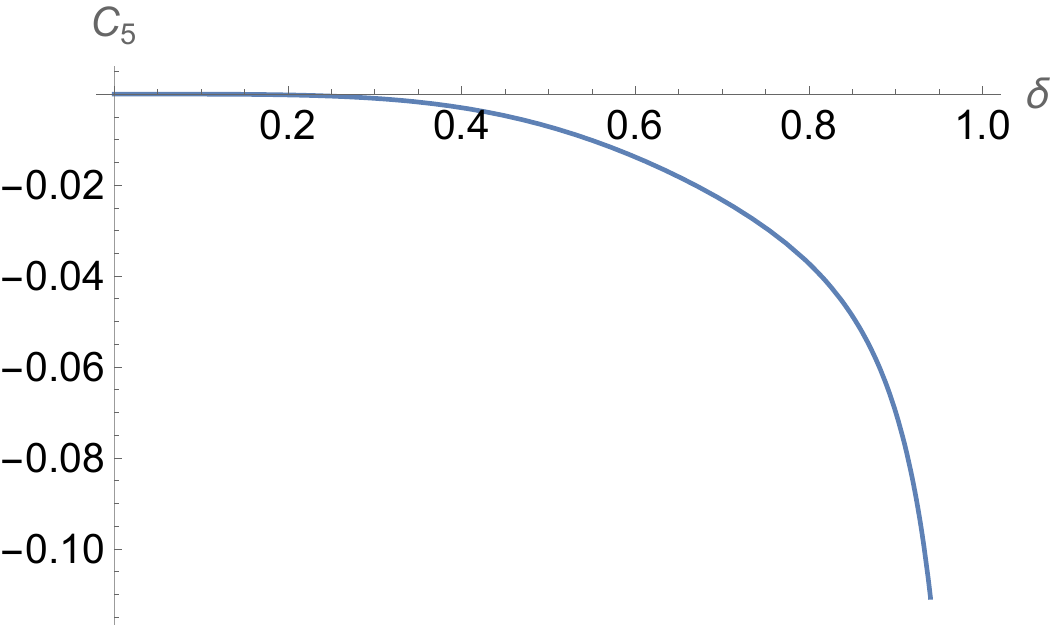}
            \caption{}
            \label{fig:5biggerpoint}
        \end{subfigure}
        \caption{Value of the fifth-order coefficient $C_5$ as a function of $0 < \delta < 1$ for (a) the left-hand and (b) the right-hand degenerate Turing bifurcations of the Brusselator model in Fig.~\ref{fig:2transitions}}
        \label{fig:fifthorder}
    \end{figure}

    One advantage of the Brusselator example is that the Turing bifurcation curve and its criticality are independent of $0 < \delta < 1$. It is then straightforward to get a closed-form \evs{expression} of the fifth-order coefficient at each of the codimension-two points. The expression is cumbersome however, so we simply choose to plot graphs of function $C_5(\delta)$ for the two codimension-two points; see Fig.~\ref{fig:fifthorder}. Note that for both, the fifth-order coefficient is indeed negative for each $0 < \delta < 1$ but \evs{have} a vertical asymptote at $\delta = 1$, just like the third-order coefficient.
    
    Finally, given the sign of $C_5$, then, according to the theory in \ref{sec:maxwell}, emanating from each of the degenerate Turing bifurcation points\evs{,} there must be a curve in the parameter plane at which there is a Maxwell point. \evs{That point corresponds to a  heteroclinic connection between the flat state and the periodic pattern within the normal form. Then, according to the theory of homoclinic snaking (see \cite{Woods_Intro}), terms which break the normal-form symmetry will then generically lead to homoclinic snaking occurring in a parameter wedge for which this Maxwell curve forms the ``backbone''.}

\begin{figure}[]
        \centering
        \includegraphics[scale=0.35]{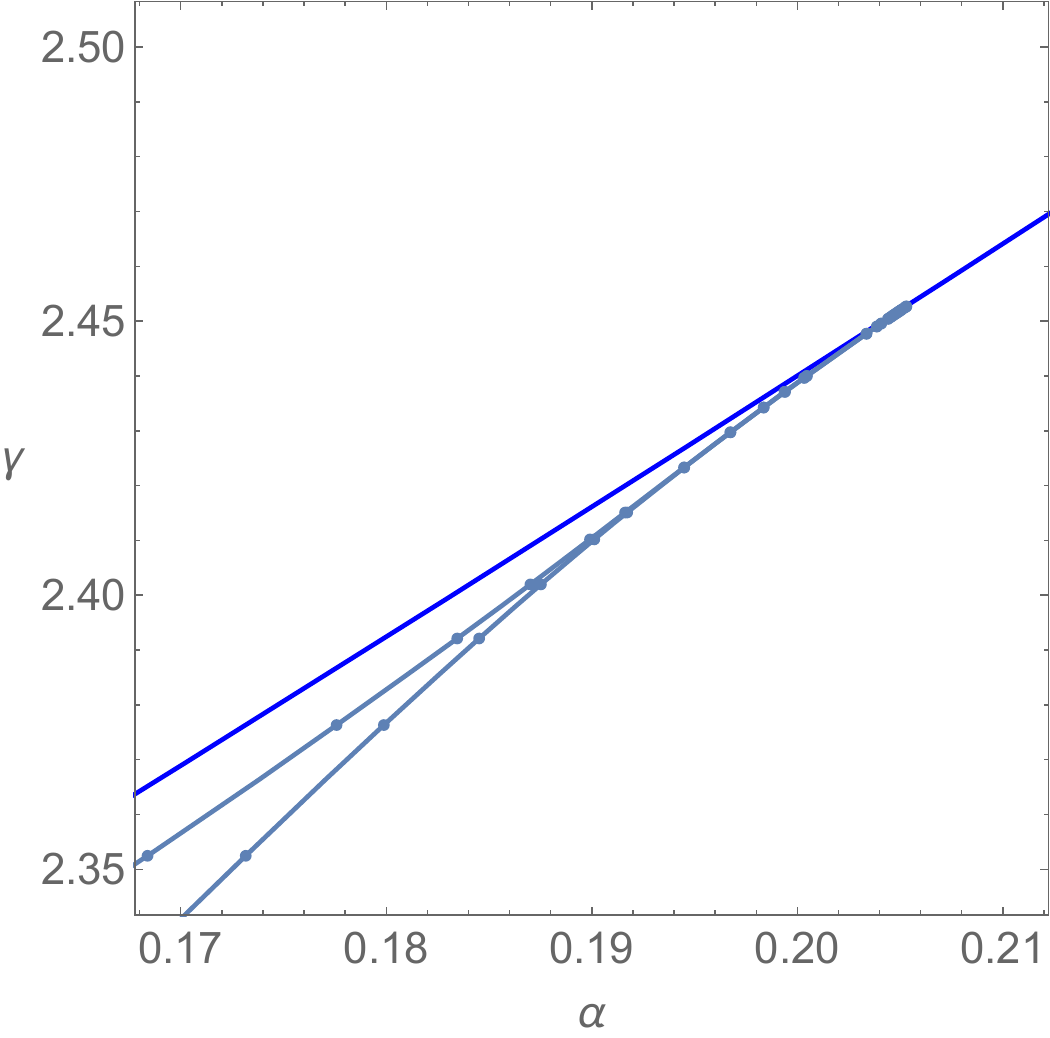}
        \caption{Two-parameter diagram in $\alpha$ and $\gamma$ for $\delta = 0.1$, close to the codimension-two point at $\alpha = 0.206762124$ of the Brusselator model. Depicted is the Turing bifurcation curve (blue) and the outermost fold points (interpolated dots) of the homoclinic snake depicted in Fig.~\ref{fig:leftsnaking}.}
        \label{fig:nontranslatedmaxwell}
    \end{figure}

    Figure \ref{fig:nontranslatedmaxwell} shows the Turing bifurcation curve together with the outermost fold points obtained for different values of $\alpha$ and $\gamma$ close to the small-$\alpha$ codimension-two point. In the limit as the codimension-two point is approached\evs{,} the numerical continuation of these folds becomes very poorly conditioned and fails, although the outer limits of one-parameter continuation can be detected reliably, which we have joined using splines.

\begin{figure}
			\centering
		\begin{subfigure}[b]{0.48\textwidth}
			\centering
	        \includegraphics[width=\textwidth]{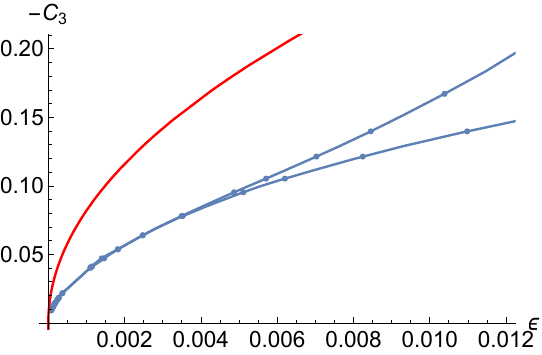}
	        \caption{ }
            \label{fig:wrongmaxwellcurve}
		\end{subfigure}
		\hfill
		\begin{subfigure}[b]{0.48\textwidth}
			\centering
            \includegraphics[width=\textwidth]{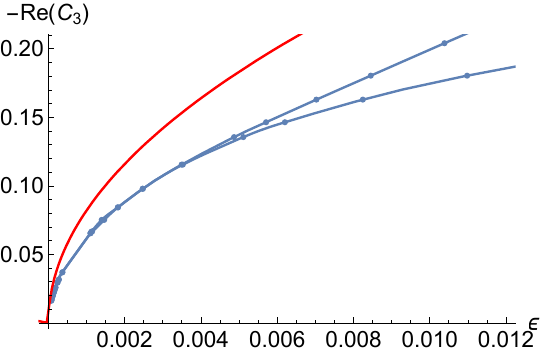}
			\caption{ }
            \label{fig:rightmaxwell}
		\end{subfigure}
		\\
		\begin{subfigure}[b]{0.5\textwidth}
		      \centering
		      \includegraphics[width=\textwidth]{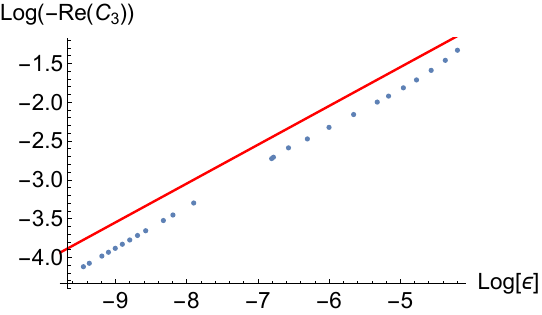}
		      \caption{ }
		      \label{fig:loglogplot}
		\end{subfigure}
	    \centering
	    \caption{(a) The blue curve shows the same information as in Fig.~\ref{fig:nontranslatedmaxwell}, with the wavenumber determined by its value at the Turing bifurcation curve, plotted as $-\Re(C_3)$ against $\varepsilon$. The red curve shows the asymptotic prediction \eqref{eq:Maxwell} for this curve as $\varepsilon\to 0$. (b) The same information as (a) but with the numerical values of $C_3$ determined using the enhanced expression \eqref{eq:newwavenumber} of the wavenumber. (c) The same information as (b) but as a log-log plot} 
	    \label{fig:maxwellcurve1} 
	\end{figure}

    To make a comparison with the theory of section \ref{sec:maxwell}, we define the parameter $\varepsilon = \left|\gamma - \gamma_0\right|$, such that $\varepsilon = 0$ represents the Turing bifurcation curve. We want to compare the behavior of $C_3$ around a codimension-two bifurcation point for the theoretical Maxwell curve given in \eqref{eq:Maxwell} and the family of fold points that delineate the outside of the homoclinic snaking region. To do this\evs{,} we fix $\alpha = \frac{1}{16}\left(21 - \sqrt{313}\right)$ to compute the theoretical prediction of the Maxwell curve, using $\gamma_0 = \left(\alpha + 1\right)^2 + 1$, while letting $\alpha$ and $\gamma$ vary for the fold curves. We expect these two curves to behave in the same way as $\varepsilon\to 0^+$. Figure \ref{fig:wrongmaxwellcurve} shows both curves obtained using this setting. At first glance, there does not seem to be good evidence that the Maxwell curve provides the true asymptotic limit of curves of folds.

    To make a better asymptotic approximation, it is helpful to consider how the wave number varies as $\varepsilon\to 0^+$. For this reason, note that \eqref{zerodet} for the Brusselator is given by
    \begin{align*}
        \delta^2 \, k^4 + \left(\alpha^2 - \gamma + 2\right) \, k^2 + \delta^{-2} \, \alpha^2 = 0,
    \end{align*}
    which implies that
    \begin{align}
        k_\pm^2 &= \frac{\gamma - \alpha^2 - 2 \pm \sqrt{\left(\gamma - (\alpha - 1)^2 - 1\right)\left(\gamma - (\alpha + 1)^2 - 1\right)}}{2}. \label{eq:newwavenumber}
    \end{align}
    However, note that the Maxwell curve exists in a region `below' the Turing bifurcation curve (that is, where the homogeneous equilibrium is stable). So, the values of $k_\pm^2$ are complex. For this reason, instead of evaluating $C_3$, we need to consider $\Re(C_3)$ inside the formula \ref{eq:Maxwell}. The graph obtained with this modification is shown in Figure \ref{fig:rightmaxwell}. It is clear now that the asymptotic behavior between these two curves is approximately the same in the limit as $\varepsilon\to 0^+$.

\begin{figure}
        \centering
        \includegraphics[scale=0.4]{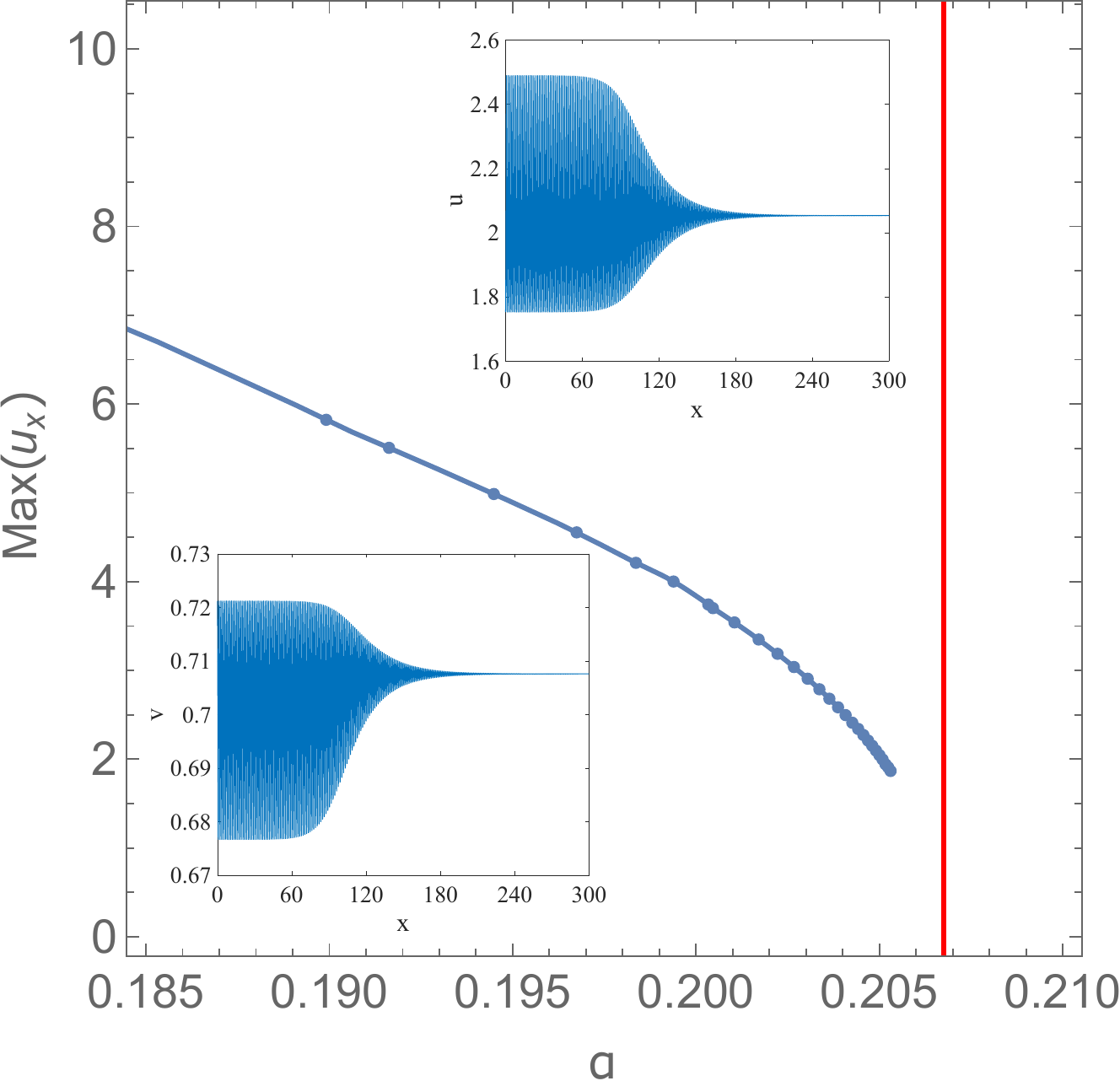}
        \caption{Numerically obtained maxima of $\max\left(\partial_x u(x)\right)$ of the periodic pattern in the centre of the localised patterns against $\alpha$ as we follow one of the outermost folds of the wedge of homoclinic snaking in the $(\alpha, \gamma)$-plane in Fig.~\ref{fig:nontranslatedmaxwell} approaching the codimension-two point $(\alpha_1, \gamma_1)$. Insets show the $u$- and $v$-components at the final computed point, closest to the vertical line $\alpha = \alpha_1$.}
        \label{fig:maxu_x}
    \end{figure}

    A particular justification of why it is necessary to use this enhanced version of the formula \eqref{eq:Maxwell} can be seen in Fig.~\ref{fig:maxu_x}, where we plot the $u$ and $v$ solutions of localised solution at the last computed vertical asymptote of a snaking curve at an $\alpha$-value approximately 0.001 away from $\alpha_1$\evs{.} Note the shear number of oscillations enclosed within this localised state --- which explains the numerical difficulty of computing much closer to $\alpha_1$\evs{.} Yet, the two-scale nature of the purely spatial problem means that the norm of \evs{the} $u$-component is still $\mathcal{O}(1)$. This effectively means that\evs{,} computationally\evs{,} we have not yet reached a sufficiently small neighbourhood of $\alpha_1$, such that the norm of the solution is sufficiently small for the normal form theory to apply. However, \evs{setting $k$ as a function of the parameters of the system} makes a sufficiently strong adjustment\evs{, allowing} a good comparison between theory and numerics.

\section{Discussion}\label{sec:8}
    This paper has expanded on and generalised the results in \cite{FahadWoods} on a broad class of activator-inhibitor reaction-diffusion equations on the real line. These models include the well-studied Schnakenberg and Brusselator models, which have been successfully used to explain observed behaviour in many naturally-occurring systems that give rise to spontaneous pattern formation.  We have sought to address the question of whether the regions in \evs{different} parameter planes in which localised pattern formation is found are\evs{,} in some sense\evs{,} generic, or whether \evs{they} were merely numerical quirks of the particular parameter values chosen there. By considering various asymptotic scalings we have been able to prove by explicit construction that\evs{,} for each \evs{model,} there is a limit in which the onset of patterns via a Turing bifurcation is always sub-critical and another limit in which it is super-critical. In the case of the Brusselator model, all calculations can be carried out explicitly, without \evs{recourse} to any asymptotic approximation. \evs{The results can, in principle, be extended to a broader class of systems by performing a Taylor expansion about the equilibrium and performing the same weakly nonlinear analysis as in Section 2; see \cite{Edgardo_NFcomp}. But, the point has been to show that there is a class of systems for which the presence of super-to-subcritical transitions can be shown explicitly.} 
    
    We believe our results to be important because they provide one of the first categorical proofs that a wide class of activator-inhibitor models naturally contain open parameter regimes in which one should expect to see localised patterns through sub-critical Turing bifurcation and homoclinic snaking. We also show how these parameter regions connect to those in which one should expect to see periodic patterns through super-critical Turing \evs{bifurcations}. This result could be of use to those trying to construct mathematical models of natural processes, in some sense echoing the conclusion of \cite{MeronIssue} that a wide class of different patterns can often be found in the same models. Moreover, transitions observed in realistic experiments can sometimes be captured by a change in the parameter regime of a simplified mathematical model, without the need to include extra confounding biological factors.
    
    There are many theoretical issues not addressed in this paper, notably finite-domain effects, extension to multiple spatial dimensions, and questions of stability and transient dynamics. \evs{See, for example, \cite{andrew},} for a summary of the state of the art. Nevertheless\evs{,} our general result that problems of this class naturally \evs{exhibit in} regimes where we would expect to \evs{see} localised patterns is likely to be of more general significance to a much wider class of mathematical problems.  Another step missing from the present paper is \evs{an evaluation of the terms that are beyond all orders of the normal form that lead to the wedge of parameters leading to homoclinic snaking around the Maxwell point curve}, as was established by Kozyreff and Chapman \cite{Kozyreff,Kozyreff2} in the context of Swift-Hohenberg-like models. That such snaking should be generic was established by Beck {\em et al} \cite{Beck}, but it remains to calculate the specific beyond-all-order coefficients to establish the asymptotics of the wedge for the class of two-component reaction-diffusion equations considered here (although we do note the work \cite{DeWitt} which provides some preliminary results in that direction). Such a calculation \evs{will form the subject of future} work.
    
    Other \evs{future} work shall consider reaction-diffusion systems with more than two components, and in multiple space dimensions \cite{Edgardo_NFcomp}. There, not only can we find Turing bifurcations, but also finite wavenumber Hopf bifurcations (so-called \evs{wave} bifurcations) (see \cite{General_Conditions}) for which normal-form theory can also establish whether rotating waves or standing waves bifurcate super- or sub-\evs{critically}. Further extension shall consider extensions to bulk-surface reaction-diffusion equations as in \cite{Anotida,Paquin}.
    
    In this paper, we have developed a new formula for the coefficients of the normal form of a Turing bifurcation up to fifth order, which helped us to characterize the Turing bifurcations we found in the general model \eqref{vectorfield}. We highlight that the fifth-order coefficient was calculated at all the codimension-two bifurcation points we found and its value was always negative.
    
    On the other hand, for the Schnakenberg model, we did not find an explicit expression of a codimension-two bifurcation point. However, taking into account that there is only one Turing bifurcation curve for all the models and, in the limit when some parameters get sufficiently large or small, there is a change of criticality in each model then, by continuity, there must exist a point $\left(a,b,c,d,h\right)\in \left(\mathbb R_+\right)^5$ in which the third-order coefficient is equal to zero, even for the Schnakenberg model. 

\subsection*{Acknowledgements}
    The work of EV-S was supported by ANID, Beca Chile Doctorado en el extranjero, \evs{number} 72210071.

	\bibliographystyle{siamplain}
	\bibliography{references}

\appendix

    \section{Normal form computation up to order 5.} 
        \paragraph{Order 3 continued.}
        To compute higher order terms, it is necessary to compute $\mbf W^{[3]}$ explicitly. To find that term, we can substitute $h^{[3]}$ into \eqref{thirdorder} to get 
		\begin{multline}
			\begin{pmatrix}
				f_{10} + D_1 \, \partial_{xx} & f_{01}
				\\
				g_{10} & g_{01} + D_2 \, \partial_{xx}
			\end{pmatrix} \mbf W^{[3]} = \begin{pmatrix}
				\phi_1
				\s 
				\phi_2
			\end{pmatrix} \cos(kx) \, h^{[3]}
			\s
			- A^3 \, \phi_1\left(\left(\alpha_1 + \frac{\beta_1}{2}\right)\cos(kx) + \frac{\beta_1}{2} \, \cos(3kx)\right)\begin{pmatrix}
				f_{20}
				\s 
				g_{20}
			\end{pmatrix}
			\s 
			- A^3 \, \phi_1 \left(\left(\alpha_2 + \frac{\beta_2}{2}\right)\cos(kx) + \frac{\beta_2}{2} \, \cos(3kx)\right)\begin{pmatrix}
				f_{11}
				\s 
				g_{11}
			\end{pmatrix}
			\s 
			- A^3 \, \phi_2 \left(\left(\alpha_1 + \frac{\beta_1}{2}\right)\cos(kx) + \frac{\beta_1}{2} \, \cos(3kx)\right)\begin{pmatrix}
				f_{11}
				\s 
				g_{11}
			\end{pmatrix}
			\s 
			- A^3 \, \phi_2 \left(\left(\alpha_2 + \frac{\beta_2}{2}\right)\cos(kx) + \frac{\beta_2}{2} \, \cos(3kx)\right)\begin{pmatrix}
				f_{02}
				\s 
				g_{02}
			\end{pmatrix}
			\s 
			- A^3 \, \mbf B \left(3 \, \cos(kx) + \cos(3kx)\right). \label{W3}
		\end{multline}
        As this equation is linear, then we know that its solution has the following form:
		\begin{align*}
			\mbf W^{[3]} = A^3\left(\begin{pmatrix}
				\gamma_1
				\s 
				\gamma_2
			\end{pmatrix} \cos(kx)+\begin{pmatrix}
				\delta_1
				\s 
				\delta_2
			\end{pmatrix} \cos(3kx)\right),
		\end{align*}
        where
		\begin{multline*}
			\begin{pmatrix}
				f_{10} - k^2 \, D_1 & f_{01}
				\s 
				g_{10} & g_{01} - k^2 \, D_2
			\end{pmatrix}\begin{pmatrix}
				\gamma_1
				\s 
				\gamma_2
			\end{pmatrix} = C_3\begin{pmatrix}
				\phi_1
				\s 
				\phi_2
			\end{pmatrix} - \phi_1 \left(\alpha_1 + \frac{\beta_1}{2}\right)\begin{pmatrix}
				f_{20}
				\s 
				g_{20}
			\end{pmatrix}
			\s 
			- \left(\phi_1 \left(\alpha_2 + \frac{\beta_2}{2}\right) + \phi_2 \left(\alpha_1 + \frac{\beta_1}{2}\right)\right)\begin{pmatrix}
				f_{11}
				\s 
				g_{11}
			\end{pmatrix}
			\s 
			- \phi_2 \left(\alpha_2 + \frac{\beta_2}{2}\right)\begin{pmatrix}
				f_{02}
				\s 
				g_{02}
			\end{pmatrix} - 3 \, \mbf B 
		\end{multline*}
        and
		\begin{multline*}
			\begin{pmatrix}
				f_{10} - 9 \, k^2 \, D_1 & f_{01}
				\s 
				g_{10} & g_{01} - 9 \, k^2 \, D_2
			\end{pmatrix}\begin{pmatrix}
				\delta_1
				\s 
				\delta_2
			\end{pmatrix} = - \frac{1}{2}\left(\phi_1 \, \beta_1\begin{pmatrix}
				f_{20}
				\s 
				g_{20}
			\end{pmatrix} + \phi_2 \, \beta_2\begin{pmatrix}
				f_{02}
				\s 
				g_{02}
			\end{pmatrix}\right.
			\s 
			\left.  + \left(\phi_1 \, \beta_2 + \phi_2 \, \beta_1\right)\begin{pmatrix}
				f_{11}
				\s 
				g_{11}
			\end{pmatrix}\right) - \mbf B.
		\end{multline*}		
        It is straightforward to find a unique solution for $(\delta_1, \delta_2)^T \in \mathbb R$. On the other hand, the system for $\gamma_1, \gamma_2$ has infinitely many solutions. In fact, if $\left(\gamma_1^*,\gamma_2^*\right)$ is a solution, then
        \begin{align*}
            \left(\gamma_1, \gamma_2\right) = \left(\gamma_1^*, \gamma_2^*\right) + \hat \gamma \left(\phi_1, \phi_2\right)
        \end{align*}
        is another solution for any $\hat \gamma\in \mathbb R$. In particular, we can choose $\hat \gamma$ such that $\left(\gamma_1, \gamma_2\right)\cdot \left(\phi_1, \phi_2\right) = 0$ since all the information in the direction of $\phi$ was already considered in the first-order equation.
		%	Therefore, we can see that
		%	
		%	\begin{multline*}
			%		\begin{pmatrix}
				%			\delta_1
				%			\s 
				%			\delta_2
				%		\end{pmatrix}=-\frac{1}{2d_3}\begin{pmatrix}
				%			g_{01}-9k^2D_2 & -f_{01}
				%			\s 
				%			-g_{10} & f_{10}-9k^2D_1
				%		\end{pmatrix}\left(\phi_1\beta_1\begin{pmatrix}
				%			f_{20}
				%			\s 
				%			g_{20}
				%		\end{pmatrix}+\phi_1\beta_2\begin{pmatrix}
				%			f_{11}
				%			\s 
				%			g_{11}
				%		\end{pmatrix}\right.
			%		\s 
			%		\left. +\phi_2\beta_1\begin{pmatrix}
				%			f_{11}
				%			\s 
				%			g_{11}
				%		\end{pmatrix} +\phi_2\beta_2\begin{pmatrix}
				%			f_{02}
				%			\s 
				%			g_{02}
				%		\end{pmatrix}+2\mbf B\right),
			%	\end{multline*}
		%	and the solution for the system of equations in variables $\gamma_1,\gamma_2$ can be chosen in any way such that it satisfies the system. In particular, one could choose $\gamma_1=0$ or $\gamma_2=0$, which will simplify the system.

    \paragraph{Order 4} At this order, equation \eqref{normalformeq} gives
		\begin{multline}
			\partial_A \mbf W^{[1]} \, h^{[4]} + \partial_A \mbf W^{[2]} \, h^{[3]} + \cancelto{0}{\partial_A \mbf W^{[3]} \, h^{[2]}} + \cancelto{0}{\partial_A \mbf W^{[4]} \, h^{[1]}}
			\s 
			= \left(\jac \mbf f(0, 0) + \mathbb D \, \partial_{xx}\right)\mbf W^{[4]} + \frac{u_2^2}{2!} \begin{pmatrix}
				f_{20}
				\s
				g_{20}
			\end{pmatrix} + \frac{2 \, u_2 \, v_2}{2!}\begin{pmatrix}
				f_{11}
				\s 
				g_{11}
			\end{pmatrix} + \frac{v_2^2}{2!}\begin{pmatrix}
				f_{02}
				\s 
				g_{02}
			\end{pmatrix}
			\s 
			+ \frac{2 \, u_1 \, u_3}{2!}\begin{pmatrix}
				f_{20}
				\s 
				g_{20}
			\end{pmatrix} + \frac{1}{2!}\left(2 \, u_1 \,v_3 + 2 \,u_3 \, v_1\right)\begin{pmatrix}
				f_{11}
				\s 
				g_{11}
			\end{pmatrix} + \frac{2 \, v_1 \, v_3}{2!}\begin{pmatrix}
				f_{02}
				\s 
				g_{02}
			\end{pmatrix}
			\s 
			+ \frac{3 \, u_1^2 \, u_2}{3!}\begin{pmatrix}
				f_{30}
				\s 
				g_{30}
			\end{pmatrix} + \frac{1}{3!}\left(3 \, u_1^2 \, v_2 + 3 \, \cdot 2 \, u_1 \, u_2 \, v_1\right)\begin{pmatrix}
				f_{21}
				\s 
				g_{21}
			\end{pmatrix}
			\s 
			+ \frac{1}{3!}\left(3 \cdot 2 \, u_1 \, v_1 \, v_2 + 3 \, u_2 \, v_1^2\right)\begin{pmatrix}
				f_{12}
				\s 
				g_{12}
			\end{pmatrix} + \frac{3 \, v_1^2 \, v_2}{3!}\begin{pmatrix}
				f_{03}
				\s 
				g_{03}
			\end{pmatrix} + \frac{u_1^4}{4!}\begin{pmatrix}
				f_{40}
				\s 
				g_{40}
			\end{pmatrix}
			\s 
			+ \frac{4 \, u_1^3 \, v_1}{4!}\begin{pmatrix}
				f_{31}
				\s 
				g_{31}
			\end{pmatrix} + \frac{6 \, u_1^2 \, v_1^2}{4!}\begin{pmatrix}
				f_{22}
				\s 
				g_{22}
			\end{pmatrix} + \frac{4 \, u_1 \, v_1^3}{4!}\begin{pmatrix}
				f_{13}
				\s 
				g_{13}
			\end{pmatrix} + \frac{v_1^4}{4!}\begin{pmatrix}
				f_{04}
				\s 
				g_{04}
			\end{pmatrix}. \label{fourthorder}
		\end{multline}
		In this case\evs{,} we will have no secular terms on the right-hand side of \eqref{fourthorder}, which lets us choose $h^{[4]} = 0$.
		\begin{multline}
			\begin{pmatrix}
				f_{10} + D_1 \, \partial_{xx} & f_{01}
				\s 
				g_{10} & g_{01} + D_2 \, \partial_{xx}
			\end{pmatrix}\mbf W^{[4]} = 2 \, A\left(\begin{pmatrix}
				\alpha_1
				\s 
				\alpha_2
			\end{pmatrix} + \begin{pmatrix}
				\beta_1
				\s 
				\beta_2
			\end{pmatrix}\cos(2kx)\right)h^{[3]}
			\s 
			- \frac{A^4}{2} \left(\alpha_1^2 + \frac{\beta_1^2}{2} + 2 \, \alpha_1 \, \beta_1 \, \cos(2kx) + \frac{\beta_1^2}{2} \, \cos(4kx)\right)\begin{pmatrix}
				f_{20}
				\s
				g_{20}
			\end{pmatrix}
			\s 
			- A^4 \left(\alpha_1 \, \alpha_2 + \frac{\beta_1 \, \beta_2}{2} + \left(\alpha_1 \, \beta_2 + \alpha_2 \, \beta_1\right)\cos(2kx) + \frac{\beta_1 \, \beta_2}{2} \cos(4kx)\right)\begin{pmatrix}
				f_{11}
				\s 
				g_{11}
			\end{pmatrix}
			\s 
			- \frac{A^4}{2}\left(\alpha_2^2 + \frac{\beta_2^2}{2} + 2 \, \alpha_2 \, \beta_2 \, \cos(2kx) + \frac{\beta_2^2}{2}\cos(4kx)\right)\begin{pmatrix}
				f_{02}
				\s 
				g_{02}
			\end{pmatrix}
			\s 
			- \frac{A^4\phi_1}{2} \left(\gamma_1 + \left(\gamma_1 + \delta_1\right) \cos(2kx) + \delta_1 \, \cos(4kx)\right)\begin{pmatrix}
				f_{20}
				\s 
				g_{20}
			\end{pmatrix}
			\s 
			- \frac{A^4 \, \phi_1}{2}\left(\gamma_2 + \left(\gamma_2 + \delta_2\right)\cos(2kx) + \delta_2 \, \cos(4kx)\right)\begin{pmatrix}
				f_{11}
				\s 
				g_{11}
			\end{pmatrix}
			\s 
			- \frac{A^4 \, \phi_2}{2}\left(\gamma_1 + \left(\gamma_1 + \delta_1\right)\cos(2kx) + \delta_1 \, \cos(4kx)\right)\begin{pmatrix}
				f_{11}
				\s 
				g_{11}
			\end{pmatrix}
			\s 
			- \frac{A^4 \, \phi_2}{2}\left(\gamma_2 + \left(\gamma_2 + \delta_2\right) \cos(2kx) + \delta_2 \, \cos(4kx)\right)\begin{pmatrix}
				f_{02}
				\s 
				g_{02}
			\end{pmatrix}
			\s 
			- \frac{A^4 \, \phi_1^2}{4}\left(\alpha_1 + \frac{\beta_1}{2} + \left(\alpha_1 + \beta_1\right)\cos(2kx) + \frac{\beta_1}{2}\cos(4kx)\right)\begin{pmatrix}
				f_{30}
				\s 
				g_{30}
			\end{pmatrix}
			\s 
			- \frac{A^4 \, \phi_1^2}{4}\left(\alpha_2 + \frac{\beta_2}{2} + \left(\alpha_2 + \beta_2\right)\cos(2kx) + \frac{\beta_2}{2} \, \cos(4kx)\right)\begin{pmatrix}
				f_{21}
				\s 
				g_{21}
			\end{pmatrix}
			\s
			- \frac{A^4 \, \phi_1 \, \phi_2}{2} \left(\alpha_1 + \frac{\beta_1}{2} + \left(\alpha_1 + \beta_1\right)\cos(2kx) + \frac{\beta_1}{2} \, \cos(4kx)\right) \begin{pmatrix}
				f_{21}
				\s 
				g_{21}
			\end{pmatrix}
			\s 
			- \frac{A^4 \, \phi_1 \, \phi_2}{2}\left(\alpha_2 + \frac{\beta_2}{2} + \left(\alpha_2 + \beta_2\right)\cos(2kx) + \frac{\beta_2}{2}\cos(4kx)\right)\begin{pmatrix}
				f_{12}
				\s 
				g_{12}
			\end{pmatrix}
			\s 
			- \frac{A^4 \, \phi_2^2}{4}\left(\alpha_1 + \frac{\beta_1}{2} + \left(\alpha_1 + \beta_1\right)\cos(2kx) + \frac{\beta_1}{2}\cos(4kx)\right)\begin{pmatrix}
				f_{12}
				\s 
				g_{12}
			\end{pmatrix}
			\s 
			- \frac{A^4 \, \phi_2^2}{4}\left(\alpha_2 + \frac{\beta_2}{2} + \left(\alpha_2 + \beta_2\right)\cos(2kx) + \frac{\beta_2}{2}\cos(4kx)\right)\begin{pmatrix}
				f_{03}
				\s 
				g_{03}
			\end{pmatrix}
			\s 
			- \frac{A^4}{8} \left(\frac{\phi_1^4}{4!}\begin{pmatrix}
				f_{40}
				\s 
				g_{40}
			\end{pmatrix} + \frac{\phi_1^3 \, \phi_2}{3!}\begin{pmatrix}
				f_{31}
				\s 
				g_{31}
			\end{pmatrix} + \frac{\phi_1^2 \, \phi_2^2}{4}\begin{pmatrix}
				f_{22}
				\s 
				g_{22}
			\end{pmatrix} + \frac{\phi_1 \, \phi_2^3}{3!}\begin{pmatrix}
				f_{13}
				\s 
				g_{13}
			\end{pmatrix}\right.
			\s 
			\left. + \frac{\phi_2^4}{4!}\begin{pmatrix}
				f_{04}
				\s 
				g_{04}
			\end{pmatrix}\right)\left(3 + 4 \, \cos(2kx) + \cos(4kx)\right). \label{W4}
		\end{multline}
		Again, this equation has infinitely many solutions since the operator on the left-hand side is non-invertible. Nonetheless, we are interested only in the solution that is orthogonal to $\ker\left(DF(0,0) + \mathbb D \partial_{xx}\right)$, \evs{to avoid the solution computed at previous orders. Thus, we find}
		\begin{align*}
			\mbf W^{[4]} = A^4\left(
			\begin{pmatrix}
				\zeta_1
				\s
				\zeta_2
			\end{pmatrix} + \begin{pmatrix}
				\theta_1
				\s 
				\theta_2
			\end{pmatrix}\cos(2kx) + \begin{pmatrix}
				\kappa_1
				\s 
				\kappa_2
			\end{pmatrix}\cos(4kx)
			\right).
		\end{align*}
		Therefore, as \eqref{W4} is a linear system of equations, we can use the superposition of solutions to find each pair of constants. In particular, when we replace the first term of $\mbf W^{[4]}$ into \eqref{W4}, we get the following system of equations:
  % HERE I AM
		\begin{multline*}
			\begin{pmatrix}
				f_{10} & f_{01}
				\s 
				g_{10} & g_{01}
			\end{pmatrix}\begin{pmatrix}
				\zeta_1
				\s 
				\zeta_2
			\end{pmatrix} = 2 \, C_3 \begin{pmatrix}
				\alpha_1
				\s 
				\alpha_2
			\end{pmatrix}
			\s 
			- \frac{1}{2} \left(\left(\alpha_1^2 + \frac{\beta_1^2}{2}\right) \begin{pmatrix}
				f_{20}
				\s 
				g_{20}
			\end{pmatrix} + 2 \, \left(\alpha_1 \, \alpha_2 + \frac{\beta_1 \, \beta_2}{2}\right) \begin{pmatrix}
				f_{11}
				\s 
				g_{11}
			\end{pmatrix} + \left(\alpha_2^2 + \frac{\beta_2^2}{2}\right)\begin{pmatrix}
				f_{02}
				\s 
				g_{02}
			\end{pmatrix}\right)
			\s 
			- \frac{1}{2} \left(\phi_1 \, \gamma_1\begin{pmatrix}
				f_{20}
				\s 
				g_{20}
			\end{pmatrix} + \left(\phi_1 \, \gamma_2 + \phi_2 \, \gamma_1\right)\begin{pmatrix}
				f_{11}
				\s 
				g_{11}
			\end{pmatrix} + \phi_2 \, \gamma_2\begin{pmatrix}
				f_{02}
				\s 
				g_{02}
			\end{pmatrix}\right)
			\s 
			- \frac{1}{4}\left(\phi_1^2\left(\alpha_1 + \frac{\beta_1}{2}\right)\begin{pmatrix}
				f_{30}
				\s 
				g_{30}
			\end{pmatrix} + \left(\phi_1^2 \left(\alpha_2 + \frac{\beta_2}{2}\right) + 2 \, \phi_1 \, \phi_2\left(\alpha_1 + \frac{\beta_1}{2}\right)\right)\begin{pmatrix}
				f_{21}
				\s 
				g_{21}
			\end{pmatrix}\right.
			\s 
			\left. + \left(2 \, \phi_1 \, \phi_2 \left(\alpha_2 + \frac{\beta_2}{2}\right) + \phi_2^2 \left(\alpha_1 + \frac{\beta_1}{2}\right)\right)\begin{pmatrix}
				f_{12}
				\s 
				g_{12}
			\end{pmatrix} + \phi_2^2\left(\alpha_2 + \frac{\beta_2}{2}\right)\begin{pmatrix}
				f_{03}
				\s 
				g_{03}
			\end{pmatrix}\right)
			\s 
			- 3 \, \mbf E,
		\end{multline*}
		where
		\begin{align*}
			\mbf E = \frac{1}{8}\left(\frac{\phi_1^4}{4!}\begin{pmatrix}
				f_{40}
				\s 
				g_{40}
			\end{pmatrix} + \frac{\phi_1^3 \, \phi_2}{3!}\begin{pmatrix}
				f_{31}
				\s 
				g_{31}
			\end{pmatrix} + \frac{\phi_1^2 \, \phi_2^2}{4}\begin{pmatrix}
				f_{22}
				\s 
				g_{22}
			\end{pmatrix} + \frac{\phi_1 \, \phi_2^3}{3!}\begin{pmatrix}
				f_{13}
				\s 
				g_{13}
			\end{pmatrix} + \frac{\phi_2^4}{4!}\begin{pmatrix}
				f_{04}
				\s 
				g_{04}
			\end{pmatrix}\right).
		\end{align*}
		On the other hand, if we replace the second term of $\mbf W^{[4]}$ into \eqref{W4}, we get the following system of equations:
		\begin{multline*}
			\begin{pmatrix}
				f_{10} - 4 \, k^2 \, D_1 & f_{01}
				\s 
				g_{10} & g_{01} - 4 \, k^2 \, D_2
			\end{pmatrix}\begin{pmatrix}
				\theta_1
				\s 
				\theta_2
			\end{pmatrix} = 2 \, C_3 \, \begin{pmatrix}
				\beta_1
				\s 
				\beta_2
			\end{pmatrix}
			\s 
			- \left(\alpha_1 \, \beta_1\begin{pmatrix}
				f_{20}
				\s 
				g_{20}
			\end{pmatrix} + \left(\alpha_1 \, \beta_2 + \alpha_2 \, \beta_1\right)\begin{pmatrix}
				f_{11}
				\s 
				g_{11}
			\end{pmatrix} + \alpha_2 \, \beta_2\begin{pmatrix}
				f_{02}
				\s 
				g_{02}
			\end{pmatrix}\right)
			\s 
			- \frac{1}{2}\left(\phi_1\left(\gamma_1 + \delta_1\right)\begin{pmatrix}
				f_{20}
				\s 
				g_{20}
			\end{pmatrix} + \phi_2 \left(\gamma_2 + \delta_2\right)\begin{pmatrix}
				f_{02}
				\s 
				g_{02}
			\end{pmatrix}\right.
			\s 
            \left. +\left(\phi_1\left(\gamma_2 + \delta_2\right) + \phi_2\left(\gamma_1 + \delta_1\right)\right)\begin{pmatrix}
				f_{11}
				\s 
				g_{11}
			\end{pmatrix}\right)
			\s 
			- \frac{1}{4}\left(\phi_1^2 \left(\alpha_1 + \beta_1\right)\begin{pmatrix}
				f_{30}
				\s 
				g_{30}
			\end{pmatrix} + \left(\phi_1^2 \left(\alpha_2 + \beta_2\right) + 2 \, \phi_1 \, \phi_2\left(\alpha_1 + \beta_1\right)\right)\begin{pmatrix}
				f_{21}
				\s 
				g_{21}
			\end{pmatrix}\right.
			\s 
			\left. + \left(2 \, \phi_1 \, \phi_2\left(\alpha_2 + \beta_2\right) + \phi_2^2 \left(\alpha_1 + \beta_1\right)\right)\begin{pmatrix}
				f_{12}
				\s 
				g_{12}
			\end{pmatrix} + \phi_2^2\left(\alpha_2 + \beta_2\right)\begin{pmatrix}
				f_{03}
				\s 
				g_{03}
			\end{pmatrix}\right) - 4 \, \mbf E.
		\end{multline*}
		Finally, when we replace the third term of $\mbf W^{[4]}$ into \eqref{W4}, we get the following system of equations:
		\begin{multline*}
			\begin{pmatrix}
				f_{10} - 16 \, k^2 \, D_1 & f_{01}
				\s 
				g_{10} & g_{01} - 16 \, k^2 \, D_2
			\end{pmatrix}\begin{pmatrix}
				\kappa_1
				\s 
				\kappa_2
			\end{pmatrix} =
			\s 
			- \frac{1}{4}\left(\beta_1^2\begin{pmatrix}
				f_{20}
				\s 
				g_{20}
			\end{pmatrix} + 2 \, \beta_1 \, \beta_2\begin{pmatrix}
				f_{11}
				\s 
				g_{11}
			\end{pmatrix} + \beta_2^2\begin{pmatrix}
				f_{02}
				\s 
				g_{02}
			\end{pmatrix}\right)
			\s 
			- \frac{1}{2}\left(\phi_1 \, \delta_1 \begin{pmatrix}
				f_{20}
				\s 
				g_{20}
			\end{pmatrix} + \left(\phi_1 \, \delta_2 + \phi_2 \, \delta_1\right)\begin{pmatrix}
				f_{11}
				\s 
				g_{11}
			\end{pmatrix} + \phi_2 \, \delta_2\begin{pmatrix}
				f_{02}
				\s 
				g_{02}
			\end{pmatrix}\right)
			\s 
			- \frac{1}{8}\left(\phi_1^2 \, \beta_1\begin{pmatrix}
				f_{30}
				\s 
				g_{30}
			\end{pmatrix} + \left(\phi_1^2 \, \beta_2 + 2 \, \phi_1 \, \phi_2 \, \beta_1\right)\begin{pmatrix}
				f_{21}
				\s 
				g_{21}
			\end{pmatrix}\right.
			\s 
			\left. + \left(2 \, \phi_1 \, \phi_2 \, \beta_2 + \phi_2^2 \, \beta_1\right)\begin{pmatrix}
				f_{12}
				\s 
				g_{12}
			\end{pmatrix} + \phi_2^2 \, \beta_2\begin{pmatrix}
				f_{03}
				\s 
				g_{30}
			\end{pmatrix}\right) - \mbf E.
		\end{multline*}
        We highlight that all of these equations have a unique solution. With this, we are ready to consider the last order we are interested in.

    \paragraph{Order 5} In this case, equation \eqref{normalformeq} is given by:
		\begin{multline}
			\partial_A \mbf W^{[1]} \, h^{[5]} + \cancelto{0}{\partial_A \mbf W^{[2]} \, h^{[4]}} + \partial_A \mbf W^{[3]} \, h^{[3]} + \cancelto{0}{\partial_A \mbf W^{[4]} \, h^{[2]}} + \cancelto{0}{\partial_A \mbf W^{[5]} \, h^{[1]}}
			\s 
			= \left(\jac \mbf f(0, 0) + \mathbb D \, \partial_{xx}\right)\mbf W^{[5]} + \frac{2 \, u_1 \, u_4}{2!}\begin{pmatrix}
				f_{20}
				\s 
				g_{20}
			\end{pmatrix} + \frac{1}{2!}\left(2 \, u_1 \, v_4 + 2 \, u_4 \, v_1\right)\begin{pmatrix}
				f_{11}
				\s 
				g_{11}
			\end{pmatrix}
			\s 
			+ \frac{2 \, v_1 \, v_4}{2!}\begin{pmatrix}
				f_{02}
				\s 
				g_{02}
			\end{pmatrix} + \frac{2 \, u_2 \, u_3}{2!}\begin{pmatrix}
				f_{20}
				\s 
				g_{20}
			\end{pmatrix} + \frac{1}{2!}\left(2 \, u_2 \, v_3 + 2 \, u_3 \, v_2\right)\begin{pmatrix}
				f_{11}
				\s 
				g_{11}
			\end{pmatrix} + \frac{2 \, v_2 \, v_3}{2!}\begin{pmatrix}
				f_{02}
				\s 
				g_{02}
			\end{pmatrix}
			\s 
			+ \frac{3 \, u_1 \, u_2^2}{3!}\begin{pmatrix}
				f_{30}
				\s 
				g_{30}
			\end{pmatrix} + \frac{3 \cdot 2 \, u_1 \, u_2 \, v_2}{3!}\begin{pmatrix}
				f_{21}
				\s 
				g_{21}
			\end{pmatrix} + \frac{3 \, u_1 \, v_2^2}{3!}\begin{pmatrix}
				f_{12}
				\s 
				g_{12}
			\end{pmatrix}
			\s 
			+ \frac{3 \, u_2^2 \, v_1}{3!}\begin{pmatrix}
				f_{21}
				\s 
				g_{21}
			\end{pmatrix} + \frac{3 \cdot 2 \, u_2 \, v_1 \, v_2}{3!}\begin{pmatrix}
				f_{12}
				\s 
				g_{12}
			\end{pmatrix} + \frac{3 \, v_1 \, v_2^2}{3!}\begin{pmatrix}
				f_{03}
				\s 
				g_{03}
			\end{pmatrix}
			\s 
			+ \frac{3 \, u_1^2 \, u_3}{3!}\begin{pmatrix}
				f_{30}
				\s 
				g_{30}
			\end{pmatrix} + \frac{1}{3!}\left(3 \, u_1^2 \, v_3 + 3 \cdot 2 \, u_1 \, u_3 \, v_1\right)\begin{pmatrix}
				f_{21}
				\s 
				g_{21}
			\end{pmatrix}
			\s 
			+ \frac{1}{3!}\left(3\cdot 2 u_1 \, v_1 \, v_3 + 3 \, u_3 \, v_1^2\right)\begin{pmatrix}
				f_{12}
				\s
				g_{12}
			\end{pmatrix} + \frac{3 \, v_1^2 \, v_3}{3!}\begin{pmatrix}
				f_{03}
				\s 
				g_{03}
			\end{pmatrix}
			\s 
			+ \frac{4 \, u_1^3 \, u_2}{4!}\begin{pmatrix}
				f_{40}
				\s 
				g_{40}
			\end{pmatrix} + \frac{1}{4!}\left(4 \, u_1^3 \, v_2 + 4 \cdot 3 \, u_1^2 \, u_2 \, v_1\right)\begin{pmatrix}
				f_{31}
				\s 
				g_{31}
			\end{pmatrix}
			\s 
			+ \frac{1}{4!}\left(6\cdot 2 \, u_1^2 \, v_1 \, v_2 + 6 \cdot 2 \, u_1 \, u_2 \, v_1^2\right)\begin{pmatrix}
				f_{22}
				\s 
				g_{22}
			\end{pmatrix} + \frac{1}{4!}\left(4\cdot 3 \, u_1 \, v_1^2 \, v_2 + 3 \, u_2 \, v_1^3\right)\begin{pmatrix}
				f_{13}
				\s 
				g_{13}
			\end{pmatrix}
			\s 
			+ \frac{4 \, v_1^3 \, v_2}{4!}\begin{pmatrix}
				f_{04}
				\s 
				g_{04}
			\end{pmatrix} + \frac{u_1^5}{5!}\begin{pmatrix}
				f_{50}
				\s 
				g_{50}
			\end{pmatrix} + \frac{5 \, u_1^4 \, v_1}{5!}\begin{pmatrix}
				f_{41}
				\s 
				g_{41}
			\end{pmatrix} + \frac{10 \, u_1^3 \, v_1^2}{5!}\begin{pmatrix}
				f_{32}
				\s 
				g_{32}
			\end{pmatrix} + \frac{10 \, u_1^2 \, v_1^3}{5!}\begin{pmatrix}
				f_{23}
				\s 
				g_{23}
			\end{pmatrix}
			\s 
			+ \frac{5 \, u_1 \, v_1^4}{5!}\begin{pmatrix}
				f_{14}
				\s 
				g_{14}
			\end{pmatrix} + \frac{v_1^5}{5!}\begin{pmatrix}
				f_{05}
				\s 
				g_{05}
			\end{pmatrix}. \label{fifthorder}
		\end{multline}
		Next, note that \eqref{fifthorder} can be written as:
		\begin{multline*}
			\left(\jac \mbf f(0, 0) + \mathbb D \, \partial_{xx}\right)\mbf W^{[5]} = \partial_A \mbf W^{[1]} \, h^{[5]} + \partial_A \mbf W^{[3]} \, h^{[3]} - u_1 \, u_4\begin{pmatrix}
				f_{20}
				\s 
				g_{20}
			\end{pmatrix}
			\s 
			- \left(u_1 \, v_4 + u_4 \, v_1\right)\begin{pmatrix}
				f_{11}
				\s 
				g_{11}
			\end{pmatrix} - v_1 \, v_4\begin{pmatrix}
				f_{02}
				\s 
				g_{02}
			\end{pmatrix} - u_2 \, u_3\begin{pmatrix}
				f_{20}
				\s 
				g_{20}
			\end{pmatrix} - \left(u_2 \, v_3 + u_3 \, v_2\right)\begin{pmatrix}
				f_{11}
				\s 
				g_{11}
			\end{pmatrix}
			\s 
			- v_2 \, v_3\begin{pmatrix}
				f_{02}
				\s 
				g_{02}
			\end{pmatrix} - \frac{u_1 \, u_2^2}{2}\begin{pmatrix}
				f_{30}
				\s 
				g_{30}
			\end{pmatrix} - u_1 \, u_2 \, v_2\begin{pmatrix}
				f_{21}
				\s 
				g_{21}
			\end{pmatrix} - \frac{u_1 \, v_2^2}{2}\begin{pmatrix}
				f_{12}
				\s 
				g_{12}
			\end{pmatrix} - \frac{u_2^2 \, v_1}{2}\begin{pmatrix}
				f_{21}
				\s 
				g_{21}
			\end{pmatrix} - u_2 \, v_1 \, v_2\begin{pmatrix}
				f_{12}
				\s 
				g_{12}
			\end{pmatrix}
			\s 
			- \frac{v_1 \, v_2^2}{2}\begin{pmatrix}
				f_{03}
				\s 
				g_{03}
			\end{pmatrix} - \frac{u_1^2 \, u_3}{2}\begin{pmatrix}
				f_{30}
				\s 
				g_{30}
			\end{pmatrix} - \left(\frac{u_1^2 \, v_3}{2} + u_1 \, u_3 \, v_1\right)\begin{pmatrix}
				f_{21}
				\s 
				g_{21}
			\end{pmatrix} - \left(u_1 \, v_1 \, v_3 + \frac{u_3 \, v_1^2}{2}\right)\begin{pmatrix}
				f_{12}
				\s
				g_{12}
			\end{pmatrix}
			\s 
			- \frac{v_1^2 \, v_3}{2}\begin{pmatrix}
				f_{03}
				\s 
				g_{03}
			\end{pmatrix} - \frac{u_1^3 \, u_2}{3!}\begin{pmatrix}
				f_{40}
				\s 
				g_{40}
			\end{pmatrix} - \left(\frac{u_1^3 \, v_2}{3!} + \frac{u_1^2 \, u_2 \, v_1}{2}\right)\begin{pmatrix}
				f_{31}
				\s 
				g_{31}
			\end{pmatrix}
			\s 
			- \left(\frac{u_1^2 \, v_1 \, v_2}{2} + \frac{u_1 \, u_2 \, v_1^2}{2}\right)\begin{pmatrix}
				f_{22}
				\s 
				g_{22}
			\end{pmatrix} - \left(\frac{u_1 \, v_1^2 \, v_2}{2} + \frac{u_2 \, v_1^3}{3!}\right)\begin{pmatrix}
				f_{13}
				\s 
				g_{13}
			\end{pmatrix} - \frac{v_1^3 \, v_2}{3!}\begin{pmatrix}
				f_{04}
				\s 
				g_{04}
			\end{pmatrix}
			\s 
			- \frac{u_1^5}{5!}\begin{pmatrix}
				f_{50}
				\s 
				g_{50}
			\end{pmatrix} - \frac{u_1^4 \, v_1}{4!}\begin{pmatrix}
				f_{41}
				\s 
				g_{41}
			\end{pmatrix} - \frac{2 \, u_1^3 \, v_1^2}{4!}\begin{pmatrix}
				f_{32}
				\s 
				g_{32}
			\end{pmatrix} - \frac{2 \, u_1^2 \, v_1^3}{4!}\begin{pmatrix}
				f_{23}
				\s 
				g_{23}
			\end{pmatrix} - \frac{u_1 \, v_1^4}{4!}\begin{pmatrix}
				f_{14}
				\s 
				g_{14}
			\end{pmatrix} - \frac{v_1^5}{5!}\begin{pmatrix}
				f_{05}
				\s 
				g_{05}
			\end{pmatrix},
		\end{multline*}

        With this, as the  right-hand side of this equation may have secular terms, we need to use the Fredholm alternative with the same vector $\bs \psi$ used to compute the coefficient $C_3$. In particular, we have that
		\begin{multline*}
			\left \langle \bs \psi, \left(\jac \mbf f(0, 0) + \mathbb D \, \partial_{xx}\right)\mbf W^{[5]} \right \rangle = \begin{pmatrix}
				\psi_1
				\s 
				\psi_2
			\end{pmatrix}\cdot\begin{pmatrix}
				\phi_1
				\s 
				\phi_2
			\end{pmatrix} \, h^{[5]} + 3 \, A^2 \begin{pmatrix}
				\psi_1
				\s 
				\psi_2
			\end{pmatrix}\cdot \begin{pmatrix}
				\gamma_1
				\s 
				\gamma_2
			\end{pmatrix} \, h^{[3]}
			\s 
			- A^5 \begin{pmatrix}
				\psi_1
				\s 
				\psi_2
			\end{pmatrix}\cdot \left(\phi_1\left(\zeta_1 + \frac{\theta_1}{2}\right)\begin{pmatrix}
				f_{20}
				\s 
				g_{20}
			\end{pmatrix}+\phi_2\left(\zeta_2 + \frac{\theta_2}{2}\right)\begin{pmatrix}
				f_{02}
				\s 
				g_{02}
			\end{pmatrix}\right.
			\s 
			\left. + \left(\phi_1\left(\zeta_2 + \frac{\theta_2}{2}\right) + \phi_2\left(\zeta_1 + \frac{\theta_1}{2}\right)\right)\begin{pmatrix}
				f_{11}
				\s 
				g_{11}
			\end{pmatrix}\right)
			\s 
			- A^5\begin{pmatrix}
				\psi_1
				\s
				\psi_2
			\end{pmatrix}\cdot \left(\left(\alpha_1 \, \gamma_1 + \frac{\beta_1 \, \gamma_1}{2}+\frac{\beta_1 \, \delta_1}{2}\right)\begin{pmatrix}
				f_{20}
				\s 
				g_{20}
			\end{pmatrix} + \left(\alpha_1 \, \gamma_2 + \frac{\beta_1 \, \gamma_2}{2} + \frac{\beta_1 \, \delta_2}{2}\right)\begin{pmatrix}
				f_{11}
				\s 
				g_{11}
			\end{pmatrix}\right.
			\s 
			\left. + \left(\alpha_2 \, \gamma_1 + \frac{\beta_2 \, \gamma_1}{2} + \frac{\beta_2 \, \delta_1}{2}\right)\begin{pmatrix}
				f_{11}
				\s 
				g_{11}
			\end{pmatrix} + \left(\alpha_2 \, \gamma_2 + \frac{\beta_2 \, \gamma_2}{2} + \frac{\beta_2 \, \delta_2}{2}\right)\begin{pmatrix}
				f_{02}
				\s 
				g_{02}
			\end{pmatrix}\right)
			\s
			- \frac{A^5}{2}\begin{pmatrix}
				\psi_1
				\s 
				\psi_2
			\end{pmatrix}\cdot \left(\phi_1 \left(\alpha_1^2 + \alpha_1 \, \beta_1 + \frac{\beta_1^2}{2}\right)\begin{pmatrix}
				f_{30}
				\s 
				g_{30}
			\end{pmatrix}\right.
			\s 
			\left. + 2 \, \phi_1\left(\alpha_1 \, \alpha_2 + \frac{\alpha_1 \, \beta_2 + \alpha_2 \, \beta_1}{2} + \frac{\beta_1 \, \beta_2}{2}\right) \begin{pmatrix}
				f_{21}
				\s 
				g_{21}
			\end{pmatrix} + \phi_1 \left(\alpha_2^2 + \alpha_2 \, \beta_2 + \frac{\beta_2^2}{2}\right)\begin{pmatrix}
				f_{12}
				\s 
				g_{12}
			\end{pmatrix}\right.
			\s 
			\left. + \phi_2\left(\alpha_1^2 + \alpha_1 \, \beta_1 + \frac{\beta_1^2}{2}\right)\begin{pmatrix}
				f_{21}
				\s 
				g_{21}
			\end{pmatrix} + 2 \, \phi_2 \left(\alpha_1 \, \alpha_2 + \frac{\alpha_1 \, \beta_2 + \alpha_2 \, \beta_1}{2} + \frac{\beta_1 \, \beta_2}{2}\right)\begin{pmatrix}
				f_{12}
				\s 
				g_{12}
			\end{pmatrix} \right.
			\s 
			\left. + \phi_2\left(\alpha_2^2 + \alpha_2 \, \beta_2 + \frac{\beta_2^2}{2}\right)\begin{pmatrix}
				f_{03}
				\s 
				g_{03}
			\end{pmatrix} \right)
			\s 
			- \frac{A^5}{8}\begin{pmatrix}
				\psi_1
				\s 
				\psi_2
			\end{pmatrix}\cdot \left(\phi_1^2 \left(3 \, \gamma_1 + \delta_1\right)\begin{pmatrix}
				f_{30}
				\s 
				g_{30}
			\end{pmatrix} + \left(\phi_1^2 \left(3 \, \gamma_2 + \delta_2\right) + 2 \, \phi_1 \, \phi_2\left(3 \, \gamma_1 + \delta_1\right)\right) \begin{pmatrix}
				f_{21}
				\s 
				g_{21}
			\end{pmatrix}\right.
			\s 
			\left. + \left(2 \, \phi_1 \, \phi_2 \left(3 \, \gamma_2 + \delta_2\right) + \phi_2^2\left(3 \, \gamma_1 + \delta_1\right)\right)\begin{pmatrix}
				f_{12}
				\s
				g_{12}
			\end{pmatrix} + \phi_2^2\left(3 \, \gamma_2 + \delta_2\right)\begin{pmatrix}
				f_{03}
				\s 
				g_{03}
			\end{pmatrix}  \right)
			\s 
			- \frac{A^5}{4!}\begin{pmatrix}
				\psi_1
				\s 
				\psi_2
			\end{pmatrix}\cdot \left(\phi_1^3 \left(3 \, \alpha_1 + 2 \, \beta_1\right)\begin{pmatrix}
				f_{40}
				\s 
				g_{40}
			\end{pmatrix} + \left(\phi_1^3 \left(3 \, \alpha_2 + 2 \, \beta_2\right) + 3 \, \phi_1^2 \, \phi_2\left(3 \, \alpha_1 + 2 \, \beta_1\right)\right)\begin{pmatrix}
				f_{31}
				\s 
				g_{31}
			\end{pmatrix}\right.
			\s 
			\left. + \left(3 \, \phi_1^2 \, \phi_2 \left(3 \, \alpha_2 + 2 \, \beta_2\right) + 3 \, \phi_1 \, \phi_2^2\left(3 \, \alpha_1 + 2 \, \beta_1\right)\right)\begin{pmatrix}
				f_{22}
				\s 
				g_{22}
			\end{pmatrix}\right.
			\s 
			\left. + \left(3 \, \phi_1 \, \phi_2^2 \left(3 \, \alpha_2 + 2 \, \beta_2\right) + \phi_2^3\left(3 \, \alpha_1 + 2 \, \beta_1\right)\right)\begin{pmatrix}
				f_{13}
				\s 
				g_{13}
			\end{pmatrix} + \phi_2^3 \left(3 \, \alpha_2 + 2 \, \beta_2\right) \begin{pmatrix}
				f_{04}
				\s 
				g_{04}
			\end{pmatrix}\right)
			\s 
			- 10 \, A^5\begin{pmatrix}
				\psi_1
				\s 
				\psi_2
			\end{pmatrix}\cdot \mbf F,
		\end{multline*}
		where
		\begin{multline*}
			\mbf F = \frac{1}{16}\left(\frac{\phi_1^5}{5!}\begin{pmatrix}
				f_{50}
				\s 
				g_{50}
			\end{pmatrix} + \frac{\phi_1^4 \, \phi_2}{4!}\begin{pmatrix}
				f_{41}
				\s 
				g_{41}
			\end{pmatrix} + \frac{2 \, \phi_1^3 \, \phi_2^2}{4!}\begin{pmatrix}
				f_{32}
				\s 
				g_{32}
			\end{pmatrix}\right.
			\s 
			\left. + \frac{2 \, \phi_1^2 \, \phi_2^3}{4!}\begin{pmatrix}
				f_{23}
				\s 
				g_{23}
			\end{pmatrix} + \frac{\phi_1 \, \phi_2^4}{4!}\begin{pmatrix}
				f_{14}
				\s 
				g_{14}
			\end{pmatrix} + \frac{\phi_2^5}{5!}\begin{pmatrix}
				f_{05}
				\s 
				g_{05}
			\end{pmatrix}\right).
		\end{multline*}
        Setting the above expression to zero, we get 
		\begin{align*}
			h^{[5]}(A) = C_5 \, A^5,
		\end{align*}
		where
		\begin{multline*}
			C_5 = - 3 \, \frac{\psi_1 \, \gamma_1 + \psi_2 \, \gamma_2}{\psi_1 \, \phi_1 + \psi_2 \, \phi_2} \, C_3
			\s 
			+ \frac{1}{\psi_1 \, \phi_1 + \psi_2 \, \phi_2}\begin{pmatrix}
				\psi_1
				\s 
				\psi_2
			\end{pmatrix}\cdot \left(\phi_1\left(\zeta_1 + \frac{\theta_1}{2}\right)\begin{pmatrix}
				f_{20}
				\s 
				g_{20}
			\end{pmatrix} + \phi_2\left(\zeta_2 + \frac{\theta_2}{2}\right)\begin{pmatrix}
				f_{02}
				\s 
				g_{02}
			\end{pmatrix}\right.
			\s 
			\left. +\left(\phi_1\left(\zeta_2 + \frac{\theta_2}{2}\right) + \phi_2\left(\zeta_1 + \frac{\theta_1}{2}\right)\right)\begin{pmatrix}
				f_{11}
				\s 
				g_{11}
			\end{pmatrix}\right)
			\s 
			+ \frac{1}{\psi_1 \, \phi_1 + \psi_2 \, \phi_2}\begin{pmatrix}
				\psi_1
				\s
				\psi_2
			\end{pmatrix}\cdot \left(\left(\alpha_1 \, \gamma_1 + \frac{\beta_1 \, \gamma_1}{2} + \frac{\beta_1 \, \delta_1}{2}\right)\begin{pmatrix}
				f_{20}
				\s 
				g_{20}
			\end{pmatrix} + \left(\alpha_1 \, \gamma_2 + \frac{\beta_1 \, \gamma_2}{2} + \frac{\beta_1 \, \delta_2}{2}\right)\begin{pmatrix}
				f_{11}
				\s 
				g_{11}
			\end{pmatrix}\right.
			\s 
			\left. + \left(\alpha_2 \, \gamma_1 + \frac{\beta_2 \, \gamma_1}{2} + \frac{\beta_2 \, \delta_1}{2}\right)\begin{pmatrix}
				f_{11}
				\s 
				g_{11}
			\end{pmatrix} + \left(\alpha_2 \, \gamma_2 + \frac{\beta_2 \, \gamma_2}{2} + \frac{\beta_2 \, \delta_2}{2}\right)\begin{pmatrix}
				f_{02}
				\s 
				g_{02}
			\end{pmatrix} \right)
			\s
			+ \frac{1}{2\left(\psi_1 \, \phi_1 + \psi_2 \, \phi_2\right)}\begin{pmatrix}
				\psi_1
				\s 
				\psi_2
			\end{pmatrix}\cdot \left(\phi_1 \left(\alpha_1^2 + \alpha_1 \, \beta_1 + \frac{\beta_1^2}{2}\right)\begin{pmatrix}
				f_{30}
				\s 
				g_{30}
			\end{pmatrix}\right.
			\s 
			\left. + 2 \, \phi_1\left(\alpha_1 \, \alpha_2 + \frac{\alpha_1 \, \beta_2 + \alpha_2 \, \beta_1}{2} + \frac{\beta_1 \, \beta_2}{2}\right) \begin{pmatrix}
				f_{21}
				\s 
				g_{21}
			\end{pmatrix} + \phi_1 \left(\alpha_2^2 + \alpha_2 \, \beta_2 + \frac{\beta_2^2}{2}\right)\begin{pmatrix}
				f_{12}
				\s 
				g_{12}
			\end{pmatrix}\right.
			\s 
			\left. + \phi_2\left(\alpha_1^2 + \alpha_1 \, \beta_1 + \frac{\beta_1^2}{2}\right)\begin{pmatrix}
				f_{21}
				\s 
				g_{21}
			\end{pmatrix} + 2 \, \phi_2\left(\alpha_1 \, \alpha_2 + \frac{\alpha_1 \, \beta_2 + \alpha_2 \, \beta_1}{2} + \frac{\beta_1 \, \beta_2}{2}\right)\begin{pmatrix}
				f_{12}
				\s 
				g_{12}
			\end{pmatrix} \right.
			\s 
			\left. +\phi_2\left(\alpha_2^2 + \alpha_2 \, \beta_2 + \frac{\beta_2^2}{2}\right)\begin{pmatrix}
				f_{03}
				\s 
				g_{03}
			\end{pmatrix}\right)
			\s 
			+ \frac{1}{8 \, \left(\psi_1 \, \phi_1 + \psi_2 \, \phi_2\right)}\begin{pmatrix}
				\psi_1
				\s 
				\psi_2
			\end{pmatrix}\cdot \left(\phi_1^2 \left(3 \, \gamma_1 + \delta_1\right)\begin{pmatrix}
				f_{30}
				\s 
				g_{30}
			\end{pmatrix}\right.
			\s 
			\left. + \left(\phi_1^2 \left(3 \, \gamma_2 + \delta_2\right) + 2 \, \phi_1 \, \phi_2\left(3 \, \gamma_1 + \delta_1\right)\right) \begin{pmatrix}
				f_{21}
				\s 
				g_{21}
			\end{pmatrix}\right.
			\s 
			\left. + \left(2 \, \phi_1 \, \phi_2 \left(3 \, \gamma_2 + \delta_2\right) + \phi_2^2\left(3 \, \gamma_1 + \delta_1\right)\right)\begin{pmatrix}
				f_{12}
				\s
				g_{12}
			\end{pmatrix} + \phi_2^2\left(3 \, \gamma_2 + \delta_2\right)\begin{pmatrix}
				f_{03}
				\s 
				g_{03}
			\end{pmatrix}\right)
			\s 
			+ \frac{1}{4! \, \left(\psi_1 \, \phi_1 + \psi_2 \, \phi_2\right)}\begin{pmatrix}
				\psi_1
				\s 
				\psi_2
			\end{pmatrix}\cdot \left(\phi_1^3 \left(3 \, \alpha_1 + 2 \, \beta_1\right)\begin{pmatrix}
				f_{40}
				\s 
				g_{40}
			\end{pmatrix} \right.
			\s 
			\left. + \left(\phi_1^3 \left(3 \, \alpha_2 + 2 \, \beta_2\right) + 3 \, \phi_1^2 \, \phi_2\left(3 \, \alpha_1 + 2 \, \beta_1\right)\right)\begin{pmatrix}
				f_{31}
				\s 
				g_{31}
			\end{pmatrix} \right.
			\s 
			\left. +\left(3 \, \phi_1^2 \, \phi_2 \left(3 \, \alpha_2 + 2 \, \beta_2\right) + 3 \, \phi_1 \, \phi_2^2\left(3 \, \alpha_1 + 2 \, \beta_1\right)\right)\begin{pmatrix}
				f_{22}
				\s 
				g_{22}
			\end{pmatrix}\right.
			\s 
			\left. +\left(3 \, \phi_1 \, \phi_2^2\left(3 \, \alpha_2 + 2 \, \beta_2\right) + \phi_2^3\left(3 \, \alpha_1 + 2 \, \beta_1\right)\right)\begin{pmatrix}
				f_{13}
				\s 
				g_{13}
			\end{pmatrix} + \phi_2^3 \left(3 \, \alpha_2 + 2 \, \beta_2\right) \begin{pmatrix}
				f_{04}
				\s 
				g_{04}
			\end{pmatrix} \right)
			\s 
			+ \frac{10}{\psi_1 \, \phi_1 + \psi_2 \, \phi_2}\begin{pmatrix}
				\psi_1
				\s 
				\psi_2
			\end{pmatrix}\cdot \mbf F.
		\end{multline*}
    \paragraph{Remark.} Just as in the case of $C_3$, different choices of $\hat \alpha$ or $\hat \beta$ may change the expression of $C_5$. In particular, if $\hat \alpha\neq 1$, then $C_5$ becomes $\hat \alpha^4 C_5$, whilst \evs{the expression is unaffected by the choice of $\hat \beta$}. Note\evs{,} therefore\evs{,} that these changes will not induce a change of sign of $C_5$.
\end{document}